\documentclass[aps,prd,showpacs,twocolumn,
superscriptaddress]{revtex4-1}

\usepackage{graphicx}
\usepackage{dcolumn}

\usepackage{amsmath}
\usepackage{amssymb}
\usepackage{bm}
\usepackage{color}
\usepackage[normalem]{ulem}
\usepackage[dvipsnames]{xcolor}
\usepackage{hyperref}
\hypersetup{
colorlinks=true, 
linkcolor=blue,  
citecolor=cyan,  
}

\newcommand{\bea}{\begin{eqnarray}}
\newcommand{\eea}{\end{eqnarray}}
\newcommand{\be}{\begin{equation}}
\newcommand{\ee}{\end{equation}}

\begin{document}
\title{
Black hole mimicker hiding in the shadow: Optical properties of the $\gamma$-metric 
}

\author{Askar B. Abdikamalov}
\email{aaskar17@fudan.edu.cn}
\affiliation{Center for Field Theory and Particle Physics and Department of Physics, Fudan University, 200438 Shanghai, China }

\author{Ahmadjon A. Abdujabbarov}
\email{ahmadjon@astrin.uz}
\affiliation{Center for Field Theory and Particle Physics and Department of Physics, Fudan University, 200438 Shanghai, China }
\affiliation{Ulugh Beg Astronomical Institute, Astronomicheskaya 33,
	Tashkent 100052, Uzbekistan }

\author{Dimitry Ayzenberg}
\email{dimitry@fudan.edu.cn}
\affiliation{Center for Field Theory and Particle Physics and Department of Physics, Fudan University, 200438 Shanghai, China }

\author{Daniele~Malafarina}
\email{daniele.malafarina@nu.edu.kz}
\affiliation{Department of Physics, Nazarbayev University, 53 Kabanbay Batyr avenue, 010000 Astana, Kazakhstan }

\author{Cosimo Bambi}
\email{bambi@fudan.edu.cn}
\affiliation{Center for Field Theory and Particle Physics and Department of Physics, Fudan University, 200438 Shanghai, China }

\author{Bobomurat Ahmedov}
\email{ahmedov@astrin.uz}
\affiliation{Ulugh Beg Astronomical Institute, Astronomicheskaya 33,
	Tashkent 100052, Uzbekistan }
\affiliation{National University of Uzbekistan, Tashkent 100174, Uzbekistan}

\date{\today}
\begin{abstract}
Can the observation of the `shadow' allow us to distinguish a black hole from a more exotic compact object? We study the motion of photons in a class of vacuum static axially-symmetric space-times that is continuously linked to the Schwarzschild metric through the value of one parameter that can be interpreted as a measure of the deformation of the source. 
We investigate the lensing effect and shadow produced by the source with the aim of comparing the expected image with the shadow of a Schwarzschild black hole. 
In the context of astrophysical black holes we found that it may not be possible to distinguish an exotic source with small deformation parameter from a black hole. However, as the deformation increases noticeable effects arise. Therefore, the future more precise measurement of the shadow of astrophysical black hole candidates would in principle allow to put constraints on the deviation of the object from spherical symmetry.

\end{abstract}


\maketitle

\section{Introduction}

It is generally believed that astrophysical black holes are well described by the Schwarzschild and Kerr space-times. However, this idea, often referred to as the {\em Kerr hypothesis}, has not yet been substantially supported by sufficient experimental evidence
\cite{Bambi17c,Berti15,Cardoso16a,Cardoso17,Krawczynski12, Krawczynski18, Yagi16, Bambi16b, Yunes13, Tripathi19}. 

Such a conviction will become robust only after well constrained observations of the gravitational field outside black hole candidates are put to the test of predictions derived from space-times that differ from the Kerr and Schwarzschild black holes.

Until now, all experimental tests of gravity (even in the `vicinity' of black hole candidates such as in Ref.~\cite{Gravity18a}) have not allowed one to ``measure'' the metric coefficients to accurately test the Kerr hypothesis. However the state of the art of black hole measurements is changing rapidly. Recently the first image of the 'shadow' of the supermassive black hole candidate in the galaxy M87 was released by the Event Horizon Telescope collaboration \cite{EHT19a,EHT19b}. The image shows the distinctive features of a black hole showing an inner edge for the accretion disk and suggesting the existence of an infinitely red-shifted surface.

On the other hand, it is indeed possible that, when other more exotic solutions are considered, some degeneracy occurs for some specific value of the parameters involved, in such a way that the exotic solution can mimic the black hole in astrophysical observations
\cite{Carballo-Rubio18}.

To this aim it is important to consider physically viable solutions that exhibit small deviations from black hole solutions. For example, investigations of the observational features of a space-time metric which slightly deviates from the Kerr solution have been mostly considered in the context of modified theories of gravity
(see for example, \cite{Takahashi05, Bambi09, Hioki09, Amarilla10, Bambi10, Amarilla12, Amarilla13, Abdujabbarov13c, Atamurotov13, Wei13, Atamurotov13b, Bambi15, Ghasemi-Nodehi15, Cunha15, Javed19, Ovgun19,Ovgun19a}) or perturbations of the Kerr metric
\cite{Johannsen11,Konoplya16,Younsi16}. The observables of the compact objects, particularly the shadow of the compact objects in modified or alternative theories of gravity have been studied in Refs.~\cite{Abdujabbarov15, Atamurotov15a, Ohgami15, Grenzebach15, Mureika17, Abdujabbarov17b, Abdujabbarov16a, Abdujabbarov16b, Mizuno18, Shaikh18b, Kogan17, Perlick17,Schee15, Schee09a, Stuchlik14, Schee09, Stuchlik10,Mishra19,Eiroa18, Giddings19}.

As of now, there is no experimental evidence in support of the need for modifications of Einstein's theory 
\cite{Bambi17c,Berti15,Cardoso16a,Cardoso17}.
Also, the relativistic space-times that are generally considered as perturbations of the Kerr black hole are not exact solutions of Einstein's vacuum field equations, nor solutions in the presence of reasonable, physically realistic, matter fields. Therefore the interpretation of the parameters that define the departure from the black hole case is not straightforward.

Indeed, it would be preferable to study black hole mimickers obtained within General Relativity from exact solutions of Einstein's equations which have a clear physical interpretation. For example, in \cite{Bambi13d}, two of us considered the observational features of solutions in the presence of matter fields and found that light emitted from accretion disks in such space-times could mimic the expected behaviour of light from accretion disks around a Kerr black hole. Similarly, in \cite{Ilyas17}, two of us considered the features of accretion disks around a Kerr black hole as compared with accretion disks immersed in a non vacuum rotating massive source.

In the present article we consider the optical properties for distant observers of a well known vacuum solution of Einstein's equations which has particular interest because it describes the gravitational field outside a static deformed body. The solution, which depends on two parameters $m$ and $\gamma$, describing the mass and deformation of the source respectively, was originally found by Zipoy and Vorhees \cite{Zipoy66,Voorhees70} and it is often referred to as the `$\gamma$-metric'. In the context of static axially symmetric space-times, there are two solutions that play an important role because of their connection with spherical symmetry: The $\gamma$-metric and the Erez-Rosen metric \cite{Erez59}. Motion in the Erez-Rosen space-time has been studied in \cite{Bini13}, while modifications of both solutions in the presence of a scalar field were studied in \cite{Turimov18a}. The Erez-Rosen solution is characterized by having only two non vanishing multipole moments, namely the monopole $M$ and quadrupole $Q$, and it reduces to Schwarzschild for $Q=0$, while in the case of the $\gamma$-metric all even multipole moments are non vanishing and depend only on $m$ and $\gamma$. The main interest for this solution comes from the fact that the parameter $\gamma$ can take any positive value and the line-element reduces to the Schwarzschild line-element in Schwarzschild coordinates for $\gamma=1$. One peculiar feature of the metric is that for $\gamma\neq 1$ it posses a curvature singularity at the surface $r=2m$ where in the Schwarzschild metric the event horizon is located. 

We can understand the singularity of the $\gamma$-metric in the context of the regime where quantum modifications to General Relativity can be expected to become important. For black holes, it is generally believed that such effects should become non negligible only at the Planck scale and thus must remain confined within the horizon. However, there is no strong justification for such belief. The `Planck scale' (be it energy, length or density) arises only from geometric arguments involving fundamental constants and there is no physical guarantee that quantum-gravity effects must not appear at other scales
\cite{Zenczykowski18}.
In fact, it has become clear in recent times, studying dynamical solutions leading to the formation of black holes, that one can not affect the behaviour of collapse close to the Planck regime (i.e. close to the classical singularity) without affecting the structure of the horizon as well
\cite{Malafarina17}.
For example, in \cite{Bambi13f} it was shown that the resolution of the singularity within the simple Oppenheimer-Snyder-Datt model, leads to a modification of the trapped surface which must affect the exterior space-time.
In this sense, it is reasonable to consider the possibility that quantum effects may become important close to the surface $r=2m$ for all values of $\gamma$, and consequently close to the horizon for $\gamma=1$, with the classical solutions being valid for $r>2m$. 

The properties of the $\gamma$-metric were studied in
\cite{Hernandez-Pastora94,Herrera99,Quevedo11} while the peculiar structure of the singular surface was investigated in \cite{Virbhadra96}.
The geodesics for test particles were considered in \cite{Herrera99, Boshkayev15}, while the properties of accretion disks were first investigated in \cite{Chowdhury12}.
Interior solutions for the $\gamma$-metric have been studied in \cite{Hernandez67,Stewart82,Herrera05}. And more recently the motion of charged particles in the $\gamma$ space-time immersed in an external magnetic field was considered in our preceding paper~\cite{Benavides-Gallego18}.

{In the present article we investigate the lensing and `shadow' properties of the $\gamma$-metric and show that large departures from spherical symmetry would lead to differences from the Schwarzschild and Kerr cases that would be measurable, at least in principle, from distant observers. On the other hand, small departures from spherical symmetry may be indistinguishable unless one is able to precisely measure the metric coefficients in the vicinity of $r=2m$, which is at present beyond our experimental capabilities.}

The paper is organized as follows:
Sect.~\ref{photmotion} is devoted to briefly review of the motion of massive and massless particles in the $\gamma$ space-time and construction of the ray tracing algorithm necessary to investigate the shadow. 
Sect~\ref{rtcs} describes the ray-tracing code used to construct the shadow of the $\gamma$-metric.
In Sect~\ref{shadow},
we consider the shadow cast by the $\gamma$ space-time for observer at infinity, while in Sect.~\ref{lensing}, we study strong lensing effects.
Finally, in Sect.~\ref{Summary} we summarize
the obtained results and discuss the possibility of distinguishing such a source from a Schwarzschild black hole via astrophysical observations.
Throughout the paper we use a space-like
signature $(-,+,+,+)$, a system of units
in which $G = c = 1$, 
and we restore them when we need to
compare our results with observational data.
Greek indices run from $0$ to $3$, Latin indices from $1$ to $3$.

\section{Photon motion \label{photmotion}}

The $\gamma$-metric is a static axially-symmetric vacuum solution of Einstein's equations belonging to Weyl's class which can be written in Erez-Rosen \cite{Erez59} coordinates as
\begin{eqnarray}\label{metric}
ds^2&=&- F dt^2 \nonumber\\ &&+ F^{-1}[G dr^2+H d\theta^2 + (r^2-2 mr)\sin^2\theta d\phi^2] ,
\end{eqnarray}
with
\begin{eqnarray}
F(r)&=&\left(1-\frac{2m}{r}\right)^\gamma\ , \\
G(r,\theta)&=&\left(\frac{r^2-2m r }{r^2-2mr +m^2\sin^2\theta}\right)^{\gamma^2-1}\ ,\\ 
H(r,\theta)&=&\frac{(r^2-2mr)^{\gamma^2}}{(r^2-2mr +m^2\sin^2\theta)^{\gamma^2-1}}\ ,
\end{eqnarray}
\\
where $\gamma$ is the dimensionless mass-density parameter, which describes the departure from spherical symmetry. The total mass of the source is given by $M=\gamma m$. The main interest in this space-time resides in the fact that for $\gamma=1$ the metric reduces to the Schwarzschild solution in Schwarzschild coordinates. However, one needs to keep in mind that for $\gamma\neq 1$ the coordinates are not spherical, as can be seen by evaluating the surfaces of revolution at $r={\rm const.}$ The most striking feature of this space-time however is the fact that for $\gamma\neq 1$ the surface $r=2m$ becomes a true curvature singularity, as can be seen from the investigation of the Kretschmann scalar~\cite{Virbhadra96}. However, the singular surface $r=2m$ still behaves as an infinitely red-shifted surface, much in the same was as in the Schwarzschild case, as could be seen by the study of radial null geodesics. For example, a stationary observer located at infinity in the $\gamma$ space-time, would measure a frequency $\nu_\infty$ for photons emitted at a fixed radius $r$ with frequency $\nu$ according to $\nu_\infty=\nu\sqrt{F(r)}$. Therefore, as $r\rightarrow 2m$ we see that $\nu_\infty\rightarrow 0$. The surface $r=2m$, is indeed infinitely red-shifted and can thus exhibit observational properties analogous to the event horizon of a black hole for observers at infinity.

\begin{figure}
	\includegraphics[width=0.5 \textwidth]{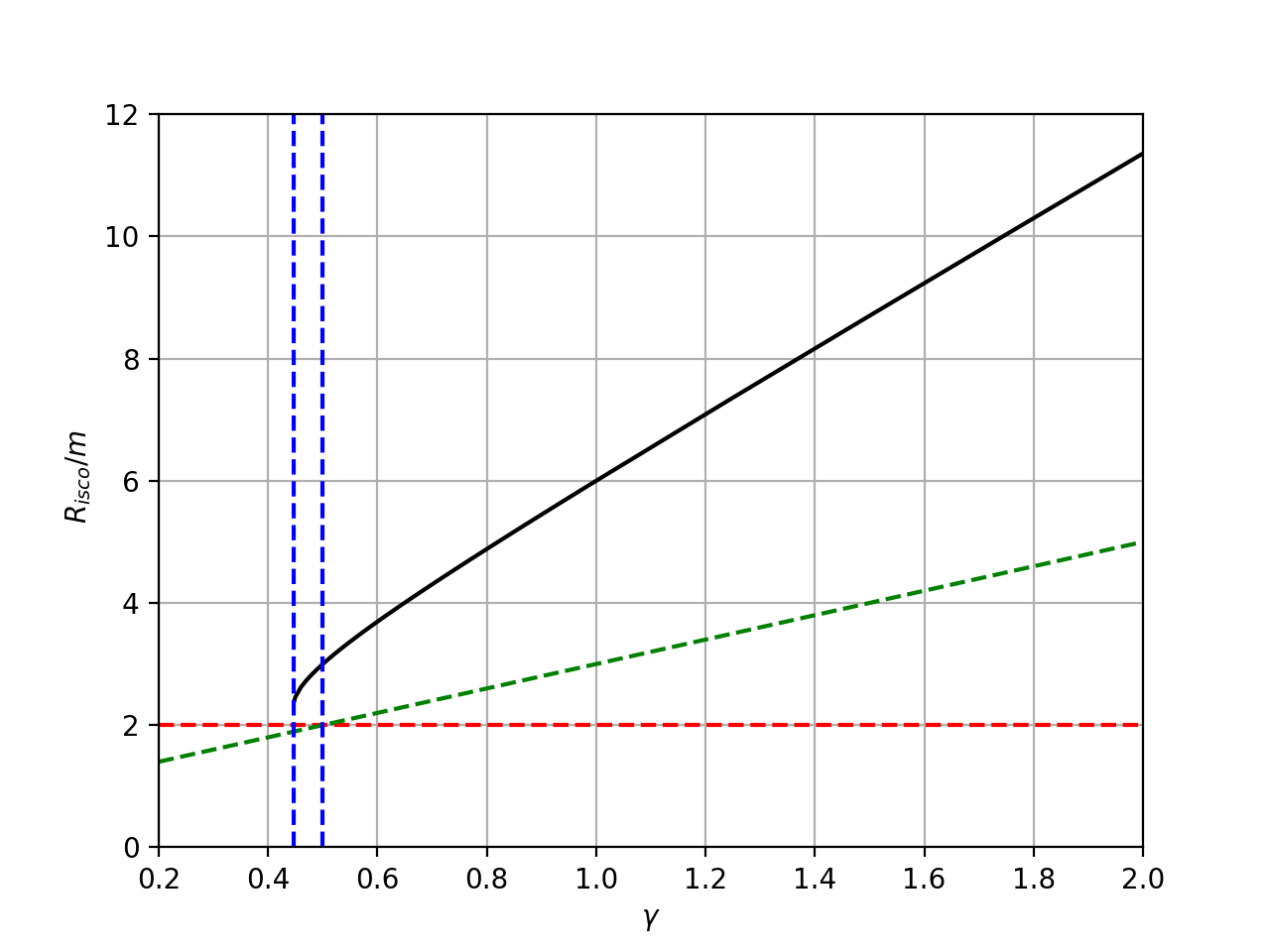}
	\caption{\label{isco_plot} The black thick line defines value of the $ISCO$ in the equatorial plane as a function of $\gamma$ in terms of $m=M/\gamma$. Vertical dashed lines define the borders of regions with different ISCO structure (see text for details) corresponding to values  $\gamma = 1/\sqrt{5}$ and $\gamma= 1/2$. The horizontal dashed red line corresponds to the radius of singular surface $r = 2m$. The dashed green line corresponds to the radius of photon sphere.}	
\end{figure}

\begin{figure*}
\includegraphics[width=0.45 \textwidth]{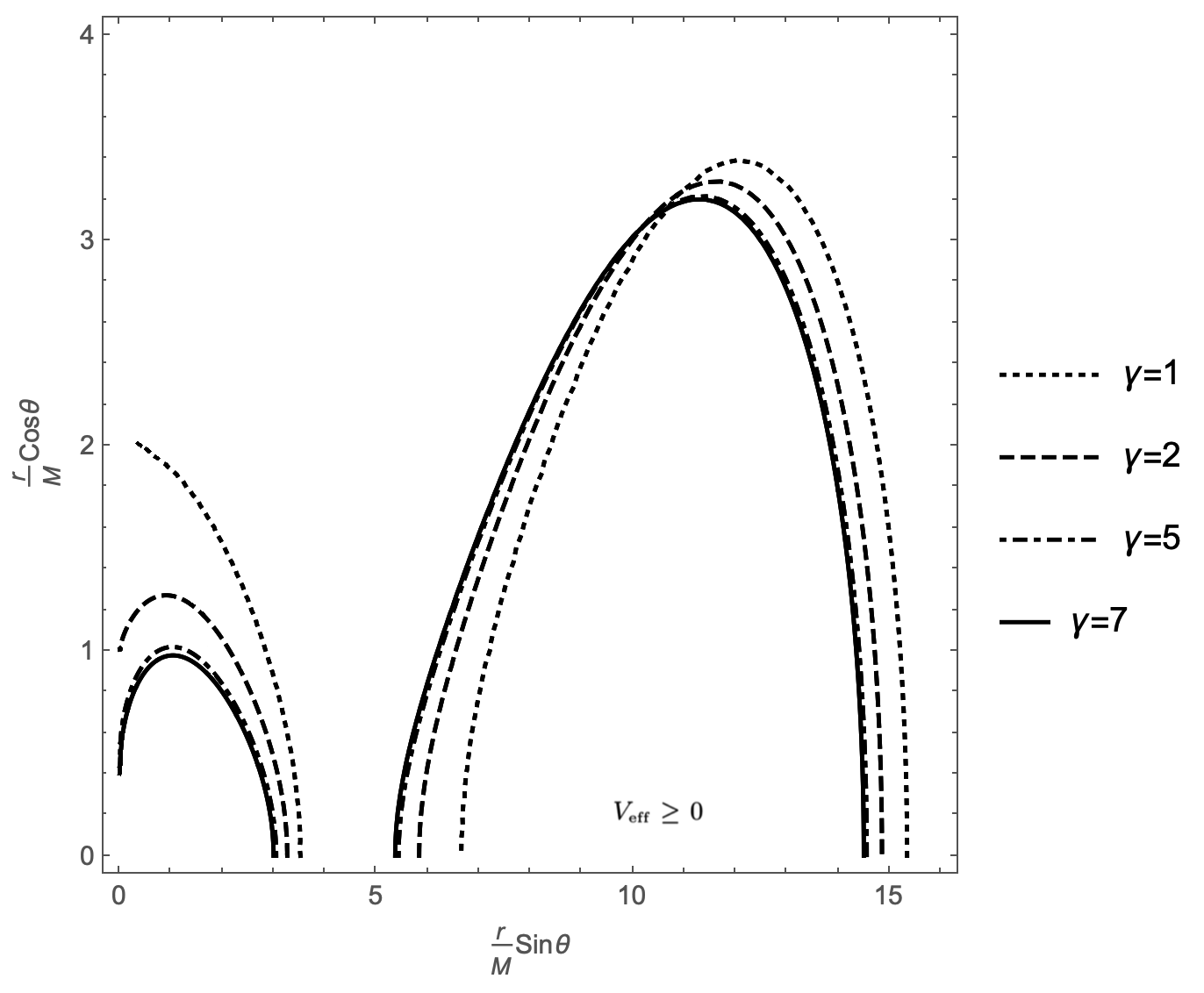}
\includegraphics[width=0.44 \textwidth]{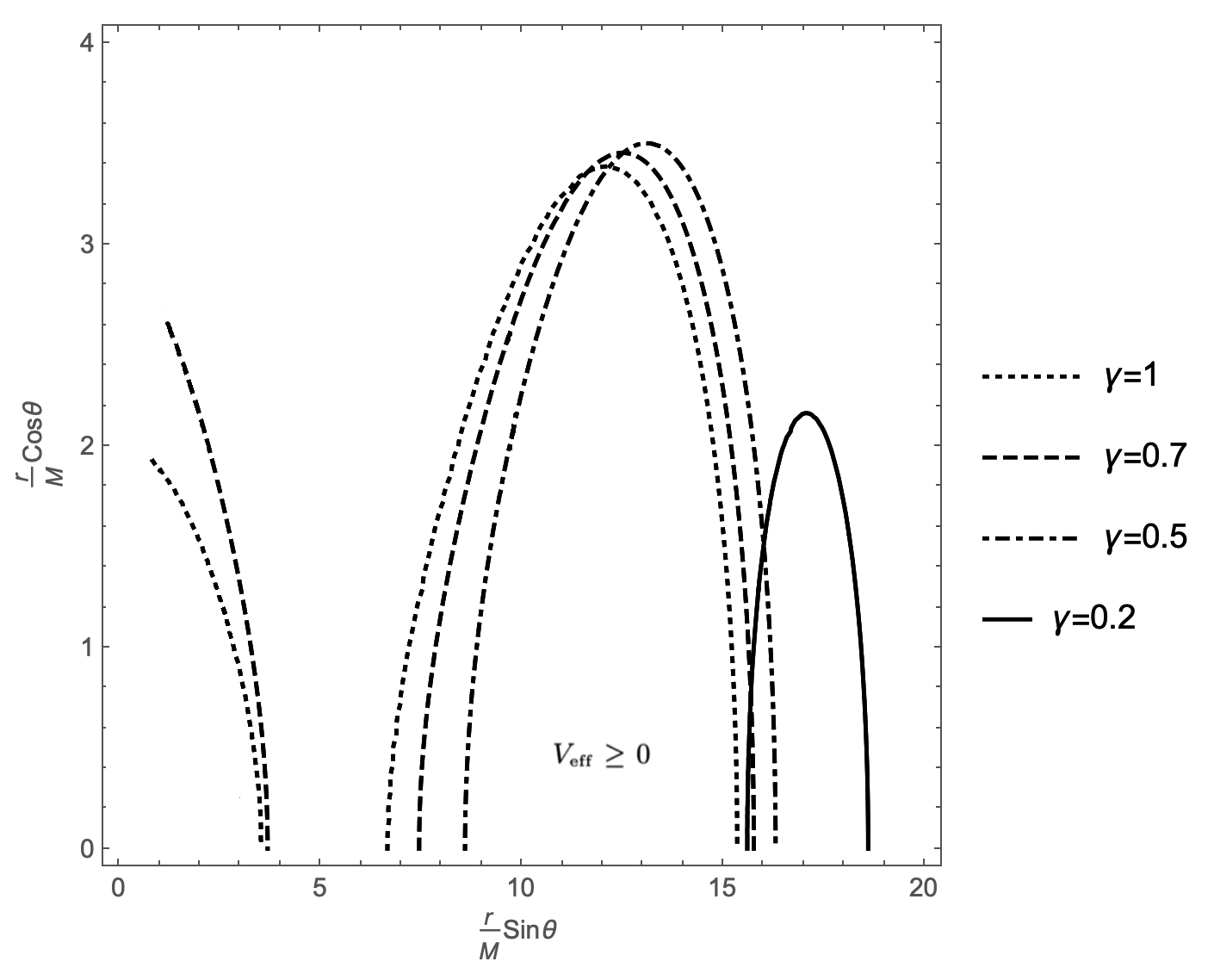}
\caption{Curves of zero velocity for $\gamma\leq 1$ (left panel) and $\gamma\geq 1$ (right panel) obtained from the effective potential ($V_{eff}=0$) given by Eq.~(\ref{veff}) for a masssive particle with $E=0.96$ and $L_z=3.75M$. The curves are obtained by taking $\dot{r}=\dot{\theta}=0$ for the given values of $E$ and $L_z$. The allowed orbits are those for which $V_{eff}\geq 0$ and in the figure orbits are allowed inside the boundary given by the curve $V_{eff}=0$. Note that the horizontal axis (for which $\cos\theta=0$) corresponds to the equatorial plane where we retrieve known results).\label{veff_plot}}
\end{figure*}

The $\gamma$ space-time has a time-like and an azimuthal Killing vector, meaning the existence of two conserved quantities: the specific energy $E$ and the z-component of the specific angular momentum $L_z$. The corresponding components of the four-momentum are $p_t =- E$ and $p_\phi = L_z$, which can be used to find the two corresponding geodesic equations:
\begin{eqnarray}\label{t_dot}
\dot{t}&=&-\frac{E}{g_{tt}},\\
\dot{\phi}&=&\frac{L_z }{g_{\phi\phi}},\label{phi_dot}
\end{eqnarray}
where a derivative with respect to the affine parameter (proper time for a massive particle) is represented by the overhead dot. We can write the equation of motion of test particles with these constants of motion. By substituting Eqs. (\ref{t_dot}) and (\ref{phi_dot}) into the normalization condition $u^\alpha u_\alpha = -1$ for a massive particle, where $u^\alpha = (\dot{t},\dot{r}, \dot{\theta}, \dot{\phi})$ is the 4-velocity, we find
\begin{equation}\label{normveff}
\begin{aligned}
&g_{rr}\dot{r}^2+g_{\theta\theta}\dot{\theta}^2 = V_{eff}(r,\theta;E,L_z),
\end{aligned}
\end{equation}
where the effective potential is 
\begin{equation}\label{veff}
\begin{aligned}
&V_{eff}\equiv-\frac{E^2g_{\phi\phi} + L_z^2g_{tt}}{g_{tt}g_{\phi\phi}}-1.
\end{aligned}
\end{equation}

If we restrict the attention to equatorial, i.e. $\theta=\pi/2$, circular orbits for massive test particles, we can then solve $V_{eff} = 0$ and $\partial V_{eff}/ \partial r = 0$ for $E$ and $L_z$ to find

\begin{equation}\label{E_comp}
\begin{aligned}
&E=-\frac{g_{tt}}{\sqrt{-(g_{tt}+g_{\phi\phi}\Omega^2)}}\ ,
\end{aligned}
\end{equation}
\begin{equation}\label{L_comp}
\begin{aligned}
&L_z=\frac{g_{\phi\phi}\Omega}{\sqrt{-(g_{tt}+g_{\phi\phi}\Omega^2)}}\ ,
\end{aligned}
\end{equation}
where 
\begin{equation}\label{Omega}
\begin{aligned}
&\Omega=\frac{d\phi}{dt}=\sqrt{-\frac{g_{tt,r}}{g_{\phi\phi,r}}},
\end{aligned}
\end{equation}
is the angular velocity of equatorial circular geodesics, i.e. the angular velocity of zero angular-momentum observers. 
The radius of the innermost stable circular orbit (ISCO) is then found by substituting Eqs.~(\ref{E_comp}) and (\ref{L_comp}) into Eq. (\ref{veff}), and then solving $ V_{eff}'=  V_{eff}'' = 0$ for $r$, where with $'$ we denote derivatives with respect to $r$. In terms of the total gravitational mass $M=m\gamma$ the ISCO as a function of $\gamma$ is given by
\be \label{isco}
r_{\rm isco}=\frac{M}{\gamma}+3M+\sqrt{5M^2-\frac{M^2}{\gamma^2}}\ .
\ee 

Similarly we can consider the orbits of photons. In the case of circular motion at a fixed value of $\theta=\theta_0> 0$ we find the effective potential to be
\be 
 V_{eff}=\frac{L^2}{r^2\sin^2\theta_0}\left(1-\frac{2m}{r}\right)^{2\gamma-1}\ .
\ee 
One can obtain the photon capture surface $r_{ps}$ requiring that $E$ and $L_z$ in Eqs.~(\ref{E_comp}) and (\ref{L_comp}) diverge in the limit $r\rightarrow r_{ps}$
\be 
r_{\rm ps}=(2\gamma+1)m = 2M +\frac{M}{\gamma}\ ,
\ee 
The first thing to notice is that, since for $\gamma\neq 1$ the coordinates are not spherical, the photon capture surface is not a sphere. The other important thing to notice is that it would seem that there is a photon capture surface also in the limit of $\gamma\rightarrow0$, which corresponds to Minkowski space-time. However, this is easily explained by remembering that, by construction, the limit of vanishing $\gamma$ implies that also $m$ must vanish.

Fig.~\ref{isco_plot} shows the ISCO, the photon capture radius and the singularity in the equatorial plane as functions of $\gamma$. 

Notice that Eq. \eqref{isco} has no real solutions for $\gamma<1/\sqrt{5}$. Therefore for $\gamma<1/\sqrt{5}$ there can be no stable circular orbits. 
The other important value is $\gamma=1/2$ where photon capture radius intersects the singularity. As can be seen from Fig.~\ref{isco_plot} there can be no photon capture orbit for $\gamma<1/2$. This already shows that a prolate source with $\gamma<1/2$ would be distinguishable from a black hole. Fig.~\ref{veff_plot} shows the curves of zero velocity for different values of $\gamma$ in the $\{r \sin(\theta)/M, r \cos(\theta)/M\}$ plane. These curves are where  $V_{eff} = 0$ in Eq.~(\ref{veff}) with $E = 0.96$, $L_z = 3.75M$. Bound orbits are allowed only if $V_{eff} \geq 0$ since the left-hand side of Eq.~(\ref{veff}) is always positive.

\section{Ray-tracing code for photons}\label{rtcs}

In order to investigate the image that the source of the $\gamma$ space-time would produce for distant observers we need to study the motion of light rays. Our ray-tracing code computes the trajectories of photons in the space-time described by $\gamma$-metric in the vicinity of the surface $r = 2m$. The code is the modified version of the one used in ~\cite{Ayzenberg18} and ~\cite{Gott19}, which follows the method developed in ~\cite{Psaltis12} to compute the trajectories of photons near black hole.

The two first-order differential Eqs. (\ref{t_comp}) and (\ref{phi_comp}) for the evolution of the $t$- and $\phi$-components of the photon's position are obtained by rewriting $p_t$ and $p_\phi$ in terms of the normalized affine parameter $\lambda'=E/\lambda$ and the impact parameter $b=L_z/E$ as

\begin{equation}\label{t_comp}
\begin{aligned}
&\frac{dt}{d\lambda'}=-\frac{1}{g_{tt}},\\
\end{aligned}
\end{equation}
\begin{equation}\label{phi_comp}
\begin{aligned}
&\frac{d\phi}{d\lambda'}=b\frac{1}{g_{\phi\phi}},
\end{aligned}
\end{equation}

The remaining geodesic equations for the $r$-component and the $\theta$-component of the photon's position in the $\gamma$ space-time are then calculated with respect to the normalized affine parameter in the standard way, through the evaluation of the Christoffel symbols $\Gamma^\sigma_{\mu\nu}$ as
\begin{equation}
\begin{aligned}\label{geod}
& \frac{d^2x^\sigma}{d\lambda'^2}+\Gamma^\sigma_{\mu\nu}\frac{dx^\mu}{d\lambda '} \frac{dx^\nu}{d\lambda'}=0. \\
\end{aligned}
\end{equation}
In this manner we obtain the system of equations that the ray-tracing code can use for this space-time.

The massive source of the $\gamma$ space-time is located at the origin of the reference frame and coordinate system when reduced to the Schwarzschild case. In the code, we set the units in such a way that the source of the $\gamma$-metric has unitary mass, $M = 1$. The reason is the mass $M$ only changes the size without affecting the shape of the shadow. The observer's screen is located at a distance of $d=1000$, the azimuthal and polar angles are $\iota$ and $0$, respectively. The celestial coordinates $(\alpha, \beta)$ on the observer's sky are related to polar coordinates $r_{scr}$ and $\phi_{scr}$ on the screen by $\alpha=r_{scr}\cos(\phi_{scr})$ and $\beta=r_{scr}\sin(\phi_{scr})$.
The system of geodesic equations is solved backwards in time since
only the final positions and momenta of the photon on the screen are known. The photons depart from some initial position on the screen with a four-momentum perpendicular to the screen. This condition imitates placing the observing screen at spatial infinity as only those photons that are moving perpendicular to the screen at a distance $d$ will also impact the screen at spatial infinity.

The initial position and four-momentum of each photon in the Erez-Rozen coordinates of the $\gamma$ space-time are given by
\begin{equation}\label{in_pos_r}
	\begin{aligned}
		& r_i =\left(d^2+\alpha^2+\beta^2\right)^{1/2}      ,\\
	\end{aligned}
\end{equation}
\begin{equation}\label{in_pos_theta}
	\begin{aligned}
		& \theta_i=\arccos\left(\frac{d\cos\iota+\beta\sin\iota}{r_i}\right),\\
	\end{aligned}
\end{equation}
\begin{equation}\label{in_pos_phi}
	\begin{aligned}
		& \phi_i = \arctan\left(\frac{\alpha}{d\sin\iota-\beta\cos\iota}\right),
	\end{aligned}
\end{equation}
and
\begin{equation}\label{in_momen_r}
	\begin{aligned}
	& \left(\frac{dr}{d\lambda'}\right)_i=\frac{d}{r_i},\\
	\end{aligned}
\end{equation}
\begin{equation}\label{in_momen_theta}
	\begin{aligned}
	& \left(\frac{d\theta}{d\lambda'}\right)_i=\frac{-\cos\iota+\frac{d}{r_i^2}(d\cos\iota+\beta\sin\iota)}{\sqrt{r_i^2-(d\cos\iota+\beta\sin\iota)^2}},\\
	\end{aligned}
\end{equation}
\begin{equation}\label{in_momen_phi}
	\begin{aligned}
	& \left(\frac{d\phi}{d\lambda'}\right)_i=\frac{-\alpha\sin\iota}{\alpha^2+(d\cos\iota+\beta\sin\iota)^2},\\
	\end{aligned}
\end{equation}
\begin{equation}\label{in_momen_t}
	\begin{aligned}
	& \left(\frac{dt}{d\lambda'}\right)_i = -\left[-g_{rr}\left(\frac{dr}{d\lambda'}\right)_i^2-g_{\theta\theta}\left(\frac{d\theta}{d\lambda'}\right)_i^2 
  - g_{\phi\phi}\left(\frac{d\phi}{d\lambda'}\right)_i^2\right]^{1/2}.
	\end{aligned}
\end{equation}

By requiring the norm of the photon four-momentum to be zero we can find the component $(dt/d\lambda')_i$. The conserved quantity $b$, which is involved in Eqs. (\ref{t_comp}) and (\ref{phi_comp}), is calculated from the initial conditions.

The code samples initial conditions on the screen in the following way.  The location of the boundary of the compact object shadow is found inside $0 \leq r_{scr} \leq 20$, for each value of $\phi_{scr}$ in the range $0 \leq \phi_{scr} \leq 2\pi$ with step of $\pi / 180$. The boundary is the border between the photons that are captured by the singularity and the photons that are able to escape to spatial infinity. The photons are considered as captured by the singularity if they cross $r = r_{surf} + \delta r$ with $\delta r = 10^{-3}$,
where $r_{surf}$ is the radius of the infinitely redshifted surface which in the present case corresponds to the location of the curvature singularity. Then the boundary is zoomed in to an accuracy of $\delta r_{scr} \thicksim 10^{-3}$ to accurately determine the value of $r_{scr}$ that corresponds to the shadow boundary for the current value of $\phi_{scr}$. This methodology lets accurately calculate the shadow produced by light traveling in the $\gamma$ space-time much more efficiently than finely sampling the entire screen.

\section{The shadow of the $\gamma$-metric \label{shadow} }

\begin{figure}
	\includegraphics[width=0.5 \textwidth]{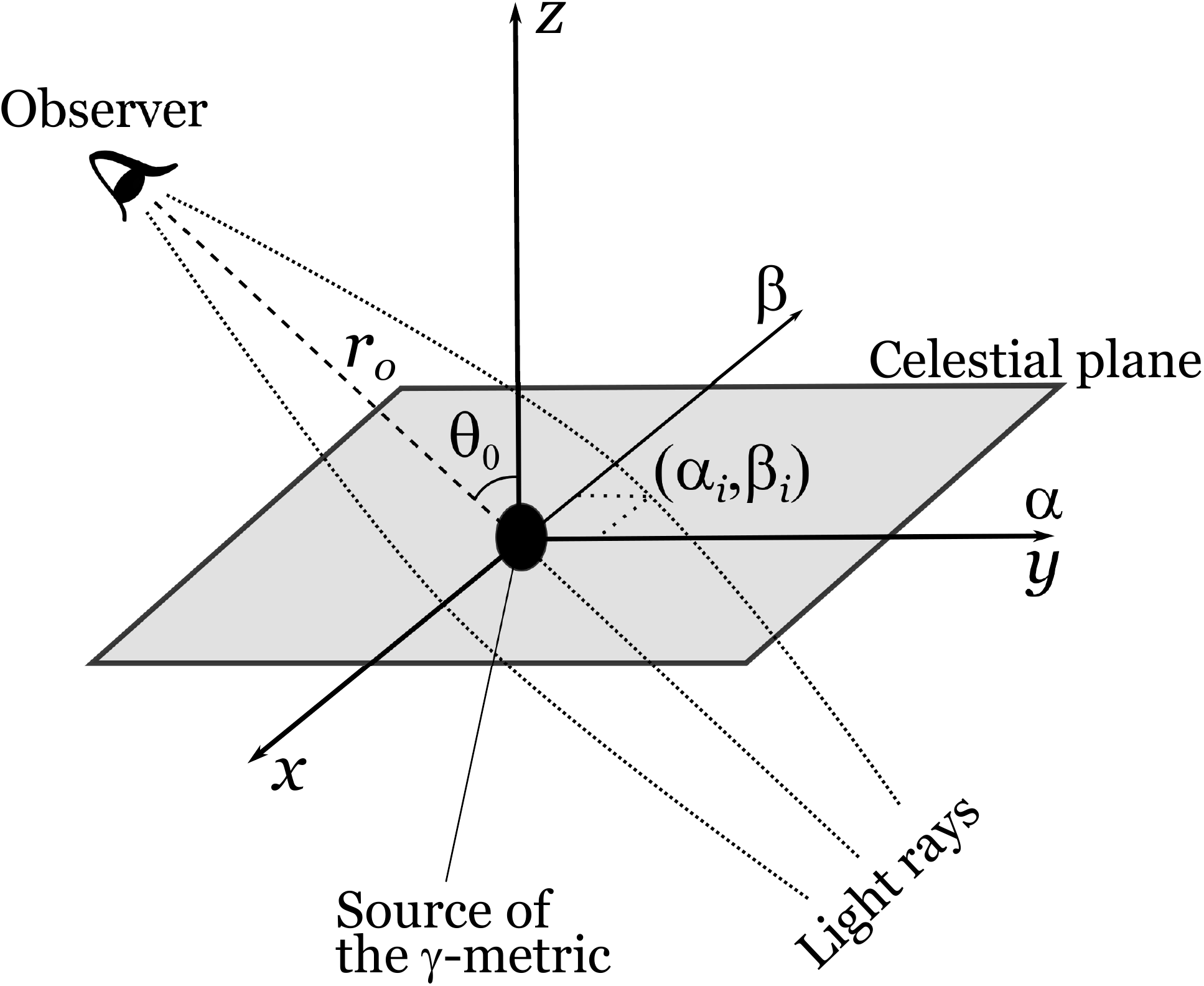}
	
	\caption{Schematic illustration of the celestial coordinates used for the ray tracing code in the $\gamma$ space-time. \label{shadsch}}
\end{figure}

In this section, we will study the apparent shape of the shadow of compact object described by $\gamma$-metric. 
To describe the shadow with better visualization one may consider the celestial coordinates $\alpha$ and $\beta$~(see \cite{Vazquez04, Abdujabbarov16b} for reference) which are defined as,
\begin{eqnarray}
\alpha &=&\underset{r_0 \rightarrow \infty}{\lim}\left(-r_0^2 \sin \theta_0 \frac{d\phi}{dr}\right) \ , \label{eq:14}\\
\beta &=&\underset{r_0 \rightarrow \infty}{\lim}\left(r_0^2 \frac{d\theta}{dr}\right) \ , \label{eq:15}
\end{eqnarray}
where $r_0$ is the distance between the observer and massive source and $\theta_0$ is the inclination angle between the normal of observer's sky plane and observer lens axis~(see Fig.~\ref{shadsch}). 

\begin{figure*}
	\includegraphics[width=0.32 \textwidth]{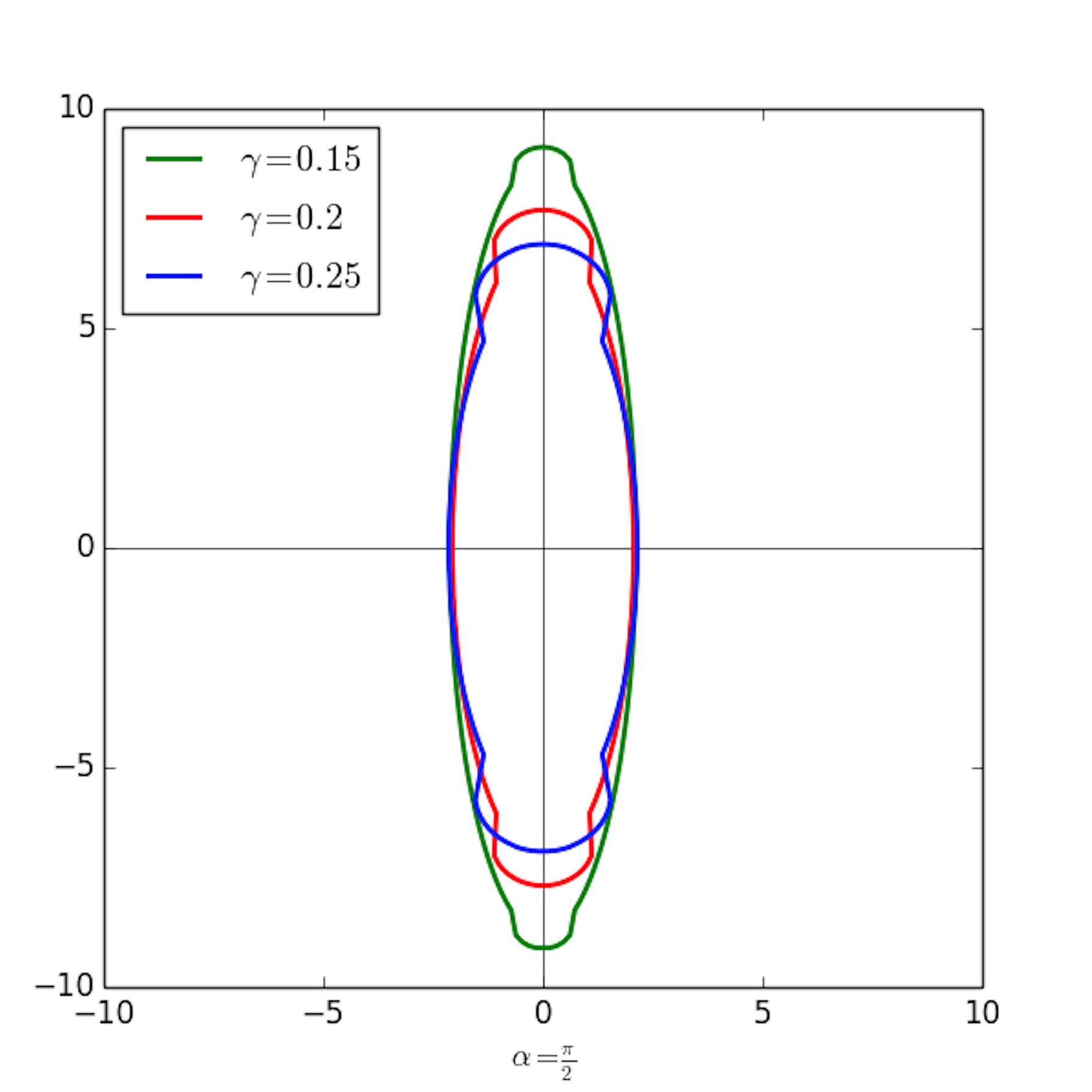}
	\includegraphics[width=0.32 \textwidth]{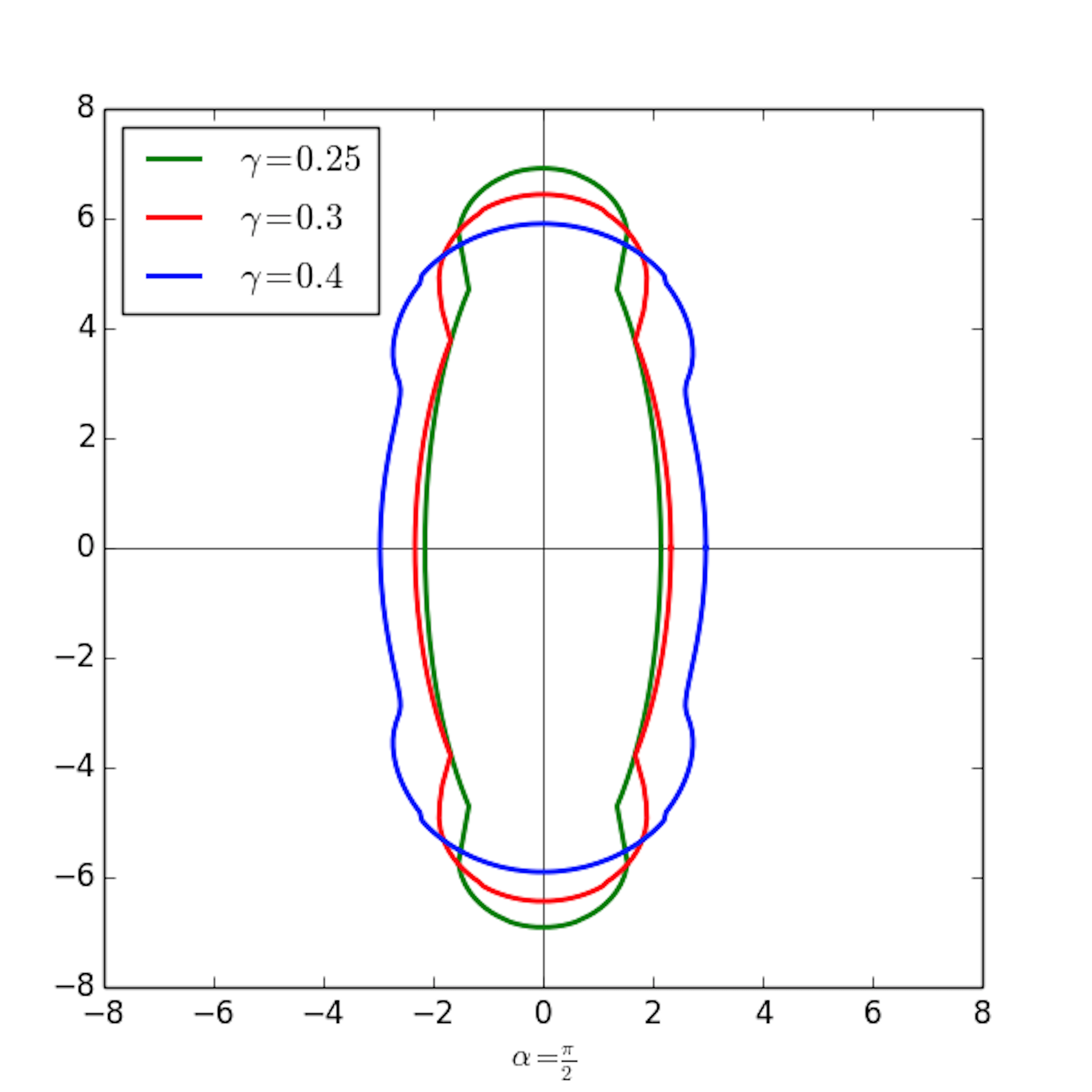}
	\includegraphics[width=0.32 \textwidth]{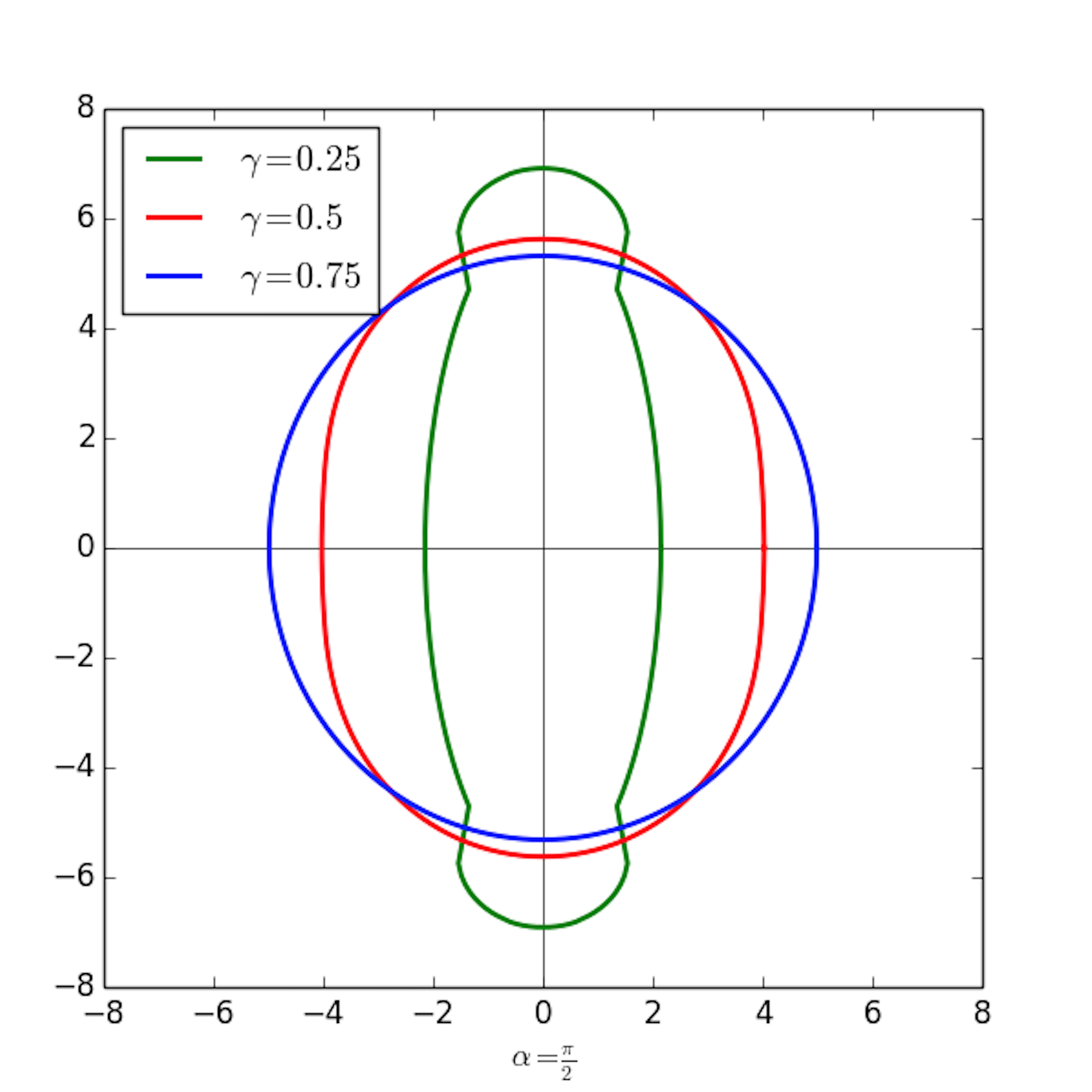}

	\includegraphics[width=0.32 \textwidth]{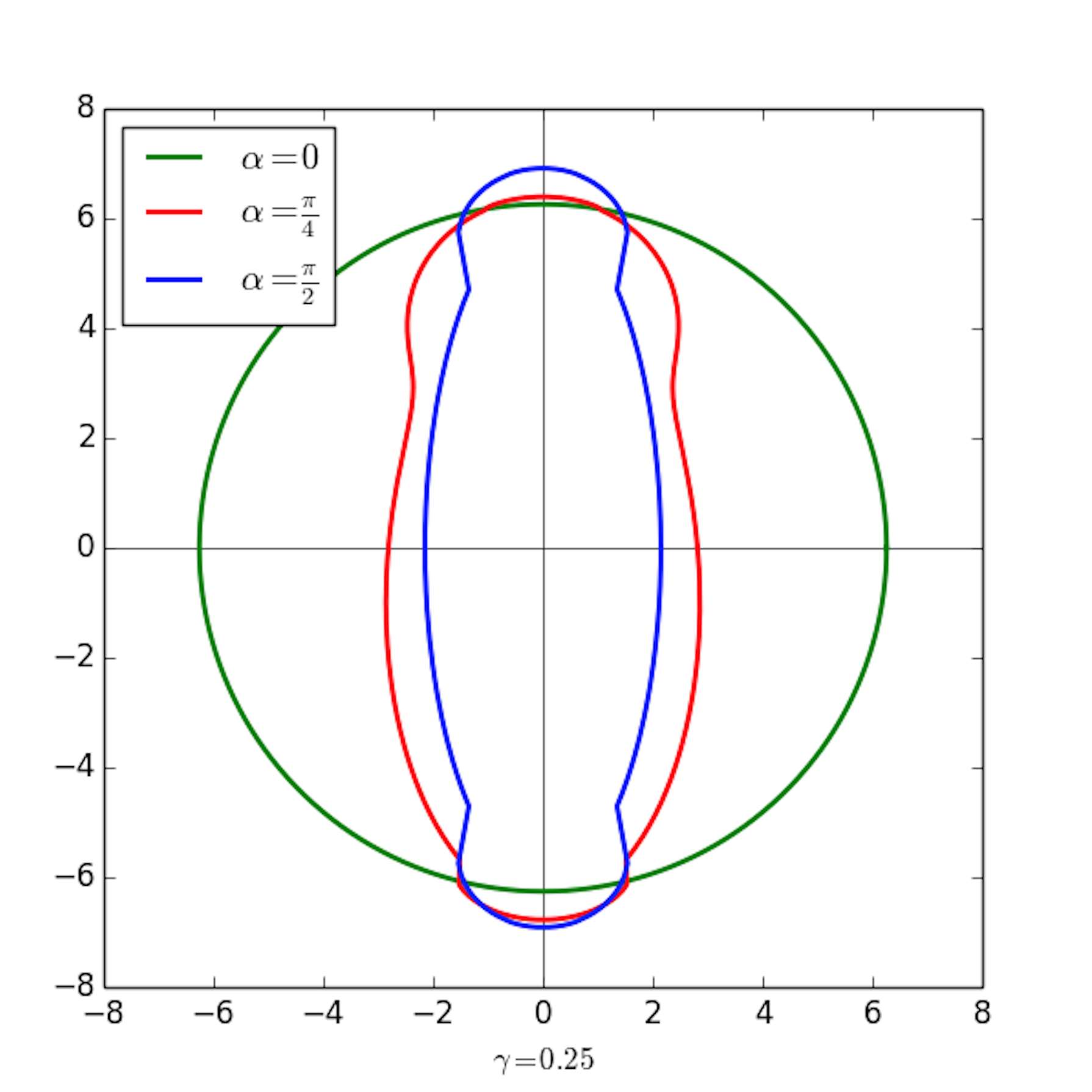}
	\includegraphics[width=0.32 \textwidth]{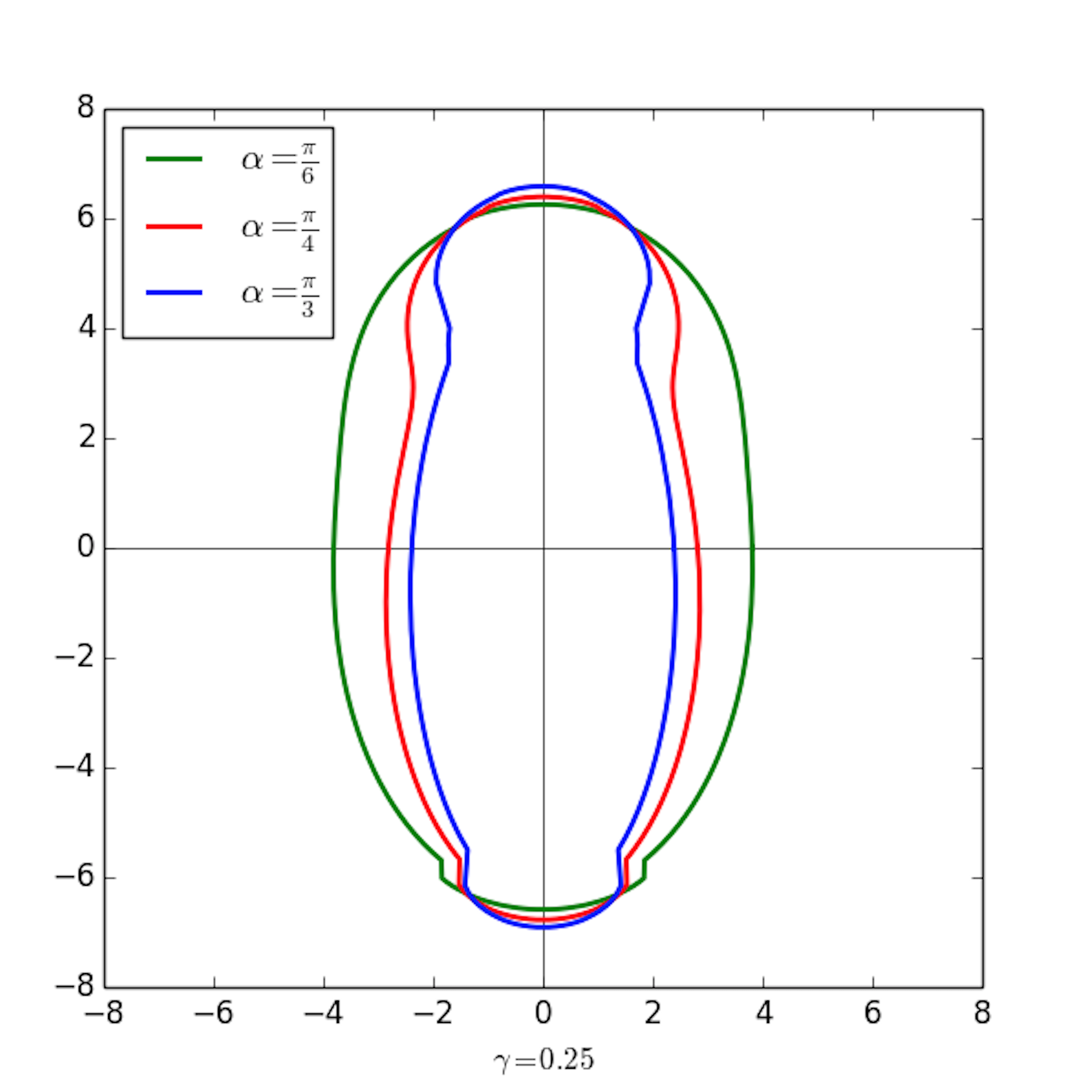}	
	\includegraphics[width=0.32 \textwidth]{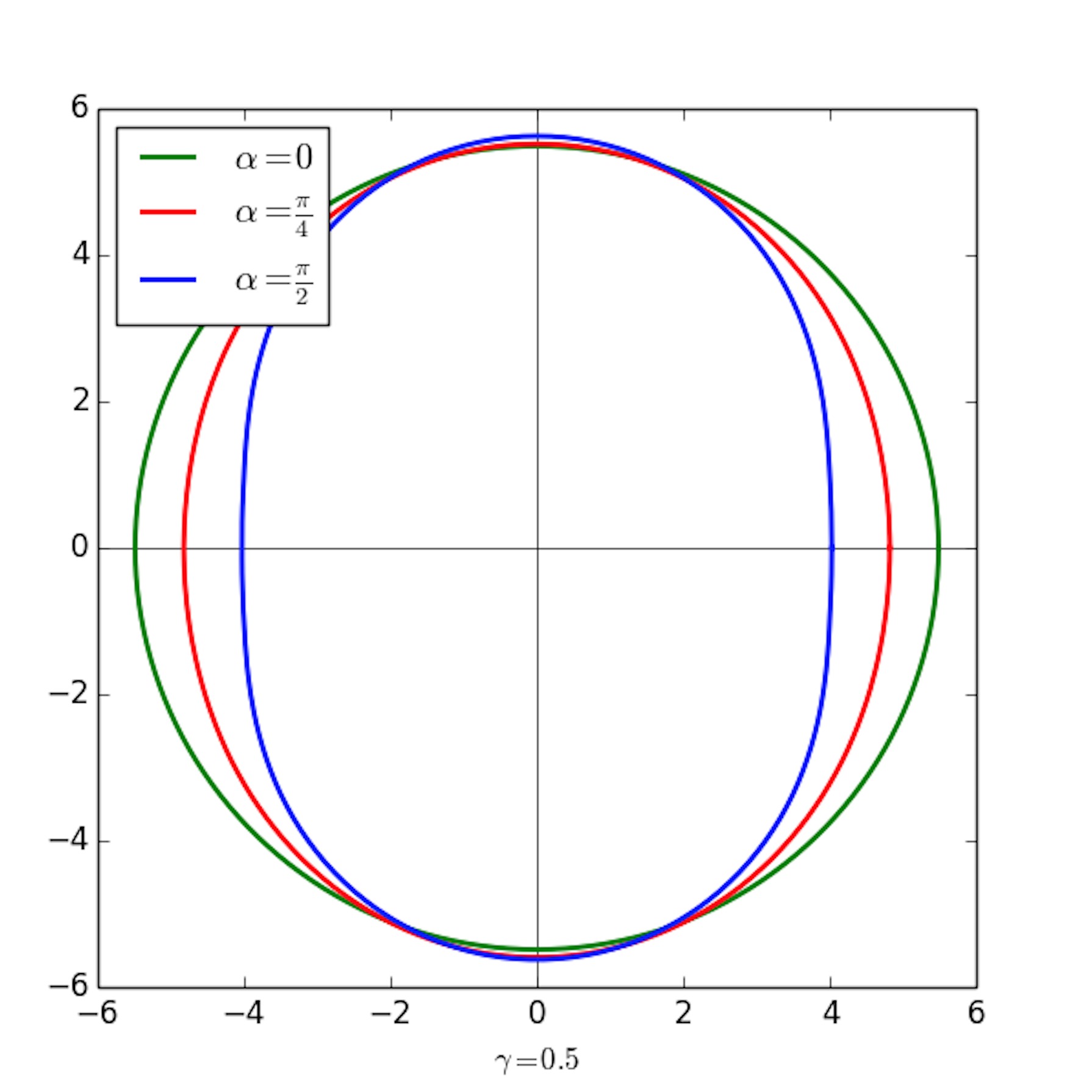}

	\includegraphics[width=0.32 \textwidth]{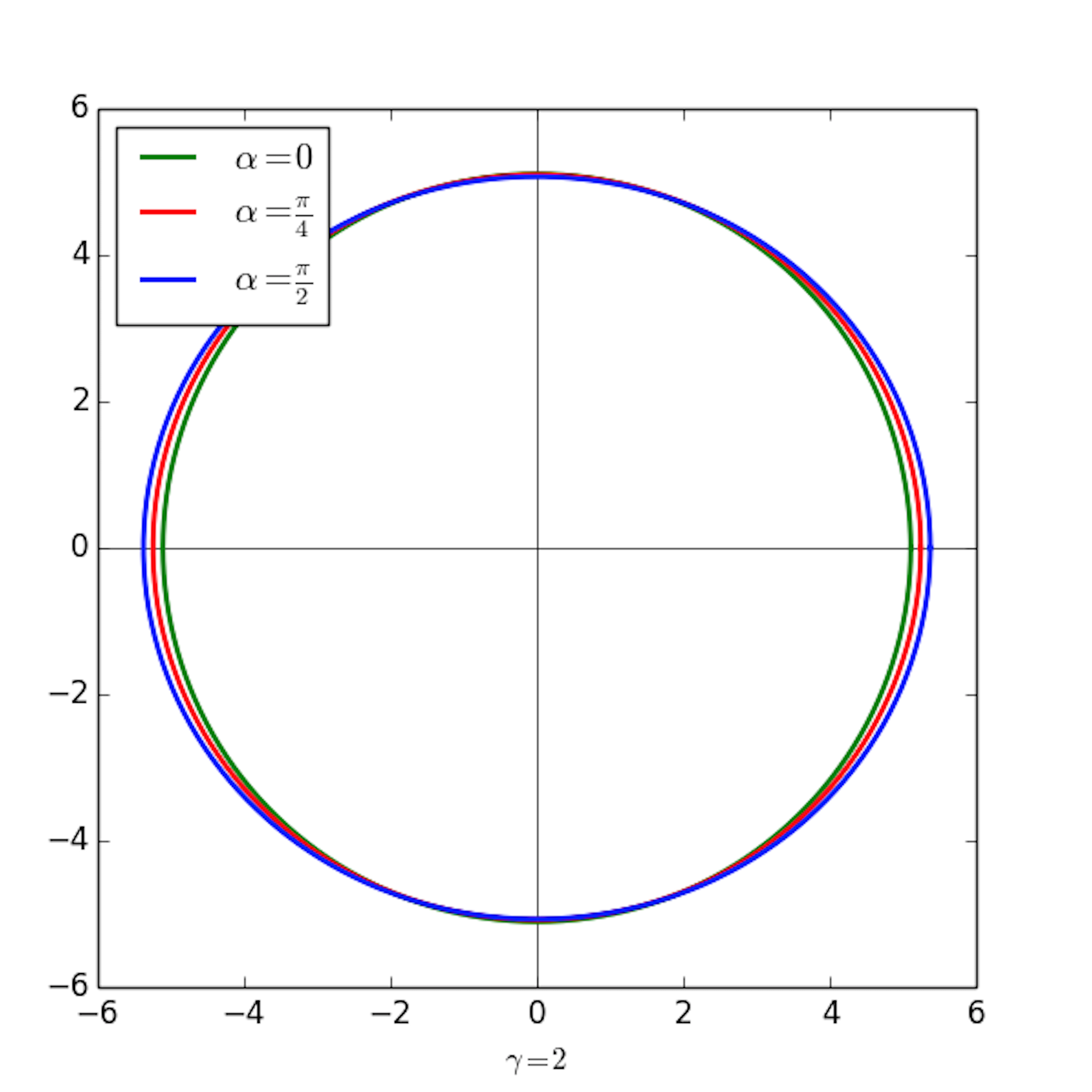}
	\includegraphics[width=0.32 \textwidth]{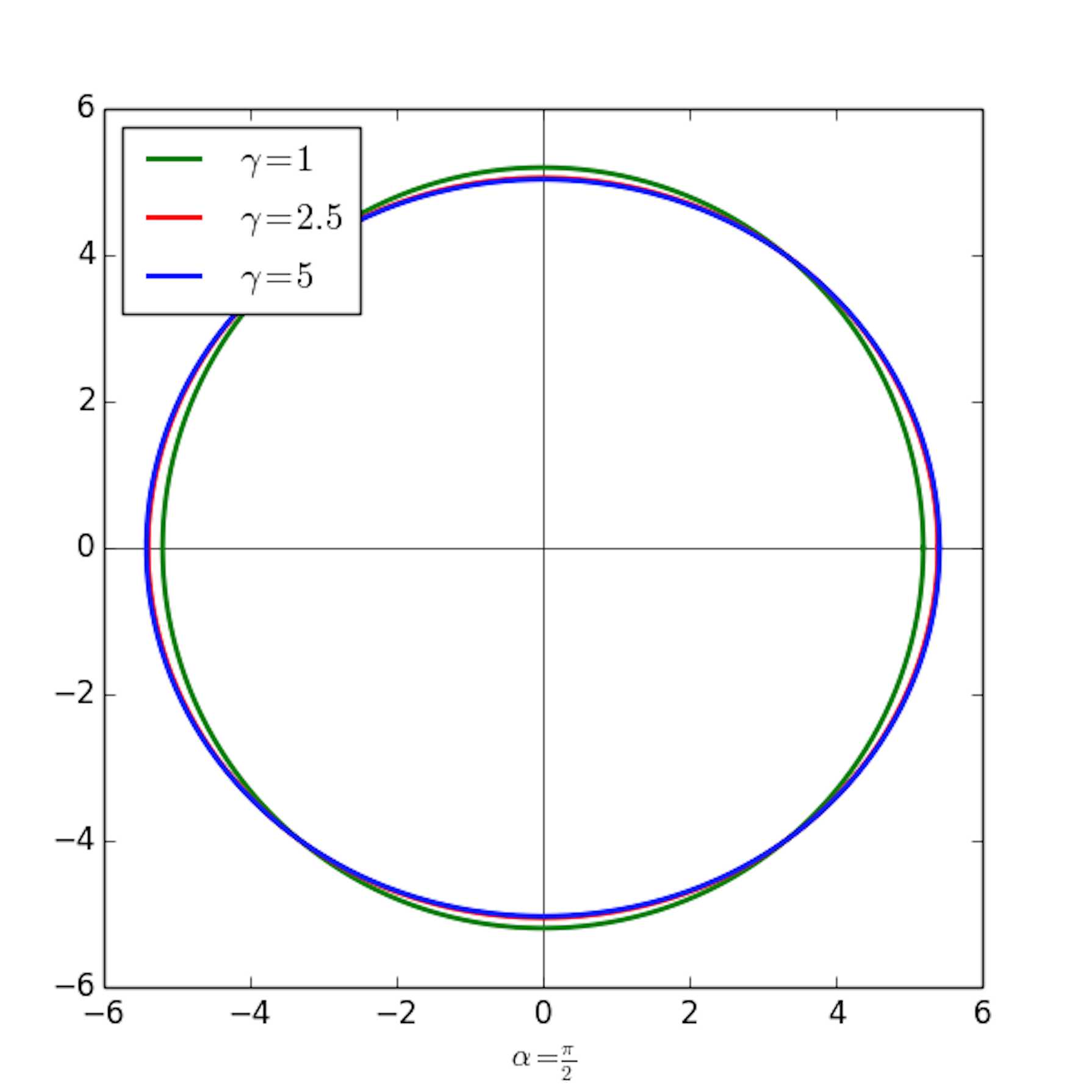}
	\includegraphics[width=0.32 \textwidth]{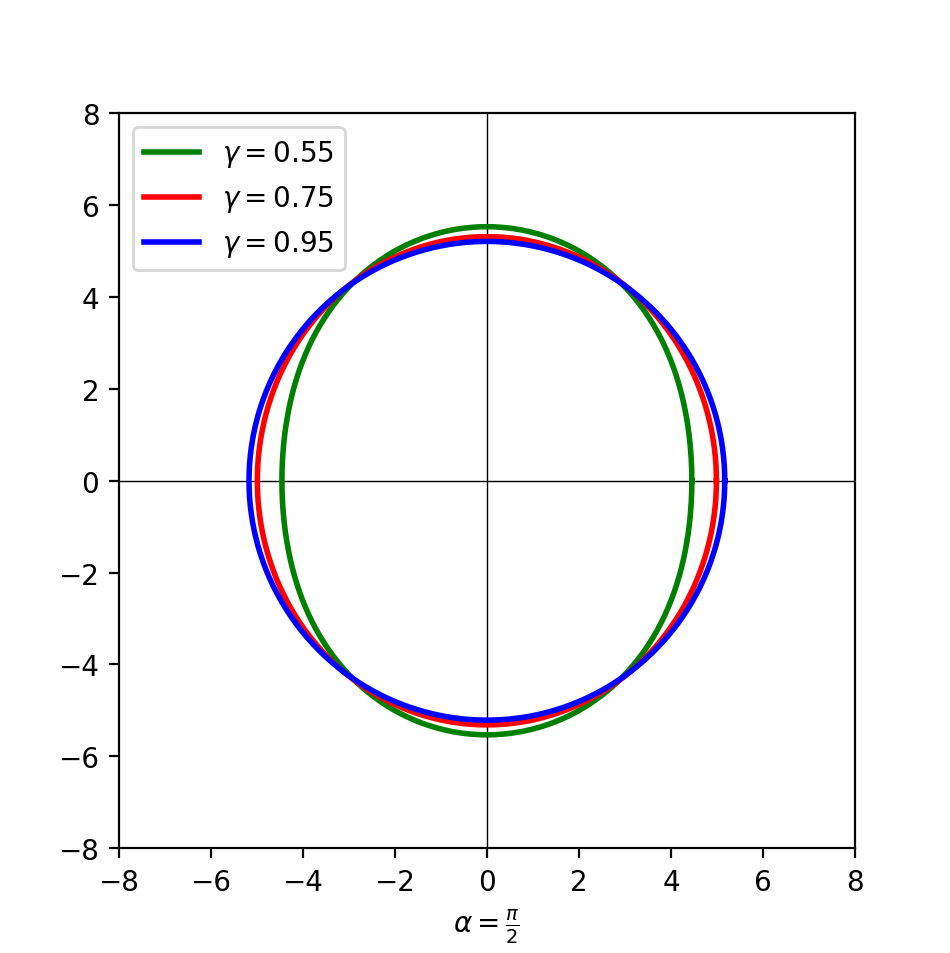}
	\caption{Ray-traced shadow images in the $\gamma$ space-time. From left to right, \textit{first row}: shadow images for $\gamma=[0.15, 0.2, 0.25]$, $\gamma=[0.25, 0.3, 0.4]$, $\gamma=[0.25, 0.5, 0.75]$ at an inclination angle $\alpha=\pi/2$, \textit{second row}: shadow images at an inclination angles $\alpha=[0,\pi/4,\pi/2]$, $\alpha=[\pi/6,\pi/4,\pi/3]$ and $\alpha=[0,\pi/4,\pi/2]$ for $\gamma=0.25$, $\gamma=0.25$ and $\gamma=0.5$, respectively, \textit{third row}: shadow images for $\gamma=2$ at an inclination angles $\alpha=[0,\pi/4,\pi/2]$, for $\gamma=[1,2.5,5]$ and $\gamma= [0.55, 0.75, 0.95]$ at an inclination angle $\alpha=\pi/2$. { It is easy to see that for values of $\gamma>1/2$ the ability to distinguish the Schwarzschild shadow from the one in the $\gamma$-metric will depend on the ability to accurately resolve the shape of the shadow. On the other hand, for $\gamma<1/2$, more distinctive features that may allow one to tell the two cases apart arise.} \label{shadimg}}
\end{figure*}

In order to describe the dependence of the shape of the shadow on the deformation parameter, we will use coordinate independent formalism proposed in~\cite{Ayzenberg18}. The shape of the shadow is parametrized in terms of the average radius of the sphere $\langle R \rangle$, and the asymmetry parameter $A$. We can safely ignore the shift of the center of the shadow from the center $D$ since in the case of the $\gamma$-metric $D$ is always identically equal to zero. There are other ways to describe the shape of the shadow (see e.g. \cite{Tsukamoto14, Abdujabbarov15}), however, the results would be similar with any chosen parametrization. The average radius $\langle R \rangle$ is the average distance of the boundary of the shadow from its center, which is defined by
\begin{eqnarray}
\langle R \rangle \equiv \frac{1}{2\pi} \int_{0}^{2\pi} R(\vartheta)d\vartheta,
\end{eqnarray}
where $R(\vartheta) \equiv \left[(\alpha-D)^2+\beta(\alpha)^2\right]^{1/2}$, $D=0$ and $ \vartheta \equiv \tan^{-1}[\beta(\alpha)/\alpha)]$. The asymmetry parameter $A$ is the distortion of the shadow from a circle. It is defined by
\begin{eqnarray}
A \equiv 2\left[\frac{1}{2\pi}\int_{0}^{2\pi} \left(R-\langle R \rangle \right)^2 d\vartheta \right]^{1/2}.
\end{eqnarray}

\begin{figure*}
\includegraphics[width=0.32 \textwidth]{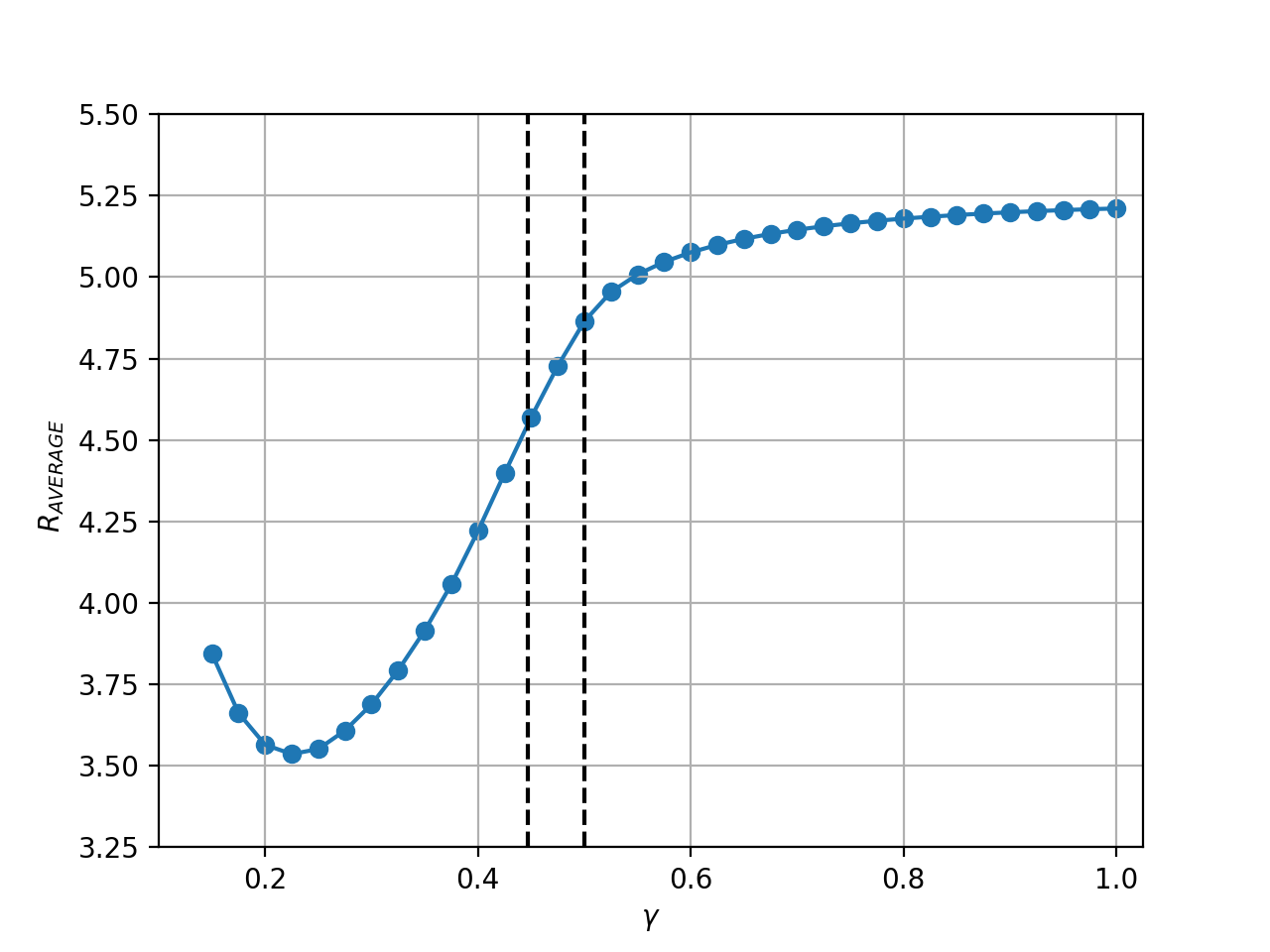}
\includegraphics[width=0.32 \textwidth]{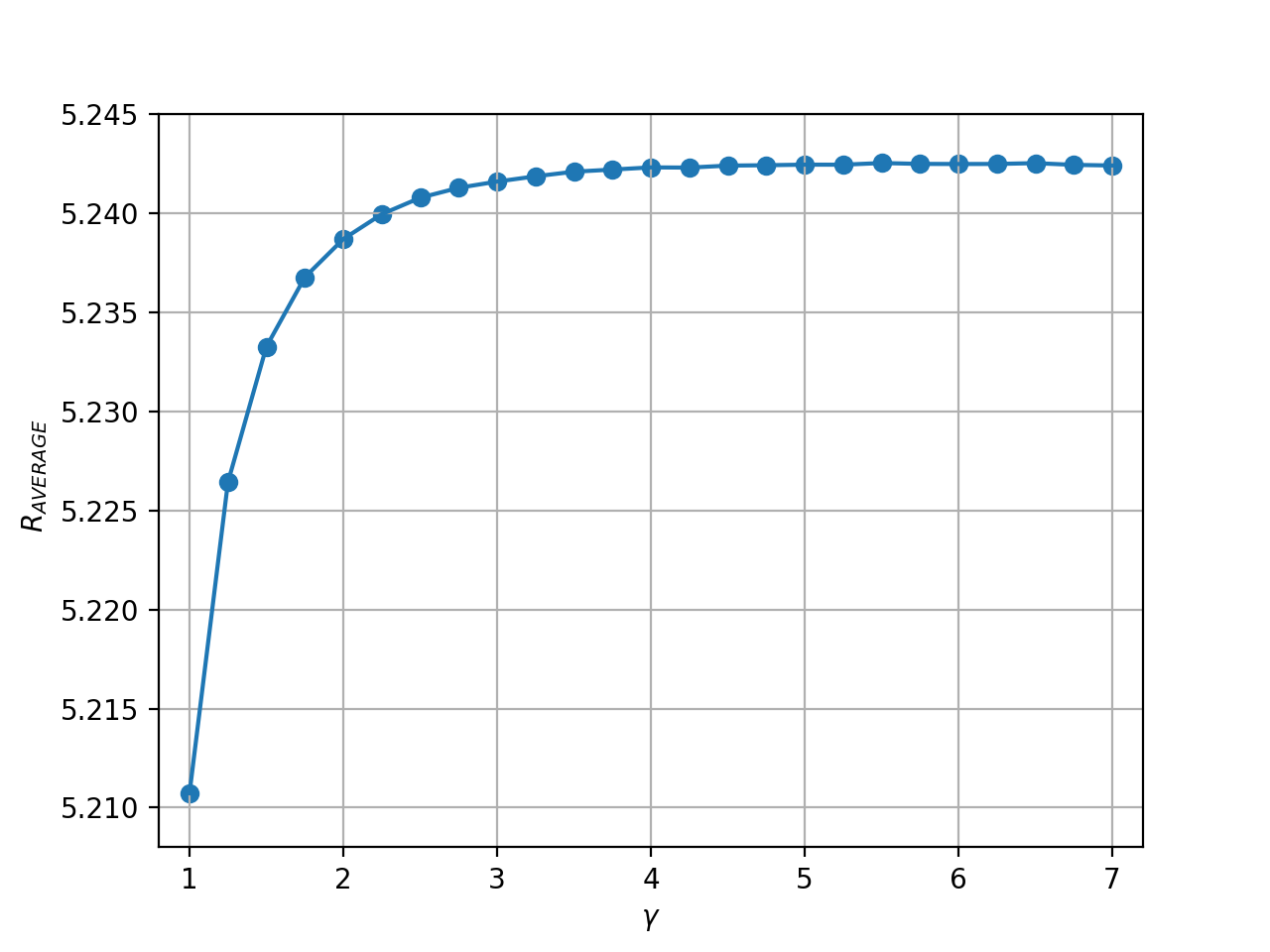}
\includegraphics[width=0.32 \textwidth]{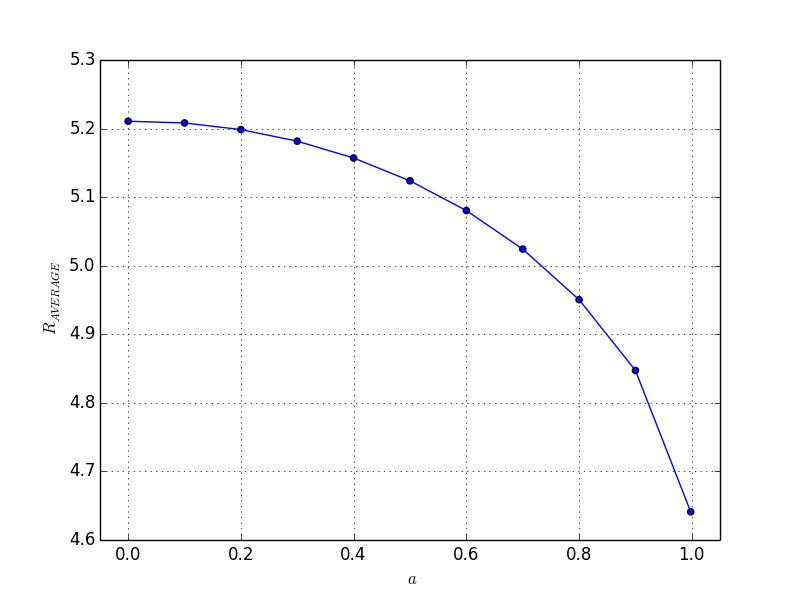}

\includegraphics[width=0.32 \textwidth]{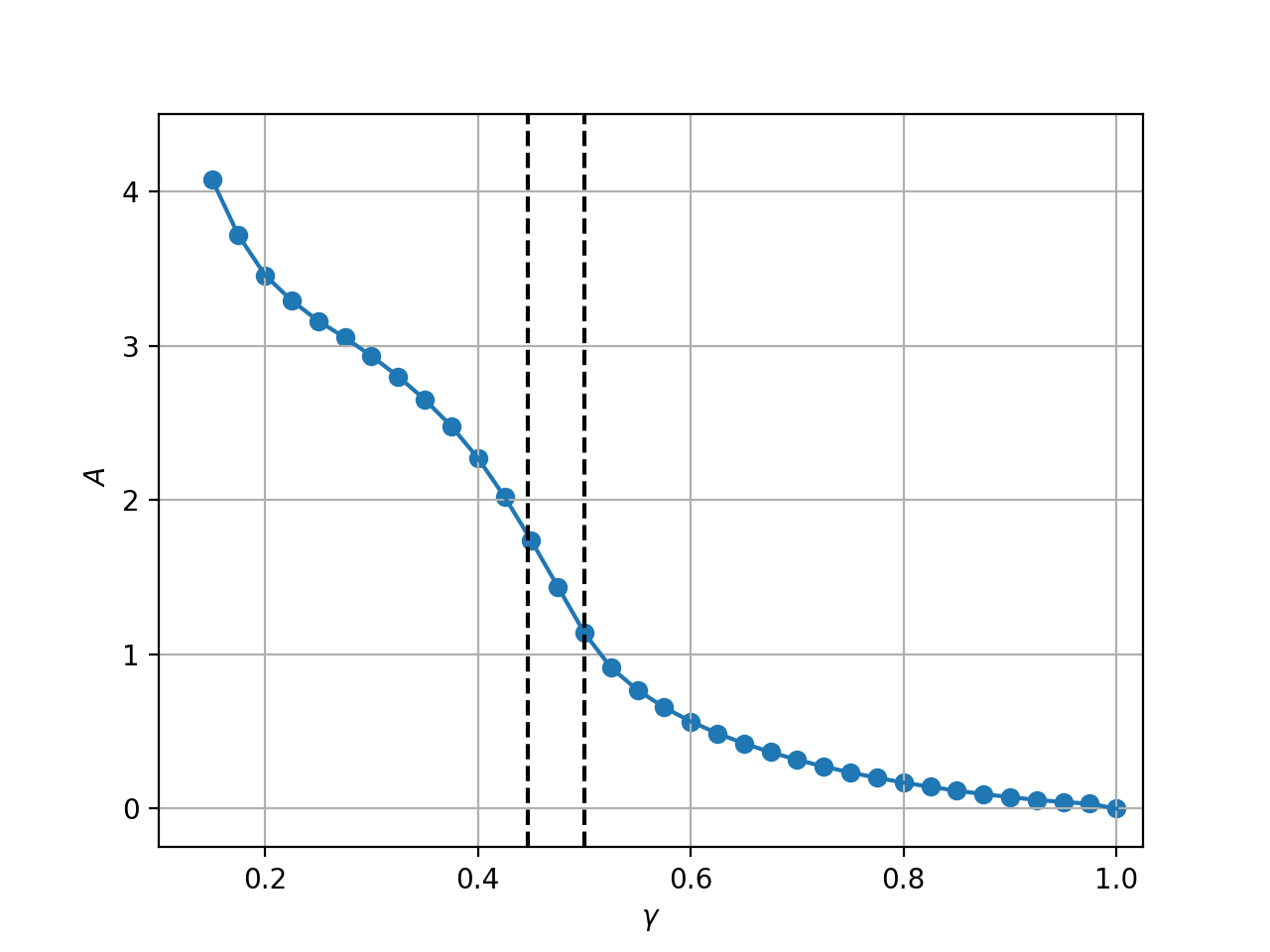}
\includegraphics[width=0.32 \textwidth]{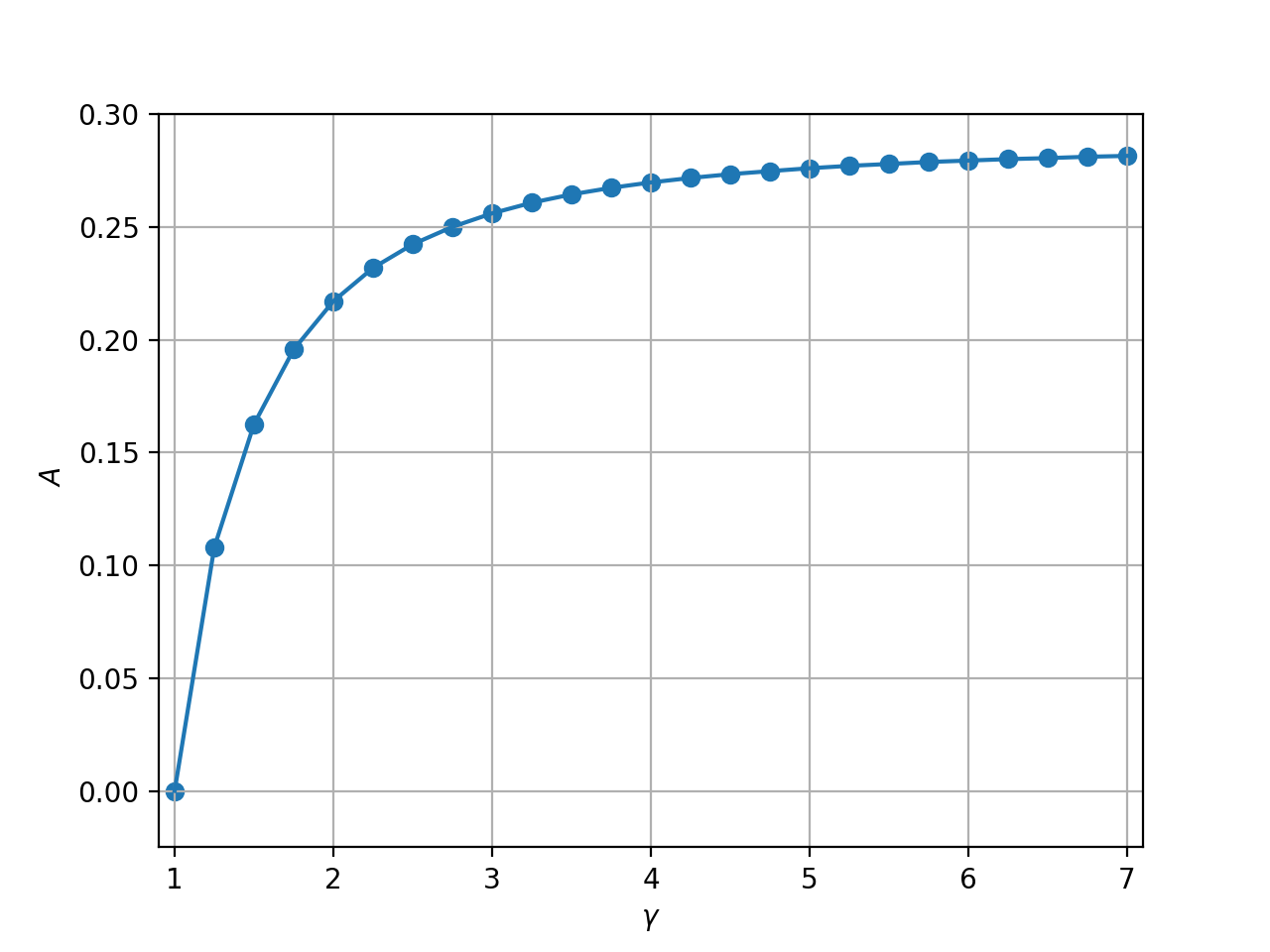}
\includegraphics[width=0.32 \textwidth]{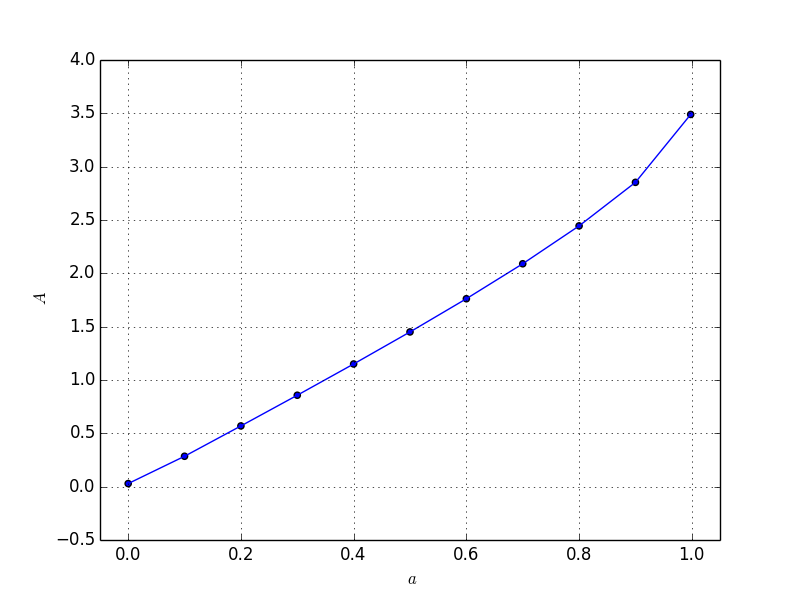}
\caption{Average radius $\langle R \rangle$ (\textit{top row}) and asymmetry parameter $A$ (\textit{bottom row}). From left ro right: $\langle R \rangle$, $A$ for $\gamma \leq 1$, (\textit{first column}), $\gamma \geq 1$ (\textit{second column}) and Kerr metric with the values of spin  $0\leq a \leq 1$ (\textit{third column}). Vertical dashed lines define the borders of regions with different ISCO structure (see text for details). \label{asymmetryplot}}

\end{figure*}

The shadow of the compact object in $\gamma$ space-time is shown in Fig.~\ref{shadimg}. One can notice that the shadow images for $\gamma<1$ and $\gamma>1$ are very different. While images for $\gamma>1$ strongly resemble Schwarzschild ones, the shadow silhouettes for $\gamma<1$ differ significantly as $\gamma$ becomes smaller. Particularly, the shadow images for $\gamma<1/2$ are clearly distinguishable from other known ones for
observers with inclination angle in the range $\pi/4\leq\iota\leq\pi/2$.
Fig.~\ref{asymmetryplot} shows $\langle R \rangle$ and $A$ as a function of $\gamma$ at an  inclination angle of $\iota=\pi/2$. For reference we include $\langle R \rangle$ and $A$ for Kerr metric as a function of spin $a$. We can see that the three cases (i.e. oblate source, prolate source and Kerr one) are substantially different.
For $\gamma=1$ the average radius $\langle R \rangle$ and asymmetry parameter $A$ are equal to Schwarzschild black hole shadow radius and zero, respectively. As $\gamma$ increases in $\gamma>1$ range, $\langle R \rangle$ and $A$ also increase and converge to their maximum values. The most interesting case is when $0<\gamma<1$. As $\gamma$ moves from $1$ to $0$, a slow downgrade of the parameter $\langle R \rangle$ starts to sharpen before $\gamma=0.5$, reaching the minimal value at around $\gamma\simeq 0.225$ followed by an increase. The asymmetry parameter $A$ increases departing from spherical symmetry for both kinds of sources. This effect for $\gamma\leq 1$ may be due to the appearance of a repulsive effect in the gravitational field close to $r=2m$ within a certain range of values of $\gamma$ as it shall be discussed below.
This suggests that, if such effects can be measured from observations, then deformations would produce features that would allow distinguishing the $\gamma$-metric from a black hole. 

\subsection{Gravitational lensing and deflection angle \label{lensing}}
We now calculate the deflection angle for photons using the same ray-tracing code discussed above. The algorithm of the process is similar to the one described in section ~\ref{rtcs}, which is the modified version of ~\cite{Ayzenberg18} and ~\cite{Gott19} that follows the method described in ~\cite{Psaltis12}.

\begin{figure*}
	\includegraphics[width=0.32 \textwidth]{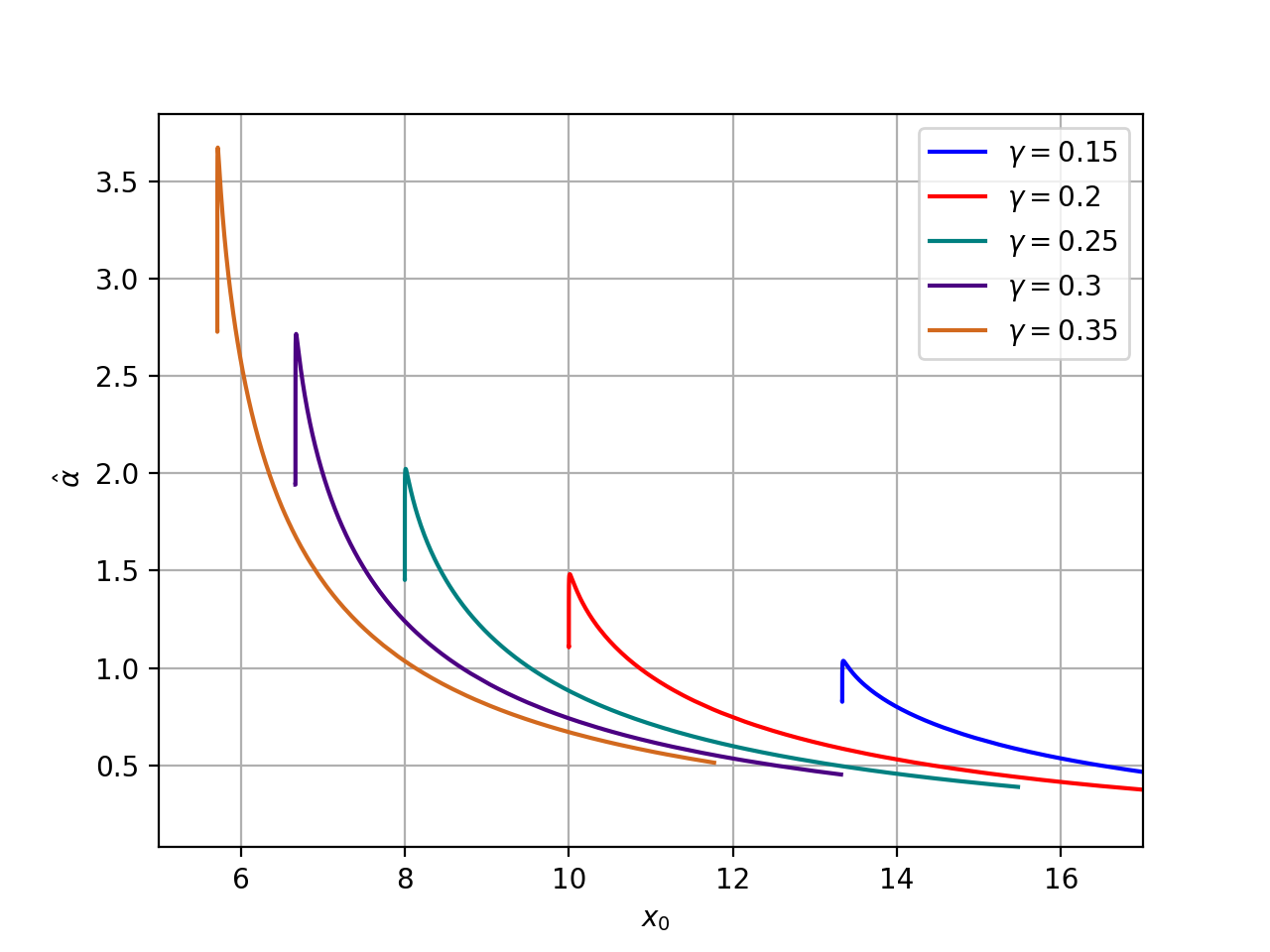}
	\includegraphics[width=0.32 \textwidth]{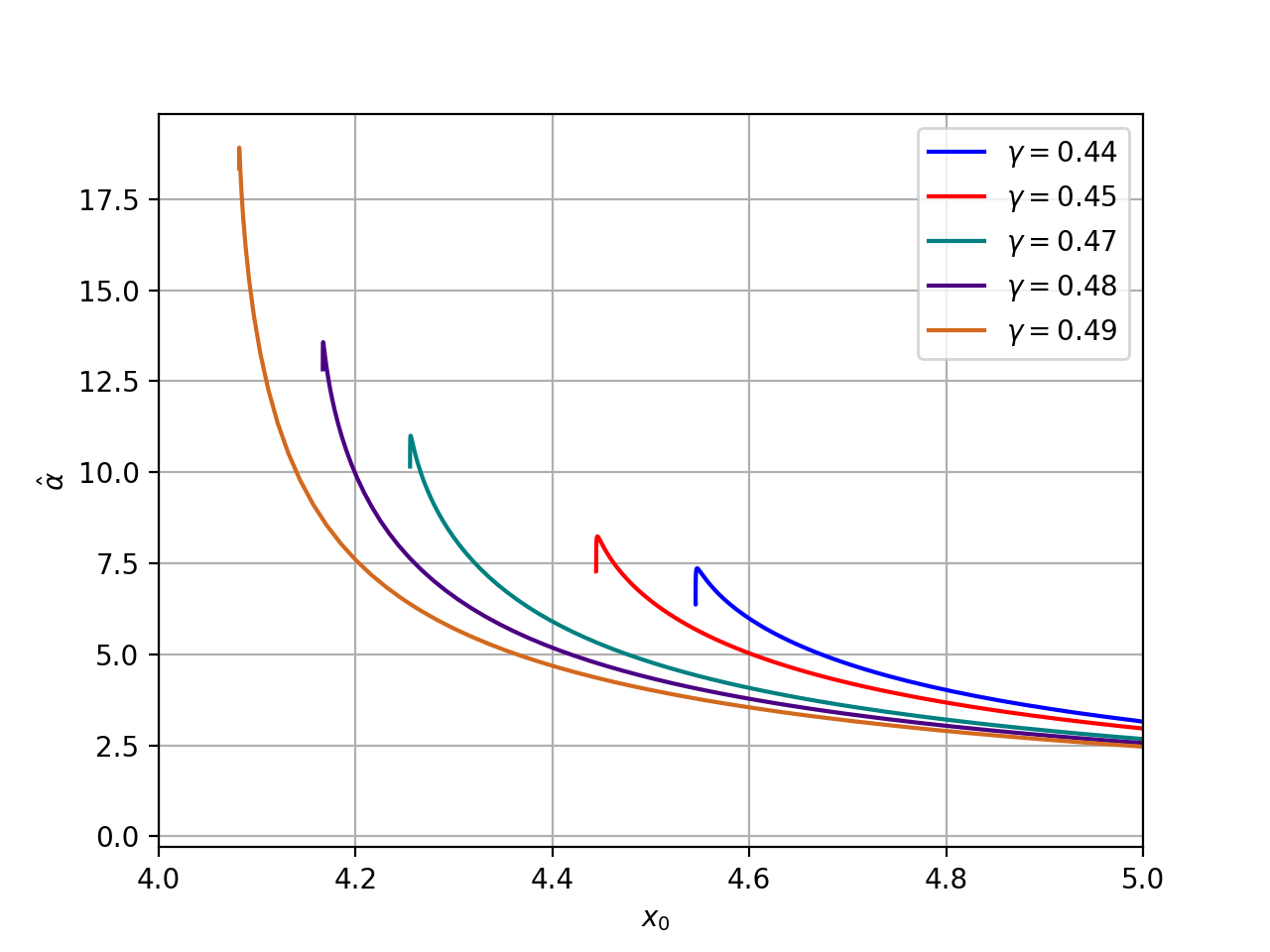}
	\includegraphics[width=0.32 \textwidth]{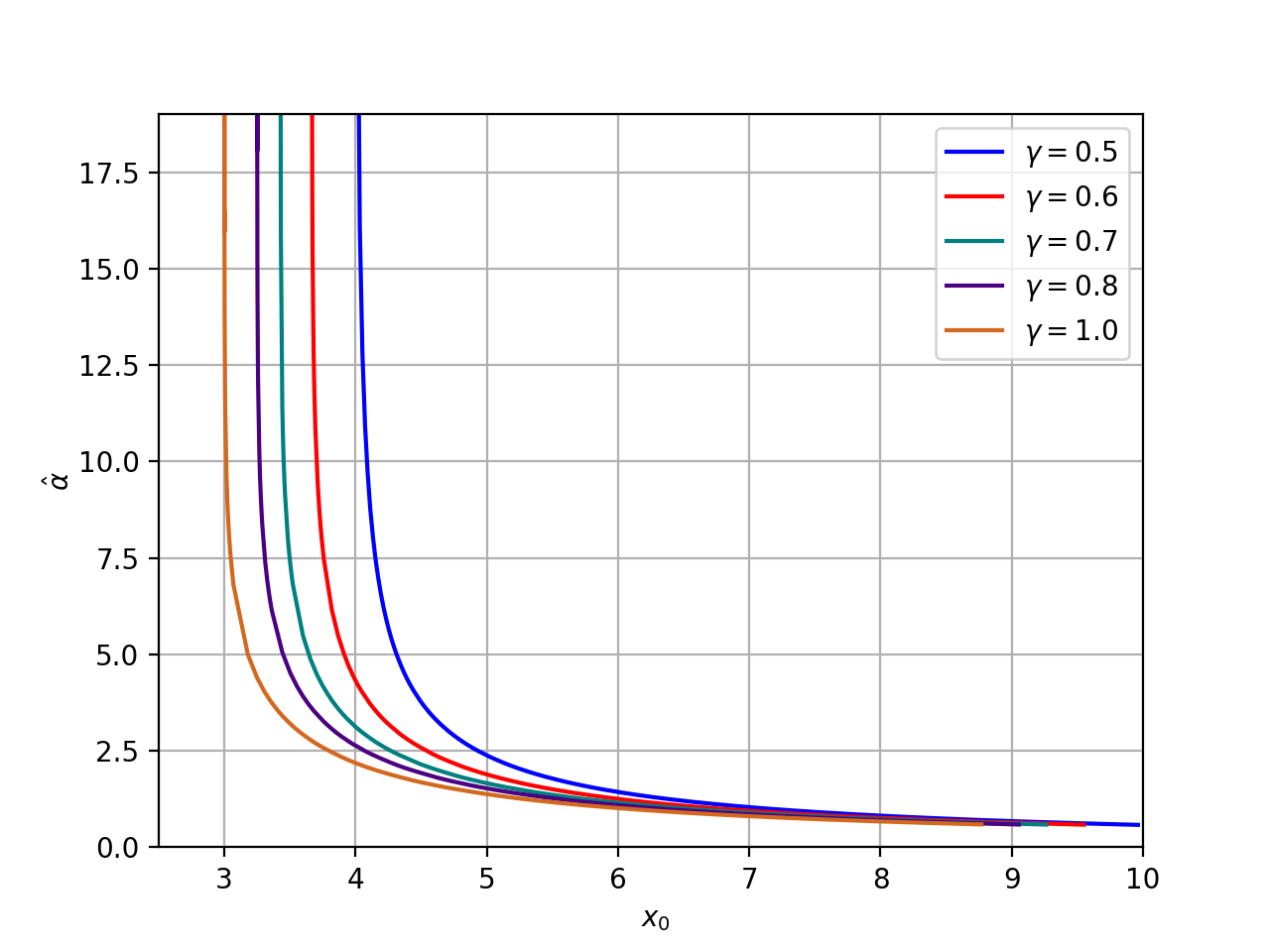}

	\includegraphics[width=0.32 \textwidth]{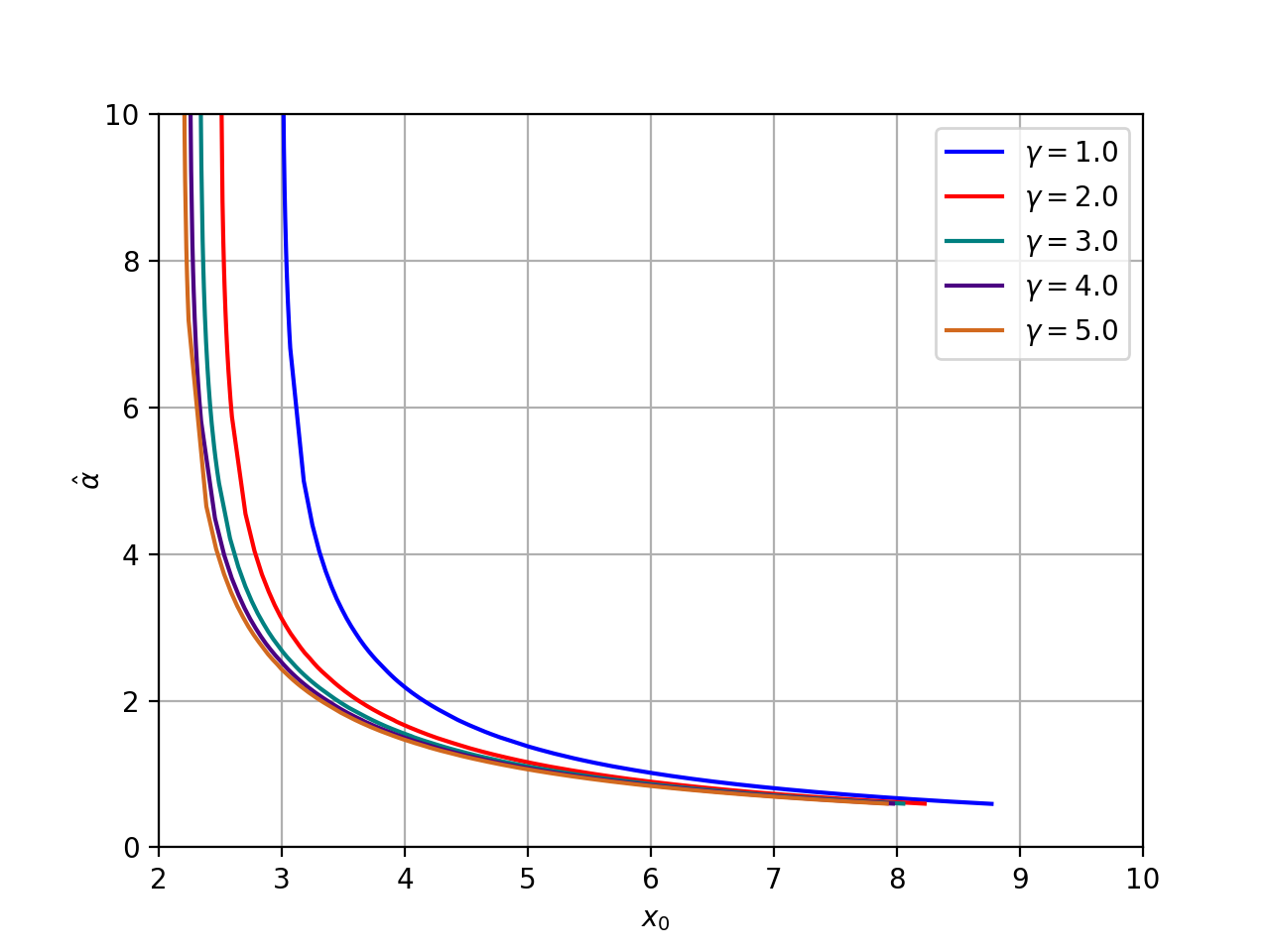}
	\includegraphics[width=0.32 \textwidth]{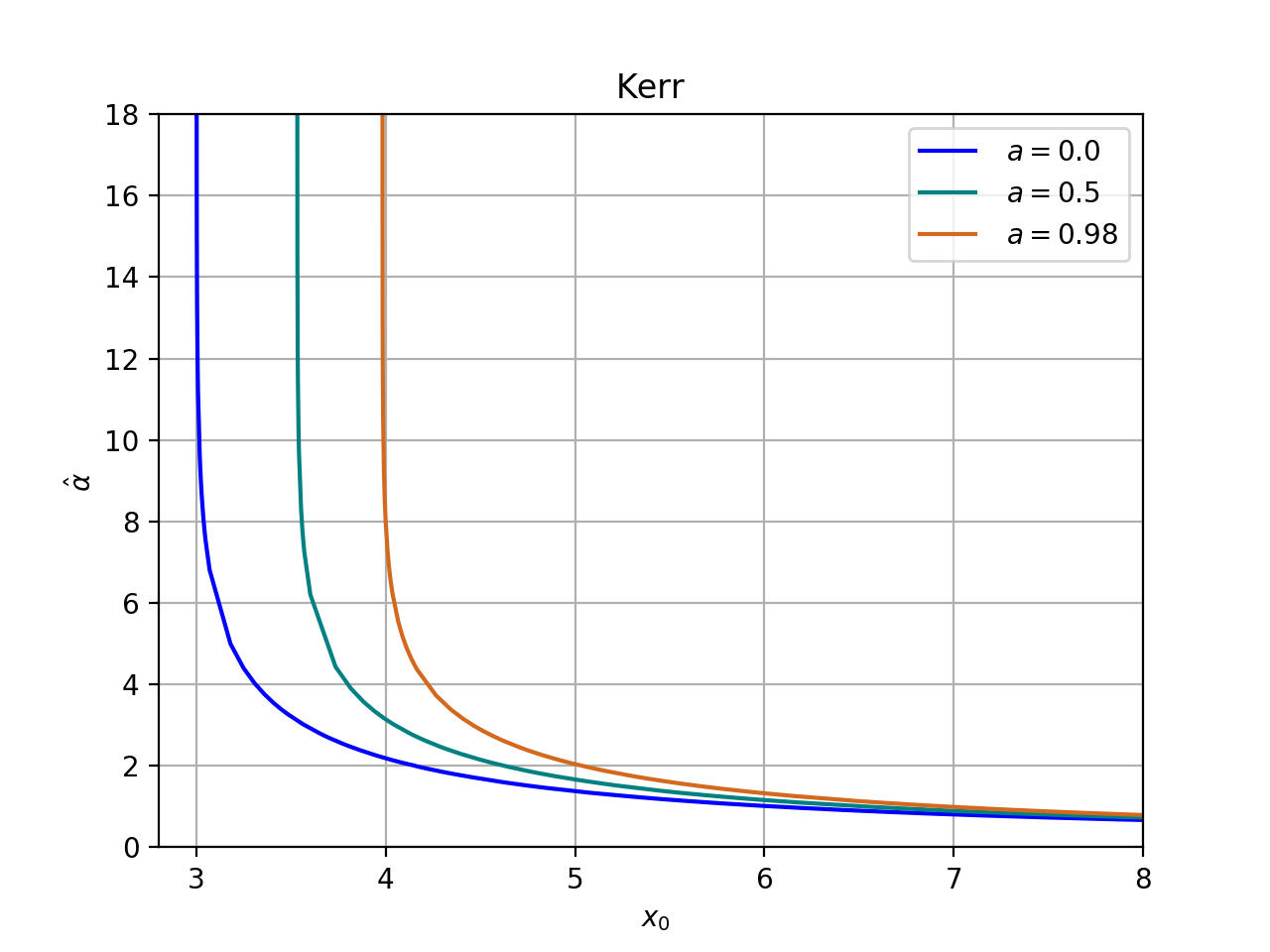}
	\caption{The deflection angle $\hat{\alpha}$ as a function of impact parameter $x_0=b$ for five different values of $\gamma$ and for Kerr case with three spin values. For $\gamma<1/2$ the deflection angle first increases to a maximum value, followed by a decrease as the impact parameter gets smaller. This can be explained by a repulsive nature of the gravitational field of the $\gamma$ space-time for this range of values for $\gamma$. The deflection angle increases as the photon approaches the photon capture surface for $\gamma \geq 1/2$. As expected, the cases when $\gamma=1$ and $a=0$ for Kerr are the same. \label{deflection_angle}}
\end{figure*}

\begin{figure*}
\includegraphics[width=0.32 \textwidth]{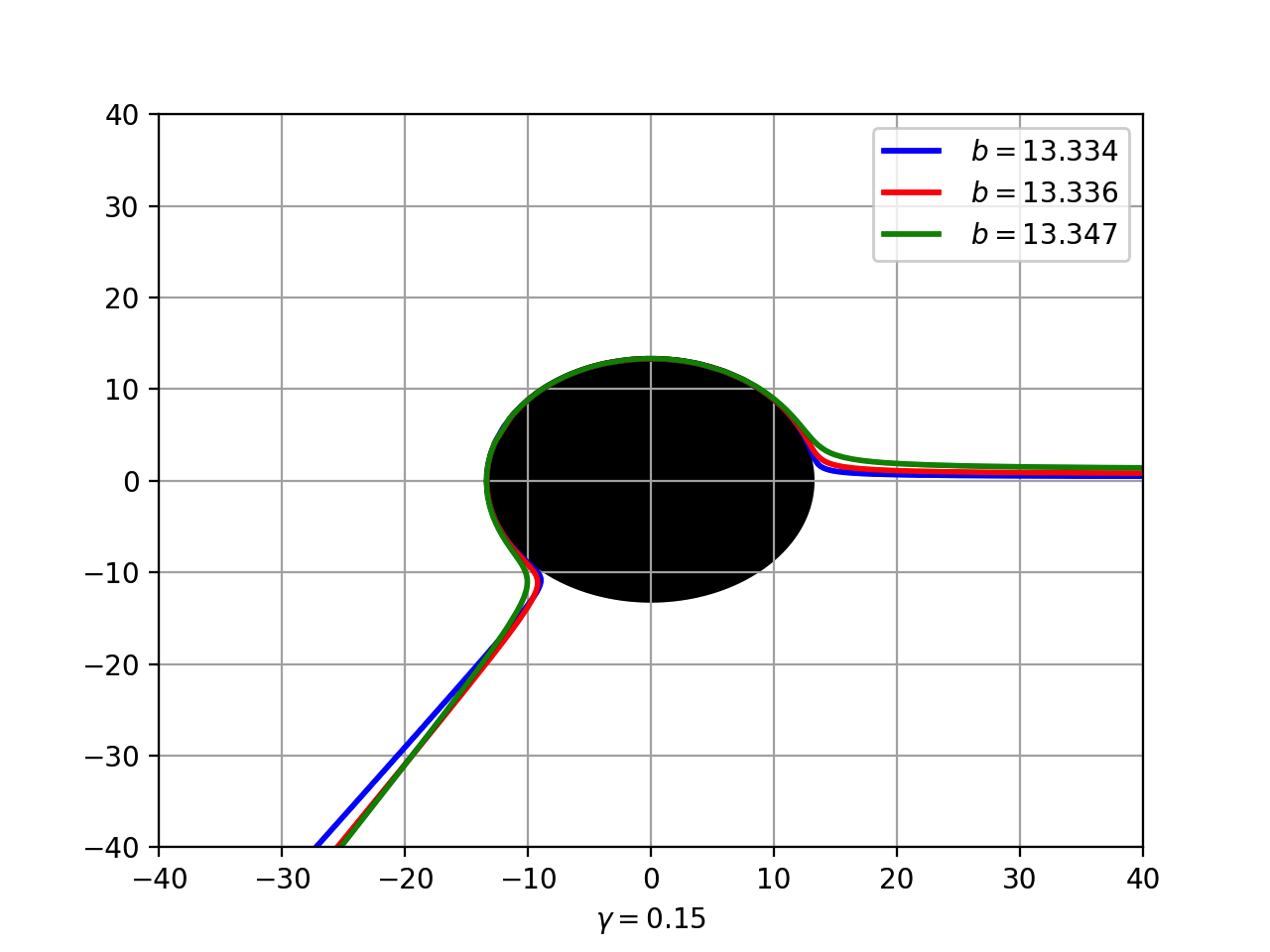}
\includegraphics[width=0.32 \textwidth]{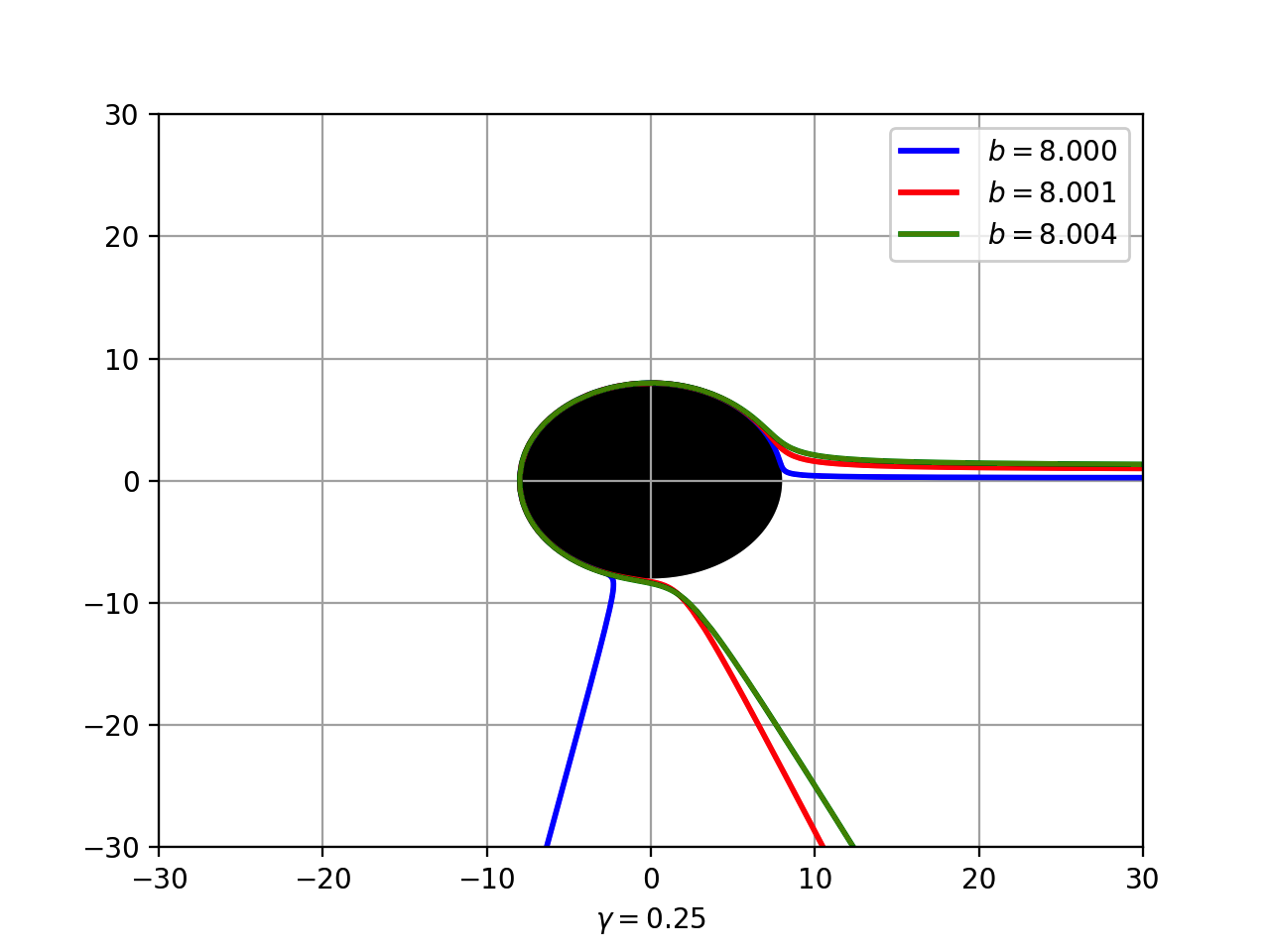}
\includegraphics[width=0.32 \textwidth]{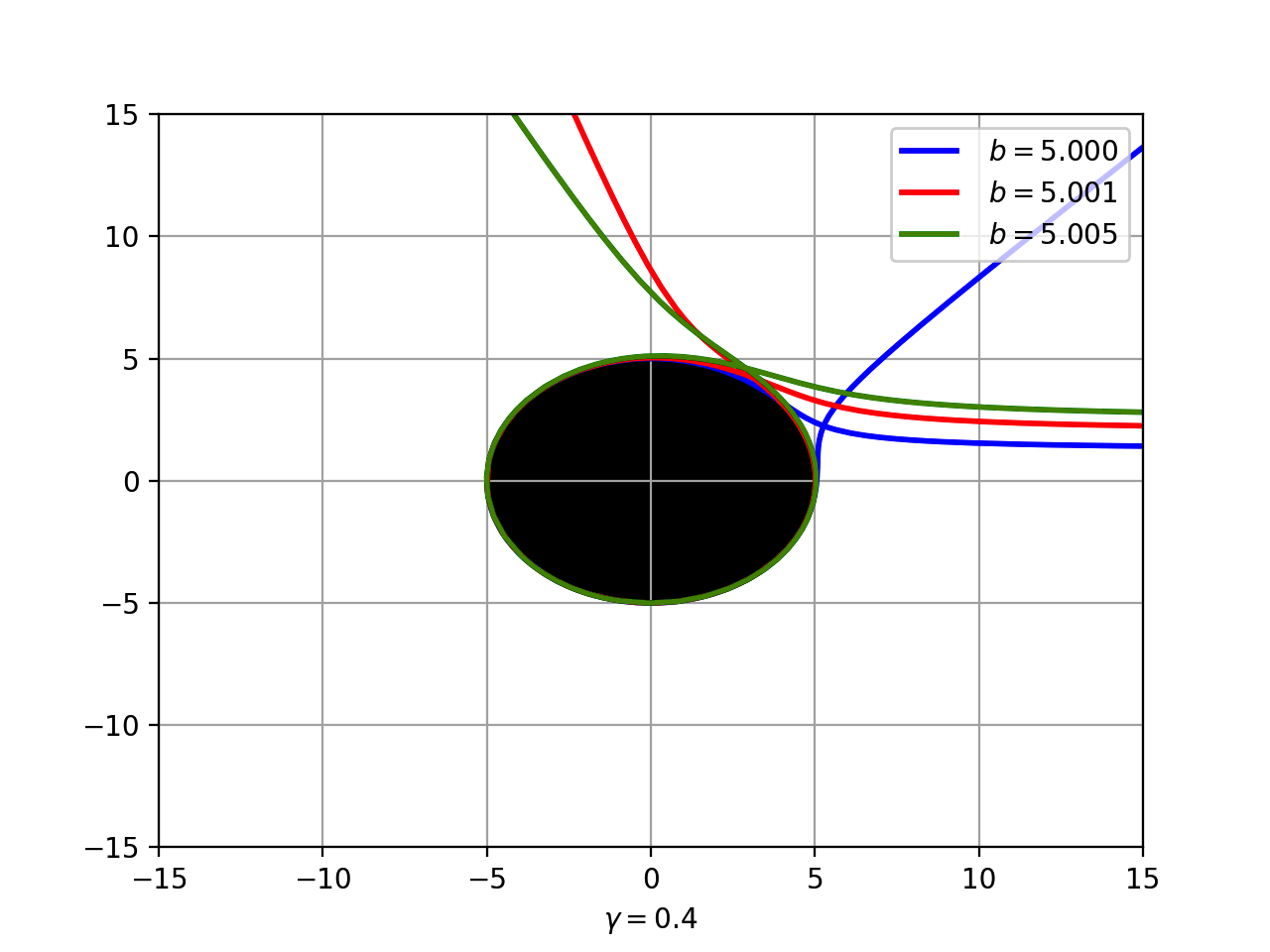}

\includegraphics[width=0.32 \textwidth]{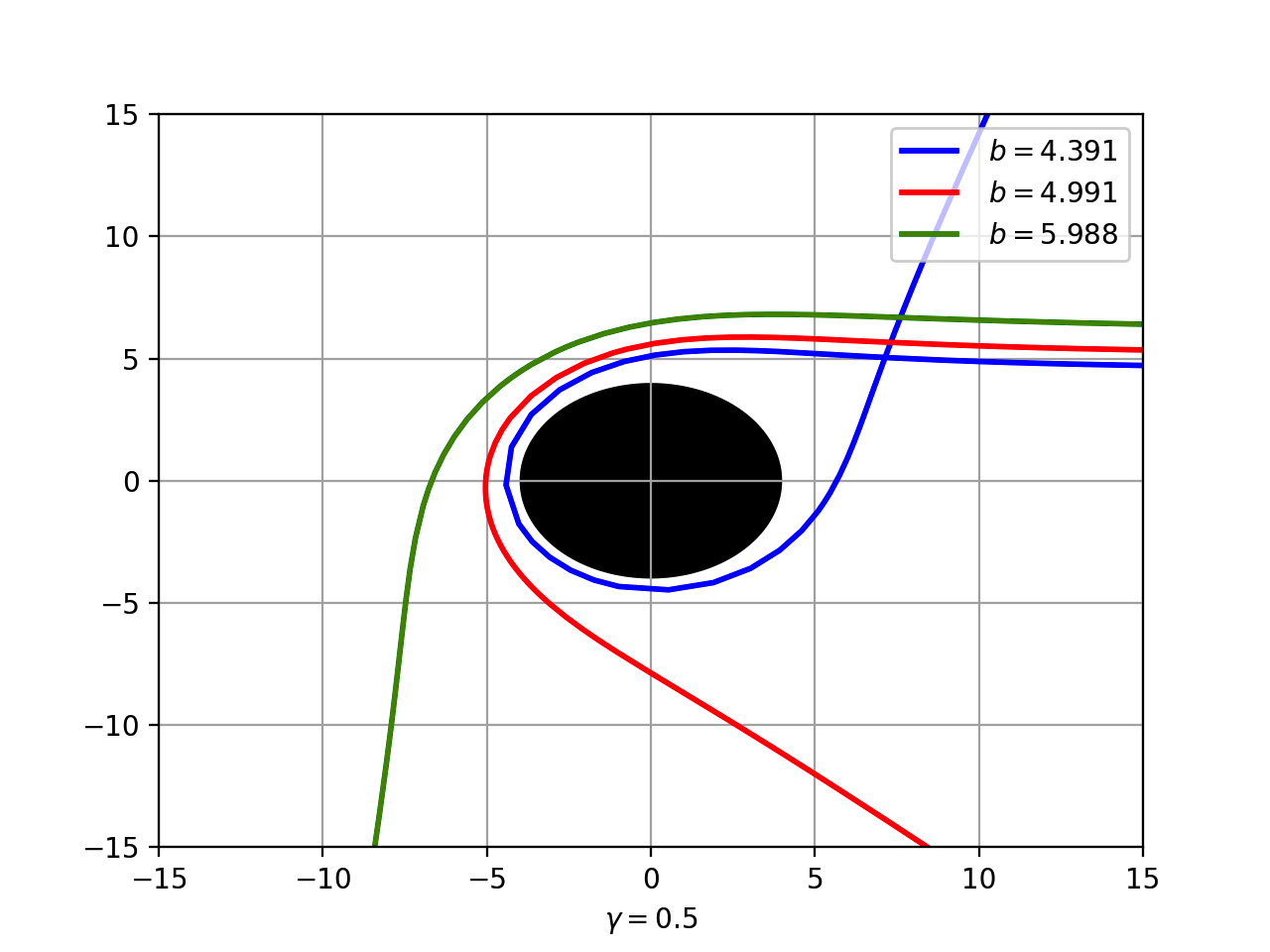}
\includegraphics[width=0.32 \textwidth]{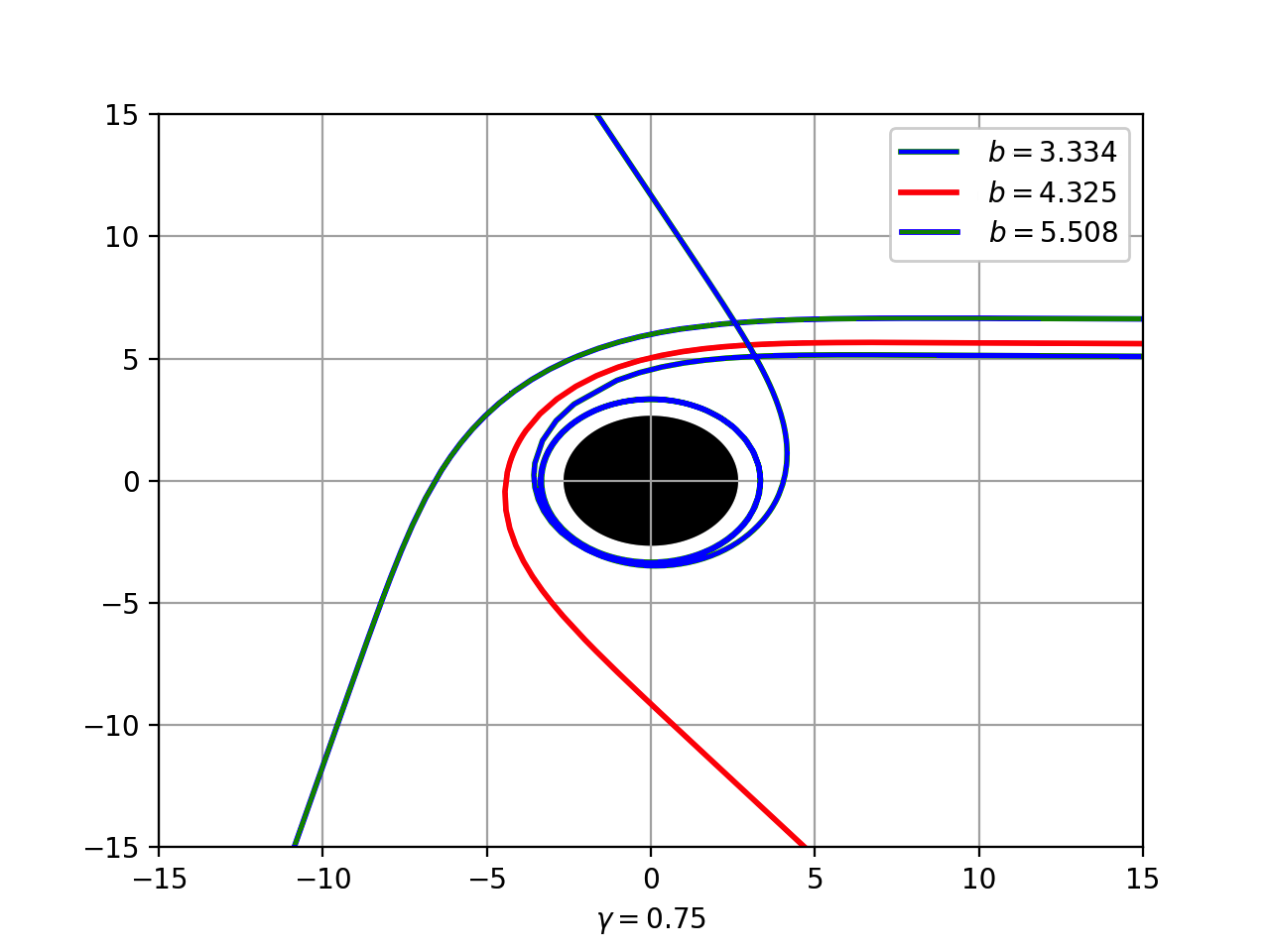}
\includegraphics[width=0.32 \textwidth]{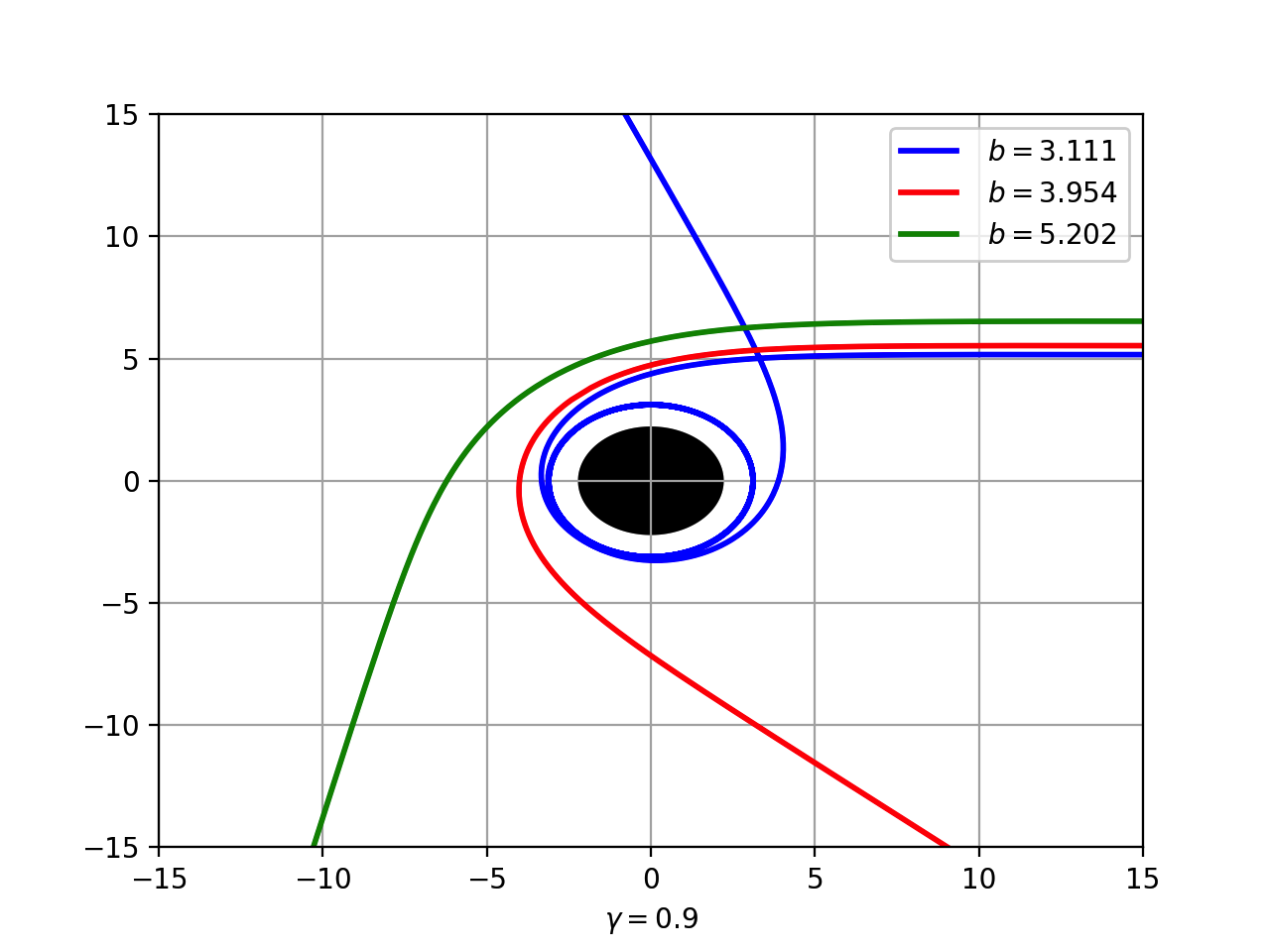}

\includegraphics[width=0.32 \textwidth]{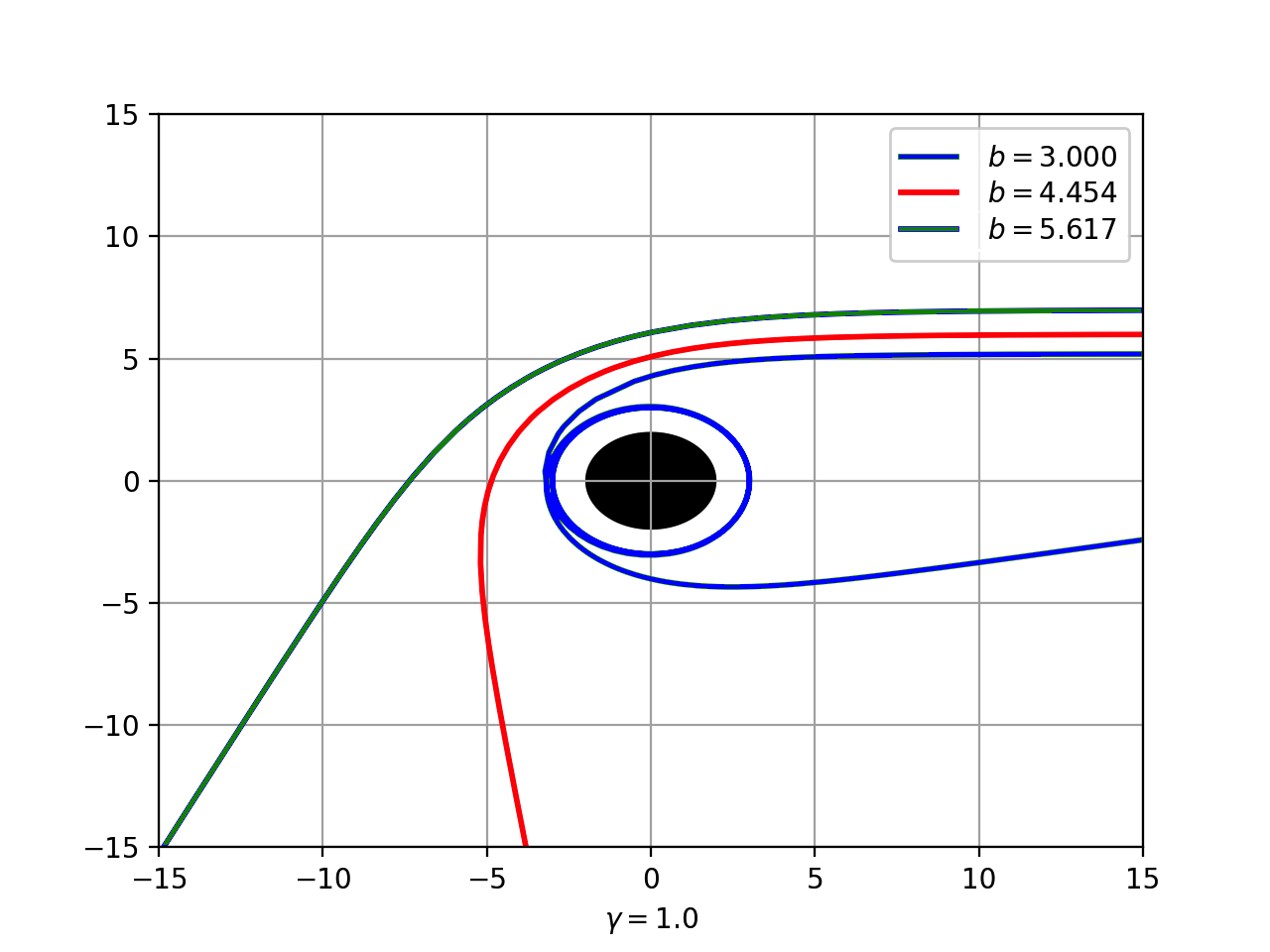}
\includegraphics[width=0.32 \textwidth]{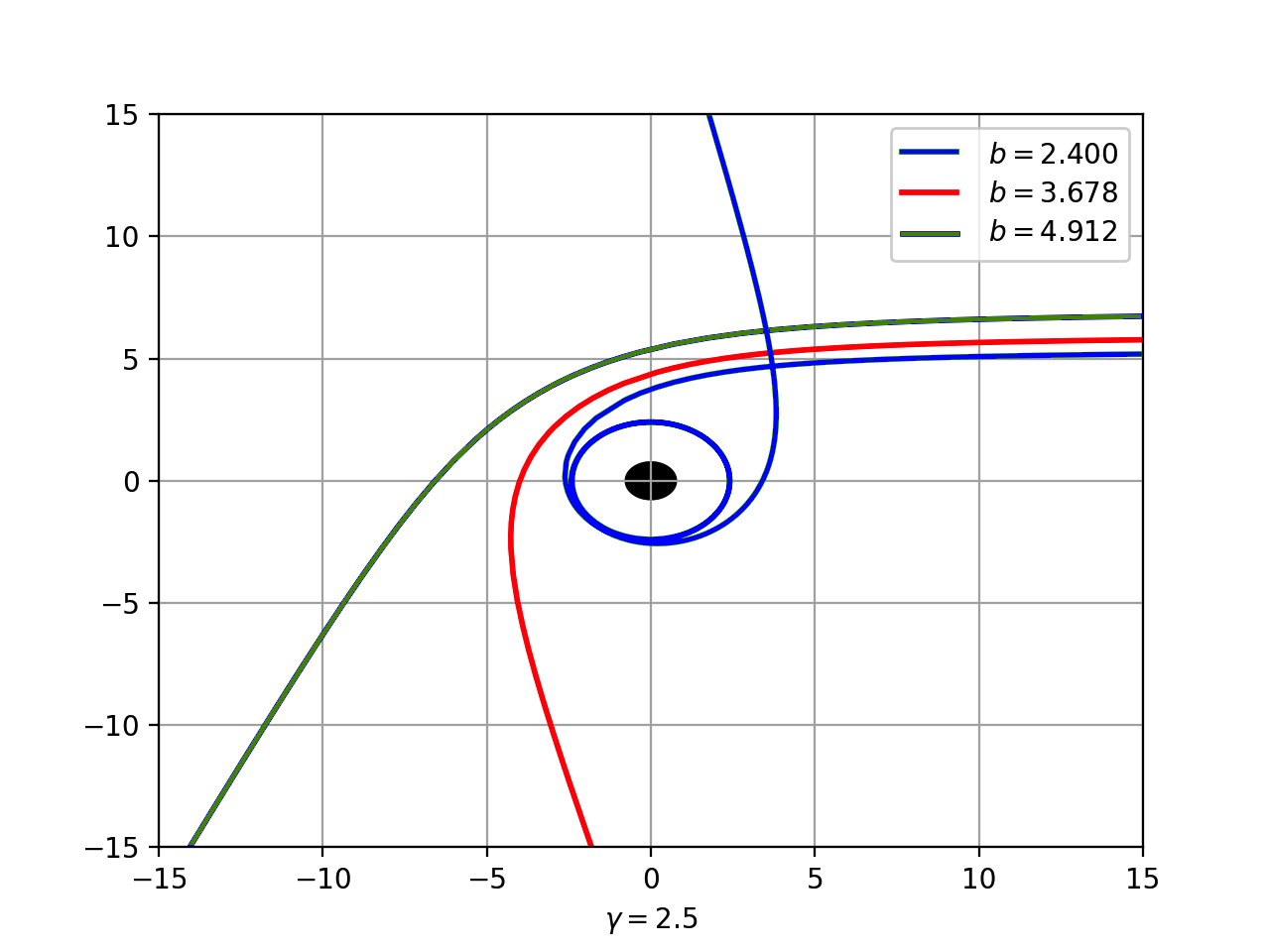}
\includegraphics[width=0.32 \textwidth]{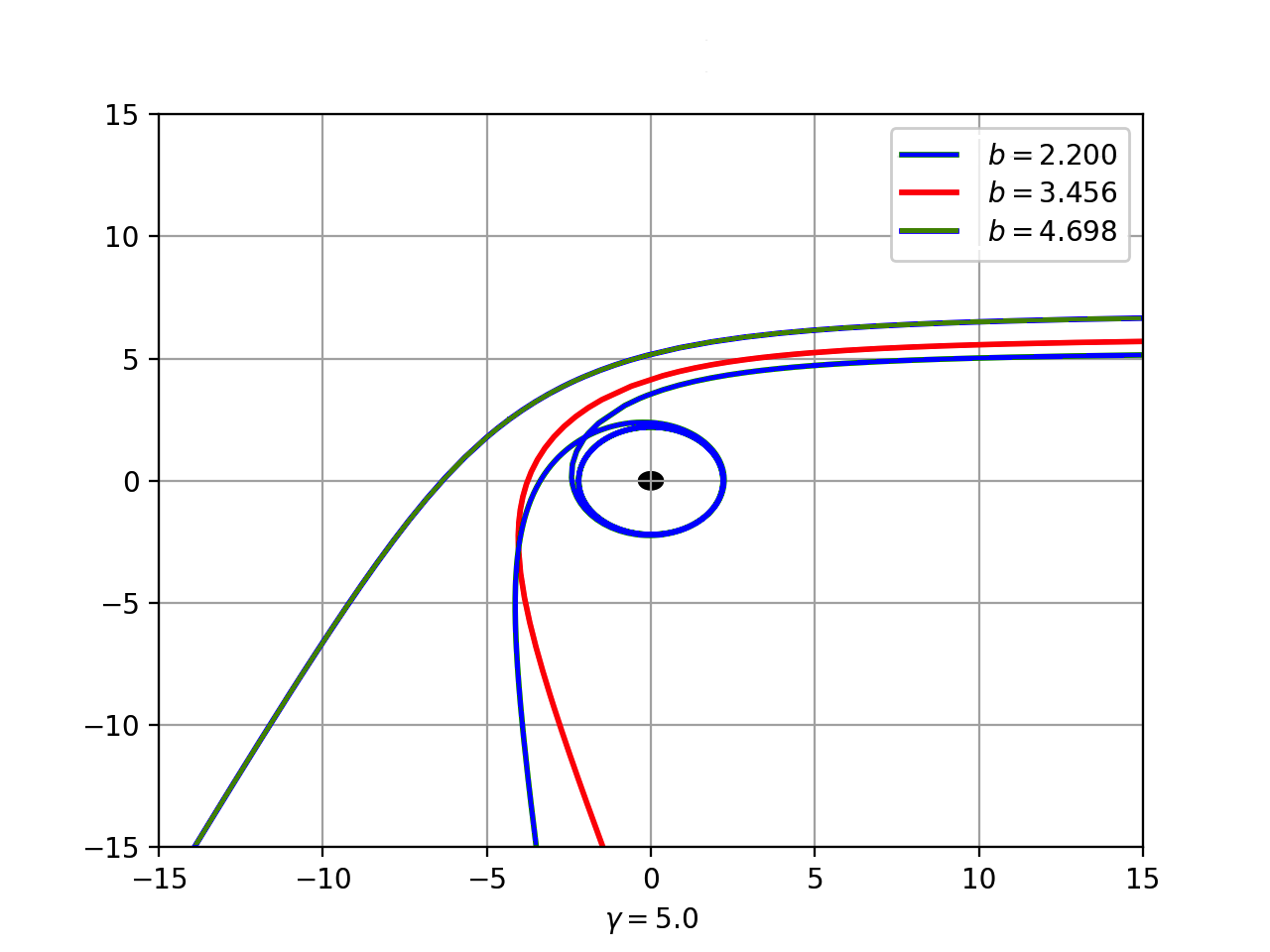}

\includegraphics[width=0.32 \textwidth]{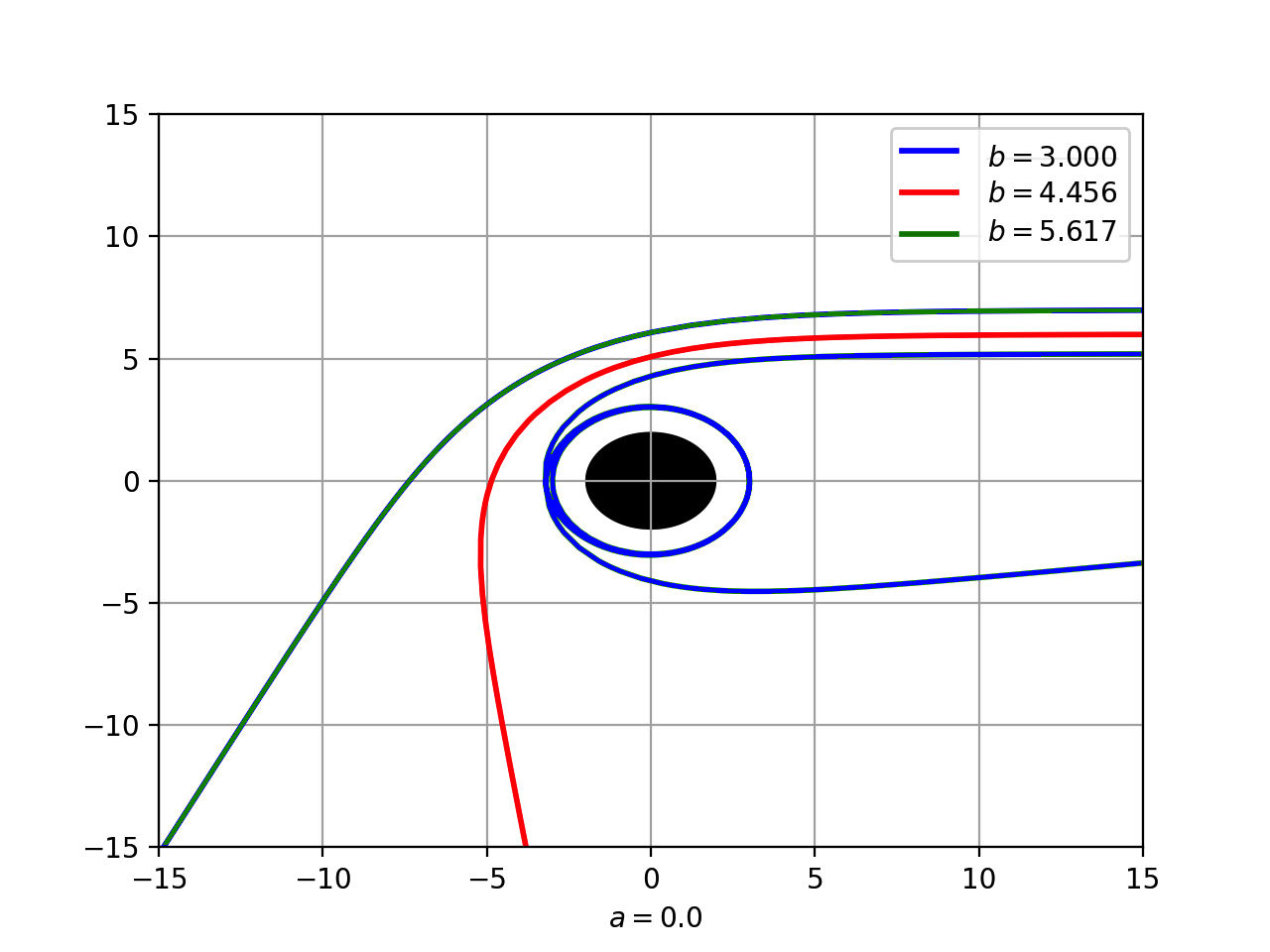}
\includegraphics[width=0.32 \textwidth]{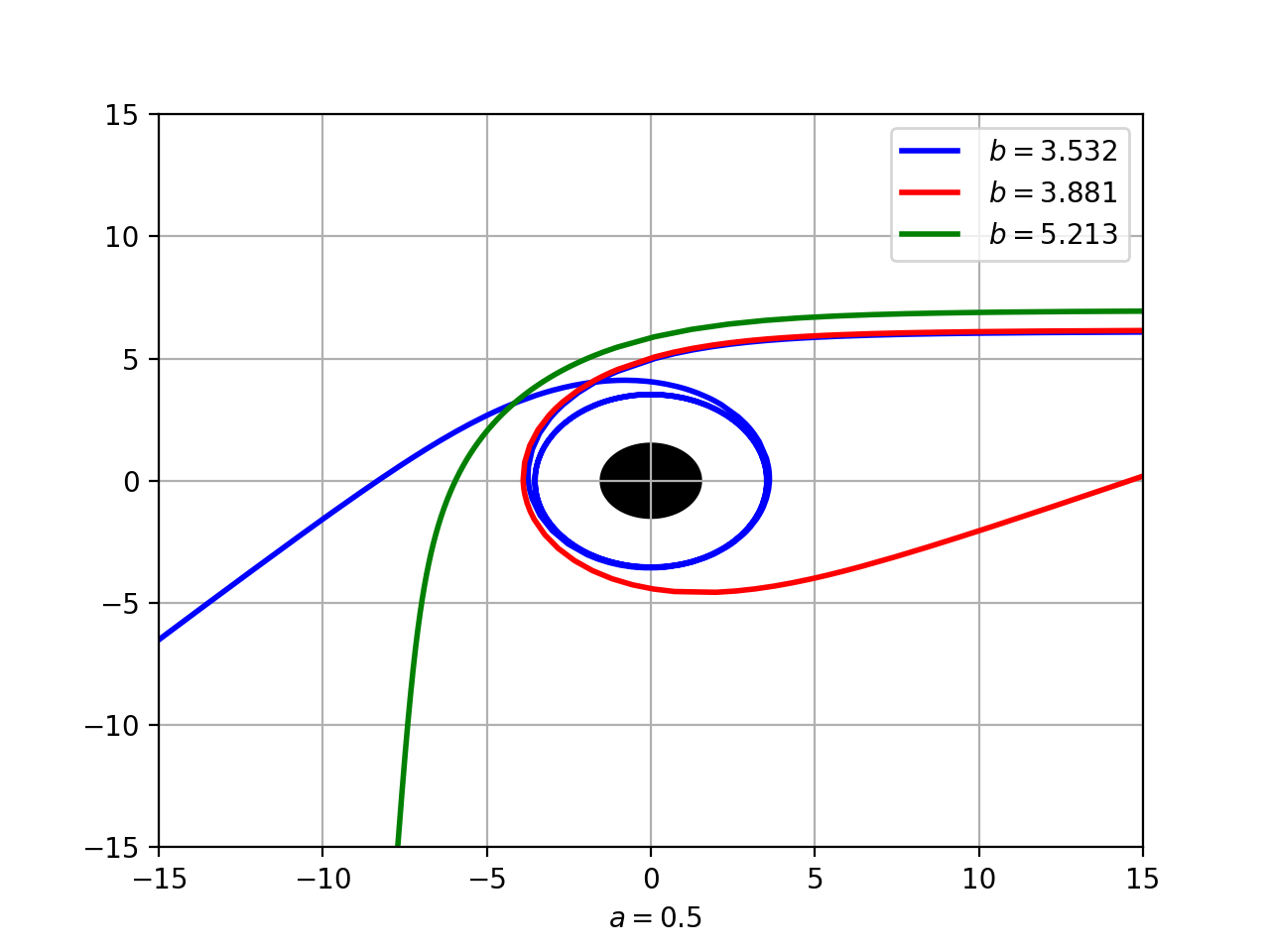}
\includegraphics[width=0.32 \textwidth]{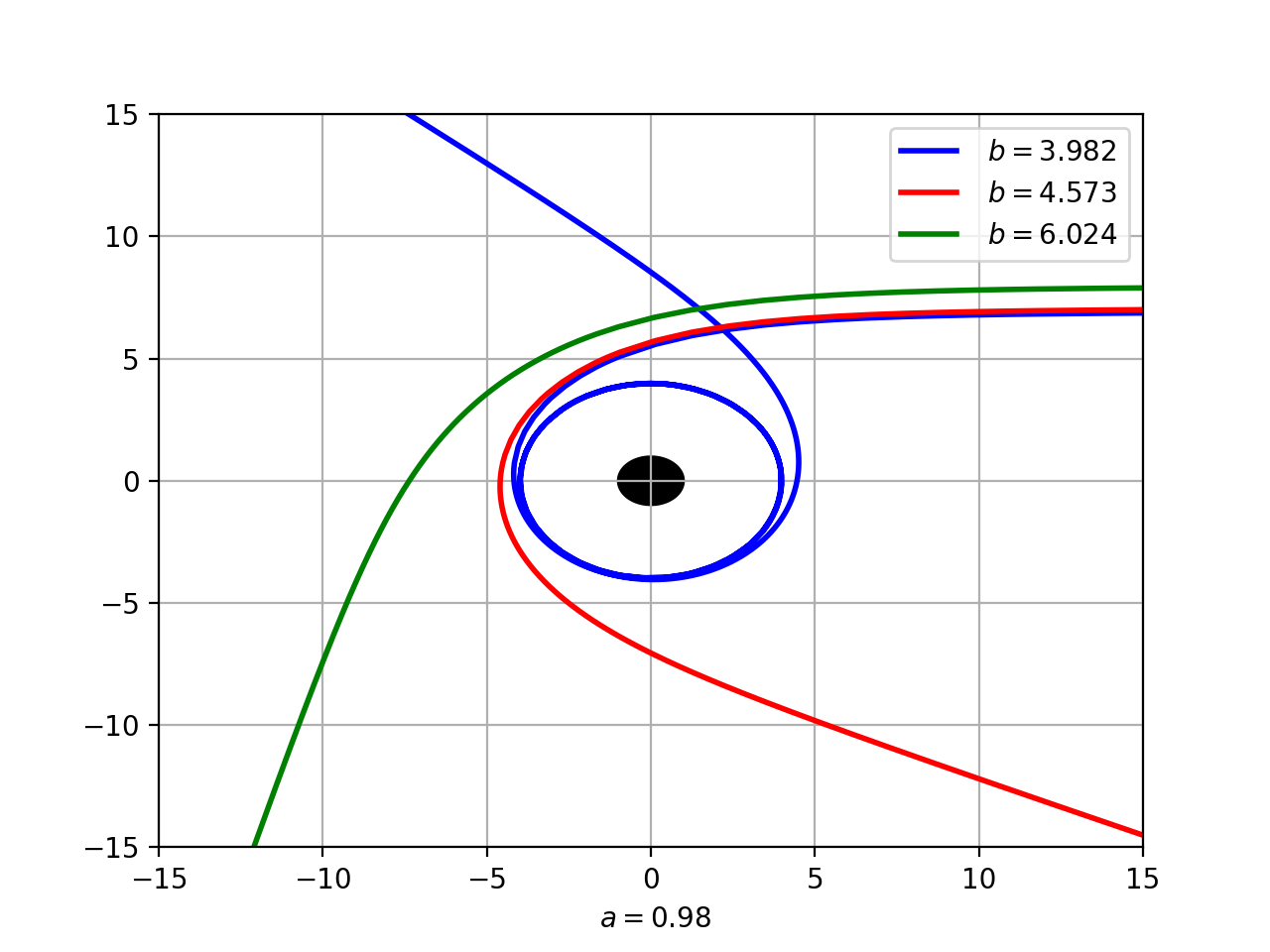}

\caption{Photon trajectories in the $\gamma$-metric in the equatorial plane. From left to right, \textit{first row}: photon trajectories for $\gamma=0.15$ (\textit{left}), $\gamma=0.25$ (\textit{middle}) and $\gamma=0.4$ (\textit{right}), \textit{second row}: photon trajectories for $\gamma=0.45$ (\textit{left}), $\gamma=0.5$ (\textit{middle}) and $\gamma=0.75$ (\textit{right}), \textit{third row}: photon trajectories for $\gamma=1$ (\textit{left}), $\gamma=2.5$ (\textit{middle}) and $\gamma=5$ (\textit{right}) with three impact parameters. \textit{fourth row}: photon trajectories around Kerr black hole with $a=0$ (\textit{left}), $a=0.5$ (\textit{middle}) and $a=0.98$ (\textit{right}). From the top row, we can see the repulsive regime for $\gamma<1/2$. The photons with the smallest impact parameters deviate to a smaller angle than those with bigger ones.\label{photon_trajectory}}

\end{figure*}

We consider the case when the whole trajectory of the photon is limited on the equatorial plane of the $\gamma$-metric. The deflection angle is calculated in the following way: Photons on the screen are initialized with some celestial coordinates $(\alpha,\beta)$. The conditions $\alpha\neq0$, $\beta=0$ and $\iota=\pi/2$ in Eqs. (\ref{in_pos_r})-(\ref{in_momen_t}) provide $\theta=\pi/2$, $\dot{\theta}=0$ with non-zero $\dot{r}$, $\dot{\phi}$ and restrict photon trajectory to lie in equatorial plane. We choose $\alpha$ in such a way that photons approach the photon ring of the $\gamma$-metric to a minimal distance $d=10^{-7}M$ with impact parameter $b$ but don't cross it. Then photons reach $r>d=1000$. We capture the initial and final positions of the photon and calculate deflection angle from a straight line. Fig.~\ref{deflection_angle} shows the calculated values of the deflection angle as a function of photon's impact parameter $b$ for different values of $\gamma$.  For the comparison the dependence of the deflection angle for the compact object described by Kerr space-time is also given. From the figures, one can see that for $\gamma \geq 1/2$ the deflection angle increases as the photon approaches the photon capture surface. However, for $\gamma < 1/2$ the deflection angle first increases then starts to decrease as photon gets closer to the photon capture surface before being caught by the central object. This can be explained by the existence of the repulsive regime depending on the value of $\gamma$. One can also notice that for some values of $\gamma$ the maximum deflection angle is less than $2\pi$. It suggests that we will not see relativistic Einstein rings for this range of $\gamma$. The repulsive effect can also be seen from Fig~\ref{photon_trajectory}, which shows trajectories of the photons around the central object on an equatorial plane with three values of photon's impact parameter $b$, for different values of $\gamma$. The photon trajectories around Kerr black hole with three different spins are included as a reference. The top row plots, which correspond to $\gamma<1/2$, show that the closest to the central object photon deviates to a smaller angle than those with bigger impact parameters. From the second row, one can notice that the deflection angle increases steadily as the impact parameter reduces. For $\gamma \geq 1/2$ we no longer see the repulsive effects. Once again, this suggests that it would be possible to qualitatively distinguish a black hole from the $\gamma$-metric for values of $\gamma\leq 1/2$. On the other hand, values of $\gamma$ close to one would not allow to easily distinguish the geometry from a black hole geometry.


Considering astrophysical objects it is interesting to check whether the shadow of a non rotating axially symmetric source could mimic a Kerr black hole. In the case of Kerr black hole the non vanishing angular momentum leads to an asymmetry in the image of the accretion disk which does not appear in the case of static objects, therefore an extreme Kerr solution can in principle be distinguished from the $\gamma$-metric for any value of $\gamma$.
Furthermore, from figure 5 we see that the average radius $\langle R \rangle$ and asymmetry parameter in the $\gamma$-metric have considerably different behaviour for oblate and prolate sources. In the oblate case ($\gamma \geq 1$) the $\langle R \rangle$ and $A$ depart only slightly from the Schwarzschild value. On the other hand, in the prolate case ($\gamma\leq1$) the values of $\langle R \rangle$ and $A$ for a given value of $1/2\leq\gamma\leq 1$ could mimic the corresponding values for a Kerr black hole.
Similarly the deflection angle for photons and the trajectory of photons in the space-time may be used to determine possible degeneracies between the $\gamma$-metric and Kerr. In figures 6 and 7 it can be seen that values of $\gamma\geq1/2$ show similar qualitative behaviour to the Kerr case, while the repulsive behaviour that appears for $\gamma\leq1/2$ does not appear in the Kerr metric.
However, it must be noted that the possible degeneracy can be broken if one is able to measure several properties of the space-time simultaneously. For example, not considering experimental uncertainties, the simultaneous determination of the ISCO and the photon capture orbit would uniquely determine whether the space-time is well described by a static or a rotating object.

\section{Conclusion\label{Summary}}

In the present work we have investigated the optical properties of the $\gamma$ space-time. Particularly, we have studied the photon motion around the gravitational source and the expected shape of the shadow that would be measured by distant observers depending on different values of the deformation parameter. 
In the limiting case when $\gamma =1$ we get the known results of Schwarzschild black hole's shadow. With the increase of $\gamma $ one may observe the increase of the size of the shadow and increase of the distortion parameter. However no significant difference arise from the black hole case, which suggest it may be difficult to distinguish the two without accurate measurements. With the decrease of the $\gamma$ parameter for  $\gamma<1$ we observe decrease of the average radius of the shadow. Noticeable differences arise for $\gamma\leq1/2$. The average radius of the shadow reaches a minimum around $\gamma\simeq 0.225$ and then increases for smaller values of $\gamma$. Also for a range of values of the deformation parameter repulsive effects appear in the vicinity of the singular surface, in striking contrast with the corresponding situations for black holes. 
Therefore, the shadow images for $\gamma<1/2$ are clearly distinguishable from other known ones, while images obtained for $\gamma$ closer to one are not easily distinguishable.

Using the ray-tracing code we have investigated the gravitational lensing, particularly studied the dependence of deflection angle on the parameter $\gamma$. 
The analysis showed that, as expected, for $\gamma>1/2$ the deflection angle increases with the decrease of impact parameter. However, for $\gamma \leq 1/2$ the deflection angle first increases then starts to decrease as photon gets closer to the photon capture surface before being caught by the central object. This is the indication of the repulsive character of the space-time for small values of the $\gamma$ parameter. 
Note that repulsive effects in exact solutions of Einstein's equations can occur. For example a similar phenomenon of gravitational repulsion in the general theory of relativity had been discussed in~\cite{Arifov81}. 
In the case of the $\gamma$-metric one could use this observation to argue for the physical validity of such geometry, in the vicinity of the singularity, for values of $\gamma\leq 1/2$ or to suggest a possible yet unobserved astrophysical effect in the vicinity of extremely prolate ($\gamma<<1$) compact objects.

Concerning the possibility to distinguish the geometry of the $\gamma$-metric from a black hole geometry through the observation of the shadow, our results show that only precise measurements of the metric coefficients obtained from observations would allow to distinguish a black hole from the $\gamma$ space-time when is close to one and $\gamma>1$.

Future observations of the the shadow of the super-massive black hole candidates in the Milky Way galaxy (Sgr-A*) and in the galaxy M87 will allow to test for the first time the validity of the hypothesis that such objects must be black holes
\cite{Goddi17,Falcke17}.
However, our work on the shadow of the $\gamma$-metric suggests that very precise measurements will be needed in order to rule out an exotic compact object described by this geometry with $\gamma\simeq 1$.


\section*{acknowledgements}

The work of A.B.A., A.A.A., D.A., and C.B. was supported by the Innovation Program of the Shanghai Municipal Education Commission, Grant No.~2019-01-07-00-07-E00035, and Fudan University, Grant No.~IDH1512060. A.B.A. also acknowledges the support from the Shanghai Government Scholarship (SGS). This research is supported by Grants No. VA-
FA-F-2-008, No. MRB-AN-2019-29  and No. YFA-Ftech-2018-8 of the Uzbek-
istan Ministry for Innovational Development, and by the
Abdus Salam International Centre for Theoretical
Physics through Grant No.~OEA-NT-01.
This research is partially supported by an
Erasmus+ exchange grant between SU and NUUz.
D.M. acknowledges support by Nazarbayev University Faculty Development
Competitive Research Grant No. 090118FD5348 and by
the Ministry of Education of Kazakhstan’s target program IRN: BR05236454.
B.A. would like to acknowledge Nazarbayev University, Astana,
Kazakhstan for the warm hospitality.
D.M. wishes to express his gratitude to Naresh Dadhich for useful discussion on the properties of the $\gamma$-metric and acknowledges the Ulugh Beg Astronomical Institute, Tashkent, Uzbekistan for the hospitality.

\bibliographystyle{apsrev4-1}  

\bibliography{gravreferences}

\begin{thebibliography}{82}%
\makeatletter
\providecommand \@ifxundefined [1]{%
 \@ifx{#1\undefined}
}%
\providecommand \@ifnum [1]{%
 \ifnum #1\expandafter \@firstoftwo
 \else \expandafter \@secondoftwo
 \fi
}%
\providecommand \@ifx [1]{%
 \ifx #1\expandafter \@firstoftwo
 \else \expandafter \@secondoftwo
 \fi
}%
\providecommand \natexlab [1]{#1}%
\providecommand \enquote  [1]{``#1''}%
\providecommand \bibnamefont  [1]{#1}%
\providecommand \bibfnamefont [1]{#1}%
\providecommand \citenamefont [1]{#1}%
\providecommand \href@noop [0]{\@secondoftwo}%
\providecommand \href [0]{\begingroup \@sanitize@url \@href}%
\providecommand \@href[1]{\@@startlink{#1}\@@href}%
\providecommand \@@href[1]{\endgroup#1\@@endlink}%
\providecommand \@sanitize@url [0]{\catcode `\\12\catcode `\$12\catcode
  `\&12\catcode `\#12\catcode `\^12\catcode `\_12\catcode `\%12\relax}%
\providecommand \@@startlink[1]{}%
\providecommand \@@endlink[0]{}%
\providecommand \url  [0]{\begingroup\@sanitize@url \@url }%
\providecommand \@url [1]{\endgroup\@href {#1}{\urlprefix }}%
\providecommand \urlprefix  [0]{URL }%
\providecommand \Eprint [0]{\href }%
\providecommand \doibase [0]{http://dx.doi.org/}%
\providecommand \selectlanguage [0]{\@gobble}%
\providecommand \bibinfo  [0]{\@secondoftwo}%
\providecommand \bibfield  [0]{\@secondoftwo}%
\providecommand \translation [1]{[#1]}%
\providecommand \BibitemOpen [0]{}%
\providecommand \bibitemStop [0]{}%
\providecommand \bibitemNoStop [0]{.\EOS\space}%
\providecommand \EOS [0]{\spacefactor3000\relax}%
\providecommand \BibitemShut  [1]{\csname bibitem#1\endcsname}%
\let\auto@bib@innerbib\@empty
\bibitem [{\citenamefont {{Bambi}}(2017)}]{Bambi17c}%
  \BibitemOpen
  \bibfield  {author} {\bibinfo {author} {\bibfnamefont {C.}~\bibnamefont
  {{Bambi}}},\ }\href {\doibase 10.1103/RevModPhys.89.025001} {\bibfield
  {journal} {\bibinfo  {journal} {Reviews of Modern Physics}\ }\textbf
  {\bibinfo {volume} {89}},\ \bibinfo {eid} {025001} (\bibinfo {year}
  {2017})}\BibitemShut {NoStop}%
\bibitem [{\citenamefont {{Berti}}\ \emph {et~al.}(2015)\citenamefont
  {{Berti}}, \citenamefont {{Barausse}}, \citenamefont {{Cardoso}},
  \citenamefont {{Gualtieri}}, \citenamefont {{Pani}}, \citenamefont
  {{Sperhake}}, \citenamefont {{Stein}}, \citenamefont {{Wex}}, \citenamefont
  {{Yagi}}, \citenamefont {{Baker}}, \citenamefont {{Burgess}}, \citenamefont
  {{Coelho}}, \citenamefont {{Doneva}}, \citenamefont {{De Felice}},
  \citenamefont {{Ferreira}}, \citenamefont {{Freire}}, \citenamefont
  {{Healy}}, \citenamefont {{Herdeiro}}, \citenamefont {{Horbatsch}},
  \citenamefont {{Kleihaus}}, \citenamefont {{Klein}}, \citenamefont
  {{Kokkotas}}, \citenamefont {{Kunz}}, \citenamefont {{Laguna}}, \citenamefont
  {{Lang}}, \citenamefont {{Li}}, \citenamefont {{Littenberg}}, \citenamefont
  {{Matas}}, \citenamefont {{Mirshekari}}, \citenamefont {{Okawa}},
  \citenamefont {{Radu}}, \citenamefont {{O'Shaughnessy}}, \citenamefont
  {{Sathyaprakash}}, \citenamefont {{Van Den Broeck}}, \citenamefont
  {{Winther}}, \citenamefont {{Witek}}, \citenamefont {{Emad Aghili}},
  \citenamefont {{Alsing}}, \citenamefont {{Bolen}}, \citenamefont
  {{Bombelli}}, \citenamefont {{Caudill}}, \citenamefont {{Chen}},
  \citenamefont {{Degollado}}, \citenamefont {{Fujita}}, \citenamefont {{Gao}},
  \citenamefont {{Gerosa}}, \citenamefont {{Kamali}}, \citenamefont {{Silva}},
  \citenamefont {{Rosa}}, \citenamefont {{Sadeghian}}, \citenamefont
  {{Sampaio}}, \citenamefont {{Sotani}},\ and\ \citenamefont
  {{Zilhao}}}]{Berti15}%
  \BibitemOpen
  \bibfield  {author} {\bibinfo {author} {\bibfnamefont {E.}~\bibnamefont
  {{Berti}}}, \bibinfo {author} {\bibfnamefont {E.}~\bibnamefont {{Barausse}}},
  \bibinfo {author} {\bibfnamefont {V.}~\bibnamefont {{Cardoso}}}, \bibinfo
  {author} {\bibfnamefont {L.}~\bibnamefont {{Gualtieri}}}, \bibinfo {author}
  {\bibfnamefont {P.}~\bibnamefont {{Pani}}}, \bibinfo {author} {\bibfnamefont
  {U.}~\bibnamefont {{Sperhake}}}, \bibinfo {author} {\bibfnamefont {L.~C.}\
  \bibnamefont {{Stein}}}, \bibinfo {author} {\bibfnamefont {N.}~\bibnamefont
  {{Wex}}}, \bibinfo {author} {\bibfnamefont {K.}~\bibnamefont {{Yagi}}},
  \bibinfo {author} {\bibfnamefont {T.}~\bibnamefont {{Baker}}}, \bibinfo
  {author} {\bibfnamefont {C.~P.}\ \bibnamefont {{Burgess}}}, \bibinfo {author}
  {\bibfnamefont {F.~S.}\ \bibnamefont {{Coelho}}}, \bibinfo {author}
  {\bibfnamefont {D.}~\bibnamefont {{Doneva}}}, \bibinfo {author}
  {\bibfnamefont {A.}~\bibnamefont {{De Felice}}}, \bibinfo {author}
  {\bibfnamefont {P.~G.}\ \bibnamefont {{Ferreira}}}, \bibinfo {author}
  {\bibfnamefont {P.~C.~C.}\ \bibnamefont {{Freire}}}, \bibinfo {author}
  {\bibfnamefont {J.}~\bibnamefont {{Healy}}}, \bibinfo {author} {\bibfnamefont
  {C.}~\bibnamefont {{Herdeiro}}}, \bibinfo {author} {\bibfnamefont
  {M.}~\bibnamefont {{Horbatsch}}}, \bibinfo {author} {\bibfnamefont
  {B.}~\bibnamefont {{Kleihaus}}}, \bibinfo {author} {\bibfnamefont
  {A.}~\bibnamefont {{Klein}}}, \bibinfo {author} {\bibfnamefont
  {K.}~\bibnamefont {{Kokkotas}}}, \bibinfo {author} {\bibfnamefont
  {J.}~\bibnamefont {{Kunz}}}, \bibinfo {author} {\bibfnamefont
  {P.}~\bibnamefont {{Laguna}}}, \bibinfo {author} {\bibfnamefont {R.~N.}\
  \bibnamefont {{Lang}}}, \bibinfo {author} {\bibfnamefont {T.~G.~F.}\
  \bibnamefont {{Li}}}, \bibinfo {author} {\bibfnamefont {T.}~\bibnamefont
  {{Littenberg}}}, \bibinfo {author} {\bibfnamefont {A.}~\bibnamefont
  {{Matas}}}, \bibinfo {author} {\bibfnamefont {S.}~\bibnamefont
  {{Mirshekari}}}, \bibinfo {author} {\bibfnamefont {H.}~\bibnamefont
  {{Okawa}}}, \bibinfo {author} {\bibfnamefont {E.}~\bibnamefont {{Radu}}},
  \bibinfo {author} {\bibfnamefont {R.}~\bibnamefont {{O'Shaughnessy}}},
  \bibinfo {author} {\bibfnamefont {B.~S.}\ \bibnamefont {{Sathyaprakash}}},
  \bibinfo {author} {\bibfnamefont {C.}~\bibnamefont {{Van Den Broeck}}},
  \bibinfo {author} {\bibfnamefont {H.~A.}\ \bibnamefont {{Winther}}}, \bibinfo
  {author} {\bibfnamefont {H.}~\bibnamefont {{Witek}}}, \bibinfo {author}
  {\bibfnamefont {M.}~\bibnamefont {{Emad Aghili}}}, \bibinfo {author}
  {\bibfnamefont {J.}~\bibnamefont {{Alsing}}}, \bibinfo {author}
  {\bibfnamefont {B.}~\bibnamefont {{Bolen}}}, \bibinfo {author} {\bibfnamefont
  {L.}~\bibnamefont {{Bombelli}}}, \bibinfo {author} {\bibfnamefont
  {S.}~\bibnamefont {{Caudill}}}, \bibinfo {author} {\bibfnamefont
  {L.}~\bibnamefont {{Chen}}}, \bibinfo {author} {\bibfnamefont {J.~C.}\
  \bibnamefont {{Degollado}}}, \bibinfo {author} {\bibfnamefont
  {R.}~\bibnamefont {{Fujita}}}, \bibinfo {author} {\bibfnamefont
  {C.}~\bibnamefont {{Gao}}}, \bibinfo {author} {\bibfnamefont
  {D.}~\bibnamefont {{Gerosa}}}, \bibinfo {author} {\bibfnamefont
  {S.}~\bibnamefont {{Kamali}}}, \bibinfo {author} {\bibfnamefont {H.~O.}\
  \bibnamefont {{Silva}}}, \bibinfo {author} {\bibfnamefont {J.~G.}\
  \bibnamefont {{Rosa}}}, \bibinfo {author} {\bibfnamefont {L.}~\bibnamefont
  {{Sadeghian}}}, \bibinfo {author} {\bibfnamefont {M.}~\bibnamefont
  {{Sampaio}}}, \bibinfo {author} {\bibfnamefont {H.}~\bibnamefont {{Sotani}}},
  \ and\ \bibinfo {author} {\bibfnamefont {M.}~\bibnamefont {{Zilhao}}},\
  }\href {\doibase 10.1088/0264-9381/32/24/243001} {\bibfield  {journal}
  {\bibinfo  {journal} {Classical and Quantum Gravity}\ }\textbf {\bibinfo
  {volume} {32}},\ \bibinfo {eid} {243001} (\bibinfo {year} {2015})},\ \Eprint
  {http://arxiv.org/abs/1501.07274} {arXiv:1501.07274 [gr-qc]} \BibitemShut
  {NoStop}%
\bibitem [{\citenamefont {{Cardoso}}\ and\ \citenamefont
  {{Gualtieri}}(2016)}]{Cardoso16a}%
  \BibitemOpen
  \bibfield  {author} {\bibinfo {author} {\bibfnamefont {V.}~\bibnamefont
  {{Cardoso}}}\ and\ \bibinfo {author} {\bibfnamefont {L.}~\bibnamefont
  {{Gualtieri}}},\ }\href {\doibase 10.1088/0264-9381/33/17/174001} {\bibfield
  {journal} {\bibinfo  {journal} {Classical and Quantum Gravity}\ }\textbf
  {\bibinfo {volume} {33}},\ \bibinfo {eid} {174001} (\bibinfo {year}
  {2016})},\ \Eprint {http://arxiv.org/abs/1607.03133} {arXiv:1607.03133
  [gr-qc]} \BibitemShut {NoStop}%
\bibitem [{\citenamefont {{Cardoso}}\ and\ \citenamefont
  {{Pani}}(2017)}]{Cardoso17}%
  \BibitemOpen
  \bibfield  {author} {\bibinfo {author} {\bibfnamefont {V.}~\bibnamefont
  {{Cardoso}}}\ and\ \bibinfo {author} {\bibfnamefont {P.}~\bibnamefont
  {{Pani}}},\ }\href {\doibase 10.1038/s41550-017-0225-y} {\bibfield  {journal}
  {\bibinfo  {journal} {Nature Astronomy}\ }\textbf {\bibinfo {volume} {1}},\
  \bibinfo {pages} {586} (\bibinfo {year} {2017})},\ \Eprint
  {http://arxiv.org/abs/1707.03021} {arXiv:1707.03021 [gr-qc]} \BibitemShut
  {NoStop}%
\bibitem [{\citenamefont {{Krawczynski}}(2012)}]{Krawczynski12}%
  \BibitemOpen
  \bibfield  {author} {\bibinfo {author} {\bibfnamefont {H.}~\bibnamefont
  {{Krawczynski}}},\ }\href {\doibase 10.1088/0004-637X/754/2/133} {\bibfield
  {journal} {\bibinfo  {journal} {Astrophys J.}\ }\textbf {\bibinfo {volume}
  {754}},\ \bibinfo {eid} {133} (\bibinfo {year} {2012})},\ \Eprint
  {http://arxiv.org/abs/1205.7063} {arXiv:1205.7063 [gr-qc]} \BibitemShut
  {NoStop}%
\bibitem [{\citenamefont {{Krawczynski}}(2018)}]{Krawczynski18}%
  \BibitemOpen
  \bibfield  {author} {\bibinfo {author} {\bibfnamefont {H.}~\bibnamefont
  {{Krawczynski}}},\ }\href {\doibase 10.1007/s10714-018-2419-8} {\bibfield
  {journal} {\bibinfo  {journal} {General Relativity and Gravitation}\ }\textbf
  {\bibinfo {volume} {50}},\ \bibinfo {eid} {100} (\bibinfo {year} {2018})},\
  \Eprint {http://arxiv.org/abs/1806.10347} {arXiv:1806.10347 [astro-ph.HE]}
  \BibitemShut {NoStop}%
\bibitem [{\citenamefont {{Yagi}}\ and\ \citenamefont
  {{Stein}}(2016)}]{Yagi16}%
  \BibitemOpen
  \bibfield  {author} {\bibinfo {author} {\bibfnamefont {K.}~\bibnamefont
  {{Yagi}}}\ and\ \bibinfo {author} {\bibfnamefont {L.~C.}\ \bibnamefont
  {{Stein}}},\ }\href {\doibase 10.1088/0264-9381/33/5/054001} {\bibfield
  {journal} {\bibinfo  {journal} {Classical and Quantum Gravity}\ }\textbf
  {\bibinfo {volume} {33}},\ \bibinfo {eid} {054001} (\bibinfo {year}
  {2016})},\ \Eprint {http://arxiv.org/abs/1602.02413} {arXiv:1602.02413
  [gr-qc]} \BibitemShut {NoStop}%
\bibitem [{\citenamefont {{Bambi}}\ \emph {et~al.}(2016)\citenamefont
  {{Bambi}}, \citenamefont {{Jiang}},\ and\ \citenamefont
  {{Steiner}}}]{Bambi16b}%
  \BibitemOpen
  \bibfield  {author} {\bibinfo {author} {\bibfnamefont {C.}~\bibnamefont
  {{Bambi}}}, \bibinfo {author} {\bibfnamefont {J.}~\bibnamefont {{Jiang}}}, \
  and\ \bibinfo {author} {\bibfnamefont {J.~F.}\ \bibnamefont {{Steiner}}},\
  }\href {\doibase 10.1088/0264-9381/33/6/064001} {\bibfield  {journal}
  {\bibinfo  {journal} {Classical and Quantum Gravity}\ }\textbf {\bibinfo
  {volume} {33}},\ \bibinfo {eid} {064001} (\bibinfo {year} {2016})},\ \Eprint
  {http://arxiv.org/abs/1511.07587} {arXiv:1511.07587 [gr-qc]} \BibitemShut
  {NoStop}%
\bibitem [{\citenamefont {{Yunes}}\ and\ \citenamefont
  {{Siemens}}(2013)}]{Yunes13}%
  \BibitemOpen
  \bibfield  {author} {\bibinfo {author} {\bibfnamefont {N.}~\bibnamefont
  {{Yunes}}}\ and\ \bibinfo {author} {\bibfnamefont {X.}~\bibnamefont
  {{Siemens}}},\ }\href {\doibase 10.12942/lrr-2013-9} {\bibfield  {journal}
  {\bibinfo  {journal} {Living Reviews in Relativity}\ }\textbf {\bibinfo
  {volume} {16}},\ \bibinfo {eid} {9} (\bibinfo {year} {2013})},\ \Eprint
  {http://arxiv.org/abs/1304.3473} {arXiv:1304.3473 [gr-qc]} \BibitemShut
  {NoStop}%
\bibitem [{\citenamefont {{Tripathi}}\ \emph {et~al.}(2019)\citenamefont
  {{Tripathi}}, \citenamefont {{Yan}}, \citenamefont {{Yang}}, \citenamefont
  {{Yan}}, \citenamefont {{Garnham}}, \citenamefont {{Yao}}, \citenamefont
  {{Li}}, \citenamefont {{Ding}}, \citenamefont {{Abdikamalov}}, \citenamefont
  {{Ayzenberg}}, \citenamefont {{Bambi}}, \citenamefont {{Dauser}},
  \citenamefont {{Garcia}}, \citenamefont {{Jiang}},\ and\ \citenamefont
  {{Nampalliwar}}}]{Tripathi19}%
  \BibitemOpen
  \bibfield  {author} {\bibinfo {author} {\bibfnamefont {A.}~\bibnamefont
  {{Tripathi}}}, \bibinfo {author} {\bibfnamefont {J.}~\bibnamefont {{Yan}}},
  \bibinfo {author} {\bibfnamefont {Y.}~\bibnamefont {{Yang}}}, \bibinfo
  {author} {\bibfnamefont {Y.}~\bibnamefont {{Yan}}}, \bibinfo {author}
  {\bibfnamefont {M.}~\bibnamefont {{Garnham}}}, \bibinfo {author}
  {\bibfnamefont {Y.}~\bibnamefont {{Yao}}}, \bibinfo {author} {\bibfnamefont
  {S.}~\bibnamefont {{Li}}}, \bibinfo {author} {\bibfnamefont {Z.}~\bibnamefont
  {{Ding}}}, \bibinfo {author} {\bibfnamefont {A.~B.}\ \bibnamefont
  {{Abdikamalov}}}, \bibinfo {author} {\bibfnamefont {D.}~\bibnamefont
  {{Ayzenberg}}}, \bibinfo {author} {\bibfnamefont {C.}~\bibnamefont
  {{Bambi}}}, \bibinfo {author} {\bibfnamefont {T.}~\bibnamefont {{Dauser}}},
  \bibinfo {author} {\bibfnamefont {J.~A.}\ \bibnamefont {{Garcia}}}, \bibinfo
  {author} {\bibfnamefont {J.}~\bibnamefont {{Jiang}}}, \ and\ \bibinfo
  {author} {\bibfnamefont {S.}~\bibnamefont {{Nampalliwar}}},\ }\href@noop {}
  {\bibfield  {journal} {\bibinfo  {journal} {arXiv e-prints}\ } (\bibinfo
  {year} {2019})},\ \Eprint {http://arxiv.org/abs/1901.03064} {arXiv:1901.03064
  [gr-qc]} \BibitemShut {NoStop}%
\bibitem [{\citenamefont {{Abuter \textit{et al.} (Gravity
  Collaboration)}}(2018)}]{Gravity18a}%
  \BibitemOpen
  \bibfield  {author} {\bibinfo {author} {\bibfnamefont {R.}~\bibnamefont
  {{Abuter \textit{et al.} (Gravity Collaboration)}}},\ }\href {\doibase
  10.1051/0004-6361/201833718} {\bibfield  {journal} {\bibinfo  {journal}
  {Astronomy \& Astrophysics}\ }\textbf {\bibinfo {volume} {615}},\ \bibinfo
  {eid} {L15} (\bibinfo {year} {2018})},\ \Eprint
  {http://arxiv.org/abs/1807.09409} {arXiv:1807.09409} \BibitemShut {NoStop}%
\bibitem [{\citenamefont {Akiyama}\ \emph
  {et~al.}(2019{\natexlab{a}})\citenamefont {Akiyama} \emph {et~al.}}]{EHT19a}%
  \BibitemOpen
  \bibfield  {author} {\bibinfo {author} {\bibfnamefont {K.}~\bibnamefont
  {Akiyama}} \emph {et~al.} (\bibinfo {collaboration} {Event Horizon
  Telescope}),\ }\href {\doibase 10.3847/2041-8213/ab0ec7} {\bibfield
  {journal} {\bibinfo  {journal} {Astrophys. J.}\ }\textbf {\bibinfo {volume}
  {875}},\ \bibinfo {pages} {L1} (\bibinfo {year}
  {2019}{\natexlab{a}})}\BibitemShut {NoStop}%
\bibitem [{\citenamefont {Akiyama}\ \emph
  {et~al.}(2019{\natexlab{b}})\citenamefont {Akiyama} \emph {et~al.}}]{EHT19b}%
  \BibitemOpen
  \bibfield  {author} {\bibinfo {author} {\bibfnamefont {K.}~\bibnamefont
  {Akiyama}} \emph {et~al.} (\bibinfo {collaboration} {Event Horizon
  Telescope}),\ }\href {\doibase 10.3847/2041-8213/ab1141} {\bibfield
  {journal} {\bibinfo  {journal} {Astrophys. J.}\ }\textbf {\bibinfo {volume}
  {875}},\ \bibinfo {pages} {L6} (\bibinfo {year}
  {2019}{\natexlab{b}})}\BibitemShut {NoStop}%
\bibitem [{\citenamefont {{Carballo-Rubio}}\ \emph {et~al.}(2018)\citenamefont
  {{Carballo-Rubio}}, \citenamefont {{Di Filippo}}, \citenamefont
  {{Liberati}},\ and\ \citenamefont {{Visser}}}]{Carballo-Rubio18}%
  \BibitemOpen
  \bibfield  {author} {\bibinfo {author} {\bibfnamefont {R.}~\bibnamefont
  {{Carballo-Rubio}}}, \bibinfo {author} {\bibfnamefont {F.}~\bibnamefont {{Di
  Filippo}}}, \bibinfo {author} {\bibfnamefont {S.}~\bibnamefont {{Liberati}}},
  \ and\ \bibinfo {author} {\bibfnamefont {M.}~\bibnamefont {{Visser}}},\
  }\href {\doibase 10.1103/PhysRevD.98.124009} {\bibfield  {journal} {\bibinfo
  {journal} {Phys. Rev. D}\ }\textbf {\bibinfo {volume} {98}},\ \bibinfo {eid}
  {124009} (\bibinfo {year} {2018})},\ \Eprint
  {http://arxiv.org/abs/1809.08238} {arXiv:1809.08238 [gr-qc]} \BibitemShut
  {NoStop}%
\bibitem [{\citenamefont {{Takahashi}}(2005)}]{Takahashi05}%
  \BibitemOpen
  \bibfield  {author} {\bibinfo {author} {\bibfnamefont {R.}~\bibnamefont
  {{Takahashi}}},\ }\href {\doibase 10.1093/pasj/57.2.273} {\bibfield
  {journal} {\bibinfo  {journal} {Publications of the Astronomical Society of
  Japan}\ }\textbf {\bibinfo {volume} {57}},\ \bibinfo {pages} {273} (\bibinfo
  {year} {2005})},\ \Eprint {http://arxiv.org/abs/astro-ph/0505316}
  {astro-ph/0505316} \BibitemShut {NoStop}%
\bibitem [{\citenamefont {{Bambi}}\ and\ \citenamefont
  {{Freese}}(2009)}]{Bambi09}%
  \BibitemOpen
  \bibfield  {author} {\bibinfo {author} {\bibfnamefont {C.}~\bibnamefont
  {{Bambi}}}\ and\ \bibinfo {author} {\bibfnamefont {K.}~\bibnamefont
  {{Freese}}},\ }\href {\doibase 10.1103/PhysRevD.79.043002} {\bibfield
  {journal} {\bibinfo  {journal} {Phys. Rev. D}\ }\textbf {\bibinfo {volume}
  {79}},\ \bibinfo {eid} {043002} (\bibinfo {year} {2009})},\ \Eprint
  {http://arxiv.org/abs/0812.1328} {arXiv:0812.1328} \BibitemShut {NoStop}%
\bibitem [{\citenamefont {{Hioki}}\ and\ \citenamefont
  {{Maeda}}(2009)}]{Hioki09}%
  \BibitemOpen
  \bibfield  {author} {\bibinfo {author} {\bibfnamefont {K.}~\bibnamefont
  {{Hioki}}}\ and\ \bibinfo {author} {\bibfnamefont {K.-I.}\ \bibnamefont
  {{Maeda}}},\ }\href {\doibase 10.1103/PhysRevD.80.024042} {\bibfield
  {journal} {\bibinfo  {journal} {Phys. Rev. D}\ }\textbf {\bibinfo {volume}
  {80}},\ \bibinfo {eid} {024042} (\bibinfo {year} {2009})}\BibitemShut
  {NoStop}%
\bibitem [{\citenamefont {{Amarilla}}\ \emph {et~al.}(2010)\citenamefont
  {{Amarilla}}, \citenamefont {{Eiroa}},\ and\ \citenamefont
  {{Giribet}}}]{Amarilla10}%
  \BibitemOpen
  \bibfield  {author} {\bibinfo {author} {\bibfnamefont {L.}~\bibnamefont
  {{Amarilla}}}, \bibinfo {author} {\bibfnamefont {E.~F.}\ \bibnamefont
  {{Eiroa}}}, \ and\ \bibinfo {author} {\bibfnamefont {G.}~\bibnamefont
  {{Giribet}}},\ }\href {\doibase 10.1103/PhysRevD.81.124045} {\bibfield
  {journal} {\bibinfo  {journal} {Phys. Rev. D}\ }\textbf {\bibinfo {volume}
  {81}},\ \bibinfo {eid} {124045} (\bibinfo {year} {2010})}\BibitemShut
  {NoStop}%
\bibitem [{\citenamefont {{Bambi}}\ and\ \citenamefont
  {{Yoshida}}(2010)}]{Bambi10}%
  \BibitemOpen
  \bibfield  {author} {\bibinfo {author} {\bibfnamefont {C.}~\bibnamefont
  {{Bambi}}}\ and\ \bibinfo {author} {\bibfnamefont {N.}~\bibnamefont
  {{Yoshida}}},\ }\href {\doibase 10.1088/0264-9381/27/20/205006} {\bibfield
  {journal} {\bibinfo  {journal} {Classical and Quantum Gravity}\ }\textbf
  {\bibinfo {volume} {27}},\ \bibinfo {eid} {205006} (\bibinfo {year}
  {2010})},\ \Eprint {http://arxiv.org/abs/1004.3149} {arXiv:1004.3149 [gr-qc]}
  \BibitemShut {NoStop}%
\bibitem [{\citenamefont {{Amarilla}}\ and\ \citenamefont
  {{Eiroa}}(2012)}]{Amarilla12}%
  \BibitemOpen
  \bibfield  {author} {\bibinfo {author} {\bibfnamefont {L.}~\bibnamefont
  {{Amarilla}}}\ and\ \bibinfo {author} {\bibfnamefont {E.~F.}\ \bibnamefont
  {{Eiroa}}},\ }\href {\doibase 10.1103/PhysRevD.85.064019} {\bibfield
  {journal} {\bibinfo  {journal} {Phys. Rev. D}\ }\textbf {\bibinfo {volume}
  {85}},\ \bibinfo {eid} {064019} (\bibinfo {year} {2012})}\BibitemShut
  {NoStop}%
\bibitem [{\citenamefont {{Amarilla}}\ and\ \citenamefont
  {{Eiroa}}(2013)}]{Amarilla13}%
  \BibitemOpen
  \bibfield  {author} {\bibinfo {author} {\bibfnamefont {L.}~\bibnamefont
  {{Amarilla}}}\ and\ \bibinfo {author} {\bibfnamefont {E.~F.}\ \bibnamefont
  {{Eiroa}}},\ }\href {\doibase 10.1103/PhysRevD.87.044057} {\bibfield
  {journal} {\bibinfo  {journal} {Phys. Rev. D}\ }\textbf {\bibinfo {volume}
  {87}},\ \bibinfo {eid} {044057} (\bibinfo {year} {2013})}\BibitemShut
  {NoStop}%
\bibitem [{\citenamefont {{Abdujabbarov}}\ \emph {et~al.}(2013)\citenamefont
  {{Abdujabbarov}}, \citenamefont {{Atamurotov}}, \citenamefont {{Kucukakca}},
  \citenamefont {{Ahmedov}},\ and\ \citenamefont {{Camci}}}]{Abdujabbarov13c}%
  \BibitemOpen
  \bibfield  {author} {\bibinfo {author} {\bibfnamefont {A.}~\bibnamefont
  {{Abdujabbarov}}}, \bibinfo {author} {\bibfnamefont {F.}~\bibnamefont
  {{Atamurotov}}}, \bibinfo {author} {\bibfnamefont {Y.}~\bibnamefont
  {{Kucukakca}}}, \bibinfo {author} {\bibfnamefont {B.}~\bibnamefont
  {{Ahmedov}}}, \ and\ \bibinfo {author} {\bibfnamefont {U.}~\bibnamefont
  {{Camci}}},\ }\href {\doibase 10.1007/s10509-012-1337-6} {\bibfield
  {journal} {\bibinfo  {journal} {Astrophys. Space Sci.}\ }\textbf {\bibinfo
  {volume} {344}},\ \bibinfo {pages} {429} (\bibinfo {year}
  {2013})}\BibitemShut {NoStop}%
\bibitem [{\citenamefont {{Atamurotov}}\ \emph
  {et~al.}(2013{\natexlab{a}})\citenamefont {{Atamurotov}}, \citenamefont
  {{Abdujabbarov}},\ and\ \citenamefont {{Ahmedov}}}]{Atamurotov13}%
  \BibitemOpen
  \bibfield  {author} {\bibinfo {author} {\bibfnamefont {F.}~\bibnamefont
  {{Atamurotov}}}, \bibinfo {author} {\bibfnamefont {A.}~\bibnamefont
  {{Abdujabbarov}}}, \ and\ \bibinfo {author} {\bibfnamefont {B.}~\bibnamefont
  {{Ahmedov}}},\ }\href {\doibase 10.1007/s10509-013-1548-5} {\bibfield
  {journal} {\bibinfo  {journal} {Astrophys Space Sci}\ }\textbf {\bibinfo
  {volume} {348}},\ \bibinfo {pages} {179} (\bibinfo {year}
  {2013}{\natexlab{a}})}\BibitemShut {NoStop}%
\bibitem [{\citenamefont {{Wei}}\ and\ \citenamefont {{Liu}}(2013)}]{Wei13}%
  \BibitemOpen
  \bibfield  {author} {\bibinfo {author} {\bibfnamefont {S.-W.}\ \bibnamefont
  {{Wei}}}\ and\ \bibinfo {author} {\bibfnamefont {Y.-X.}\ \bibnamefont
  {{Liu}}},\ }\href {\doibase 10.1088/1475-7516/2013/11/063} {\bibfield
  {journal} {\bibinfo  {journal} {Journal of Cosmology and Astroparticles}\
  }\textbf {\bibinfo {volume} {11}},\ \bibinfo {eid} {063} (\bibinfo {year}
  {2013})},\ \Eprint {http://arxiv.org/abs/1311.4251} {arXiv:1311.4251 [gr-qc]}
  \BibitemShut {NoStop}%
\bibitem [{\citenamefont {{Atamurotov}}\ \emph
  {et~al.}(2013{\natexlab{b}})\citenamefont {{Atamurotov}}, \citenamefont
  {{Abdujabbarov}},\ and\ \citenamefont {{Ahmedov}}}]{Atamurotov13b}%
  \BibitemOpen
  \bibfield  {author} {\bibinfo {author} {\bibfnamefont {F.}~\bibnamefont
  {{Atamurotov}}}, \bibinfo {author} {\bibfnamefont {A.}~\bibnamefont
  {{Abdujabbarov}}}, \ and\ \bibinfo {author} {\bibfnamefont {B.}~\bibnamefont
  {{Ahmedov}}},\ }\href {\doibase 10.1103/PhysRevD.88.064004} {\bibfield
  {journal} {\bibinfo  {journal} {Phys. Rev. D}\ }\textbf {\bibinfo {volume}
  {88}},\ \bibinfo {eid} {064004} (\bibinfo {year}
  {2013}{\natexlab{b}})}\BibitemShut {NoStop}%
\bibitem [{\citenamefont {{Bambi}}(2015)}]{Bambi15}%
  \BibitemOpen
  \bibfield  {author} {\bibinfo {author} {\bibfnamefont {C.}~\bibnamefont
  {{Bambi}}},\ }\href@noop {} {\bibfield  {journal} {\bibinfo  {journal} {arXiv
  e-prints}\ ,\ \bibinfo {eid} {arXiv:1507.05257}} (\bibinfo {year} {2015})},\
  \Eprint {http://arxiv.org/abs/1507.05257} {arXiv:1507.05257 [gr-qc]}
  \BibitemShut {NoStop}%
\bibitem [{\citenamefont {{Ghasemi-Nodehi}}\ \emph {et~al.}(2015)\citenamefont
  {{Ghasemi-Nodehi}}, \citenamefont {{Li}},\ and\ \citenamefont
  {{Bambi}}}]{Ghasemi-Nodehi15}%
  \BibitemOpen
  \bibfield  {author} {\bibinfo {author} {\bibfnamefont {M.}~\bibnamefont
  {{Ghasemi-Nodehi}}}, \bibinfo {author} {\bibfnamefont {Z.}~\bibnamefont
  {{Li}}}, \ and\ \bibinfo {author} {\bibfnamefont {C.}~\bibnamefont
  {{Bambi}}},\ }\href {\doibase 10.1140/epjc/s10052-015-3539-x} {\bibfield
  {journal} {\bibinfo  {journal} {European Physical Journal C}\ }\textbf
  {\bibinfo {volume} {75}},\ \bibinfo {eid} {315} (\bibinfo {year} {2015})},\
  \Eprint {http://arxiv.org/abs/1506.02627} {arXiv:1506.02627 [gr-qc]}
  \BibitemShut {NoStop}%
\bibitem [{\citenamefont {{Cunha}}\ \emph {et~al.}(2015)\citenamefont
  {{Cunha}}, \citenamefont {{Herdeiro}}, \citenamefont {{Radu}},\ and\
  \citenamefont {{R{\'u}narsson}}}]{Cunha15}%
  \BibitemOpen
  \bibfield  {author} {\bibinfo {author} {\bibfnamefont {P.~V.~P.}\
  \bibnamefont {{Cunha}}}, \bibinfo {author} {\bibfnamefont {C.~A.~R.}\
  \bibnamefont {{Herdeiro}}}, \bibinfo {author} {\bibfnamefont
  {E.}~\bibnamefont {{Radu}}}, \ and\ \bibinfo {author} {\bibfnamefont {H.~F.}\
  \bibnamefont {{R{\'u}narsson}}},\ }\href {\doibase
  10.1103/PhysRevLett.115.211102} {\bibfield  {journal} {\bibinfo  {journal}
  {Physical Review Letters}\ }\textbf {\bibinfo {volume} {115}},\ \bibinfo
  {eid} {211102} (\bibinfo {year} {2015})},\ \Eprint
  {http://arxiv.org/abs/1509.00021} {arXiv:1509.00021 [gr-qc]} \BibitemShut
  {NoStop}%
\bibitem [{\citenamefont {{Javed}}\ \emph {et~al.}(2019)\citenamefont
  {{Javed}}, \citenamefont {{Babar}},\ and\ \citenamefont
  {{{\"O}vg{\"u}n}}}]{Javed19}%
  \BibitemOpen
  \bibfield  {author} {\bibinfo {author} {\bibfnamefont {W.}~\bibnamefont
  {{Javed}}}, \bibinfo {author} {\bibfnamefont {R.}~\bibnamefont {{Babar}}}, \
  and\ \bibinfo {author} {\bibfnamefont {A.}~\bibnamefont {{{\"O}vg{\"u}n}}},\
  }\href {\doibase 10.1103/PhysRevD.99.084012} {\bibfield  {journal} {\bibinfo
  {journal} {Phys. Rev. D}\ }\textbf {\bibinfo {volume} {99}},\ \bibinfo {eid}
  {084012} (\bibinfo {year} {2019})},\ \Eprint
  {http://arxiv.org/abs/1903.11657} {arXiv:1903.11657 [gr-qc]} \BibitemShut
  {NoStop}%
\bibitem [{\citenamefont {{{\"O}vg{\"u}n}}\ \emph {et~al.}(2019)\citenamefont
  {{{\"O}vg{\"u}n}}, \citenamefont {{Gyulchev}},\ and\ \citenamefont
  {{Jusufi}}}]{Ovgun19}%
  \BibitemOpen
  \bibfield  {author} {\bibinfo {author} {\bibfnamefont {A.}~\bibnamefont
  {{{\"O}vg{\"u}n}}}, \bibinfo {author} {\bibfnamefont {G.}~\bibnamefont
  {{Gyulchev}}}, \ and\ \bibinfo {author} {\bibfnamefont {K.}~\bibnamefont
  {{Jusufi}}},\ }\href {\doibase 10.1016/j.aop.2019.04.007} {\bibfield
  {journal} {\bibinfo  {journal} {Annals of Physics}\ }\textbf {\bibinfo
  {volume} {406}},\ \bibinfo {pages} {152} (\bibinfo {year} {2019})},\ \Eprint
  {http://arxiv.org/abs/1806.03719} {arXiv:1806.03719 [gr-qc]} \BibitemShut
  {NoStop}%
\bibitem [{\citenamefont {{{\"O}vg{\"u}n}}(2019)}]{Ovgun19a}%
  \BibitemOpen
  \bibfield  {author} {\bibinfo {author} {\bibfnamefont {A.}~\bibnamefont
  {{{\"O}vg{\"u}n}}},\ }\href {\doibase 10.3390/universe5050115} {\bibfield
  {journal} {\bibinfo  {journal} {Universe}\ }\textbf {\bibinfo {volume} {5}},\
  \bibinfo {pages} {115} (\bibinfo {year} {2019})},\ \Eprint
  {http://arxiv.org/abs/1806.05549} {arXiv:1806.05549 [physics.gen-ph]}
  \BibitemShut {NoStop}%
\bibitem [{\citenamefont {{Johannsen}}\ and\ \citenamefont
  {{Psaltis}}(2011)}]{Johannsen11}%
  \BibitemOpen
  \bibfield  {author} {\bibinfo {author} {\bibfnamefont {T.}~\bibnamefont
  {{Johannsen}}}\ and\ \bibinfo {author} {\bibfnamefont {D.}~\bibnamefont
  {{Psaltis}}},\ }\href {\doibase 10.1103/PhysRevD.83.124015} {\bibfield
  {journal} {\bibinfo  {journal} {Phys. Rev. D}\ }\textbf {\bibinfo {volume}
  {83}},\ \bibinfo {eid} {124015} (\bibinfo {year} {2011})},\ \Eprint
  {http://arxiv.org/abs/1105.3191} {arXiv:1105.3191 [gr-qc]} \BibitemShut
  {NoStop}%
\bibitem [{\citenamefont {{Konoplya}}\ \emph {et~al.}(2016)\citenamefont
  {{Konoplya}}, \citenamefont {{Rezzolla}},\ and\ \citenamefont
  {{Zhidenko}}}]{Konoplya16}%
  \BibitemOpen
  \bibfield  {author} {\bibinfo {author} {\bibfnamefont {R.}~\bibnamefont
  {{Konoplya}}}, \bibinfo {author} {\bibfnamefont {L.}~\bibnamefont
  {{Rezzolla}}}, \ and\ \bibinfo {author} {\bibfnamefont {A.}~\bibnamefont
  {{Zhidenko}}},\ }\href {\doibase 10.1103/PhysRevD.93.064015} {\bibfield
  {journal} {\bibinfo  {journal} {Phys. Rev. D}\ }\textbf {\bibinfo {volume}
  {93}},\ \bibinfo {eid} {064015} (\bibinfo {year} {2016})},\ \Eprint
  {http://arxiv.org/abs/1602.02378} {arXiv:1602.02378 [gr-qc]} \BibitemShut
  {NoStop}%
\bibitem [{\citenamefont {{Younsi}}\ \emph {et~al.}(2016)\citenamefont
  {{Younsi}}, \citenamefont {{Zhidenko}}, \citenamefont {{Rezzolla}},
  \citenamefont {{Konoplya}},\ and\ \citenamefont {{Mizuno}}}]{Younsi16}%
  \BibitemOpen
  \bibfield  {author} {\bibinfo {author} {\bibfnamefont {Z.}~\bibnamefont
  {{Younsi}}}, \bibinfo {author} {\bibfnamefont {A.}~\bibnamefont
  {{Zhidenko}}}, \bibinfo {author} {\bibfnamefont {L.}~\bibnamefont
  {{Rezzolla}}}, \bibinfo {author} {\bibfnamefont {R.}~\bibnamefont
  {{Konoplya}}}, \ and\ \bibinfo {author} {\bibfnamefont {Y.}~\bibnamefont
  {{Mizuno}}},\ }\href {\doibase 10.1103/PhysRevD.94.084025} {\bibfield
  {journal} {\bibinfo  {journal} {Phys. Rev. D}\ }\textbf {\bibinfo {volume}
  {94}},\ \bibinfo {eid} {084025} (\bibinfo {year} {2016})},\ \Eprint
  {http://arxiv.org/abs/1607.05767} {arXiv:1607.05767 [gr-qc]} \BibitemShut
  {NoStop}%
\bibitem [{\citenamefont {{Abdujabbarov}}\ \emph {et~al.}(2015)\citenamefont
  {{Abdujabbarov}}, \citenamefont {{Rezzolla}},\ and\ \citenamefont
  {{Ahmedov}}}]{Abdujabbarov15}%
  \BibitemOpen
  \bibfield  {author} {\bibinfo {author} {\bibfnamefont {A.~A.}\ \bibnamefont
  {{Abdujabbarov}}}, \bibinfo {author} {\bibfnamefont {L.}~\bibnamefont
  {{Rezzolla}}}, \ and\ \bibinfo {author} {\bibfnamefont {B.~J.}\ \bibnamefont
  {{Ahmedov}}},\ }\href {\doibase 10.1093/mnras/stv2079} {\bibfield  {journal}
  {\bibinfo  {journal} {Mon. Not. R. Astron. Soc.}\ }\textbf {\bibinfo {volume}
  {454}},\ \bibinfo {pages} {2423} (\bibinfo {year} {2015})},\ \Eprint
  {http://arxiv.org/abs/1503.09054} {arXiv:1503.09054 [gr-qc]} \BibitemShut
  {NoStop}%
\bibitem [{\citenamefont {{Atamurotov}}\ \emph {et~al.}(2015)\citenamefont
  {{Atamurotov}}, \citenamefont {{Ahmedov}},\ and\ \citenamefont
  {{Abdujabbarov}}}]{Atamurotov15a}%
  \BibitemOpen
  \bibfield  {author} {\bibinfo {author} {\bibfnamefont {F.}~\bibnamefont
  {{Atamurotov}}}, \bibinfo {author} {\bibfnamefont {B.}~\bibnamefont
  {{Ahmedov}}}, \ and\ \bibinfo {author} {\bibfnamefont {A.}~\bibnamefont
  {{Abdujabbarov}}},\ }\href@noop {} {\bibfield  {journal} {\bibinfo  {journal}
  {Phys. Rev. D}\ }\textbf {\bibinfo {volume} {92}},\ \bibinfo {pages} {084005}
  (\bibinfo {year} {2015})},\ \Eprint {http://arxiv.org/abs/1507.08131}
  {arXiv:1507.08131 [gr-qc]} \BibitemShut {NoStop}%
\bibitem [{\citenamefont {{Ohgami}}\ and\ \citenamefont
  {{Sakai}}(2015)}]{Ohgami15}%
  \BibitemOpen
  \bibfield  {author} {\bibinfo {author} {\bibfnamefont {T.}~\bibnamefont
  {{Ohgami}}}\ and\ \bibinfo {author} {\bibfnamefont {N.}~\bibnamefont
  {{Sakai}}},\ }\href {\doibase 10.1103/PhysRevD.91.124020} {\bibfield
  {journal} {\bibinfo  {journal} {Phys. Rev. D}\ }\textbf {\bibinfo {volume}
  {91}},\ \bibinfo {eid} {124020} (\bibinfo {year} {2015})}\BibitemShut
  {NoStop}%
\bibitem [{\citenamefont {{Grenzebach}}\ \emph {et~al.}(2015)\citenamefont
  {{Grenzebach}}, \citenamefont {{Perlick}},\ and\ \citenamefont
  {{L{\"a}mmerzahl}}}]{Grenzebach15}%
  \BibitemOpen
  \bibfield  {author} {\bibinfo {author} {\bibfnamefont {A.}~\bibnamefont
  {{Grenzebach}}}, \bibinfo {author} {\bibfnamefont {V.}~\bibnamefont
  {{Perlick}}}, \ and\ \bibinfo {author} {\bibfnamefont {C.}~\bibnamefont
  {{L{\"a}mmerzahl}}},\ }\href {\doibase 10.1142/S0218271815420249} {\bibfield
  {journal} {\bibinfo  {journal} {International Journal of Modern Physics D}\
  }\textbf {\bibinfo {volume} {24}},\ \bibinfo {eid} {1542024} (\bibinfo {year}
  {2015})},\ \Eprint {http://arxiv.org/abs/1503.03036} {arXiv:1503.03036
  [gr-qc]} \BibitemShut {NoStop}%
\bibitem [{\citenamefont {{Mureika}}\ and\ \citenamefont
  {{Varieschi}}(2017)}]{Mureika17}%
  \BibitemOpen
  \bibfield  {author} {\bibinfo {author} {\bibfnamefont {J.~R.}\ \bibnamefont
  {{Mureika}}}\ and\ \bibinfo {author} {\bibfnamefont {G.~U.}\ \bibnamefont
  {{Varieschi}}},\ }\href {\doibase 10.1139/cjp-2017-0241} {\bibfield
  {journal} {\bibinfo  {journal} {Canadian Journal of Physics}\ }\textbf
  {\bibinfo {volume} {95}},\ \bibinfo {pages} {1299} (\bibinfo {year}
  {2017})},\ \Eprint {http://arxiv.org/abs/1611.00399} {arXiv:1611.00399
  [gr-qc]} \BibitemShut {NoStop}%
\bibitem [{\citenamefont {{Abdujabbarov}}\ \emph {et~al.}(2017)\citenamefont
  {{Abdujabbarov}}, \citenamefont {{Toshmatov}}, \citenamefont
  {{Stuchl{\'{\i}}k}},\ and\ \citenamefont {{Ahmedov}}}]{Abdujabbarov17b}%
  \BibitemOpen
  \bibfield  {author} {\bibinfo {author} {\bibfnamefont {A.}~\bibnamefont
  {{Abdujabbarov}}}, \bibinfo {author} {\bibfnamefont {B.}~\bibnamefont
  {{Toshmatov}}}, \bibinfo {author} {\bibfnamefont {Z.}~\bibnamefont
  {{Stuchl{\'{\i}}k}}}, \ and\ \bibinfo {author} {\bibfnamefont
  {B.}~\bibnamefont {{Ahmedov}}},\ }\href {\doibase 10.1142/S0218271817500511}
  {\bibfield  {journal} {\bibinfo  {journal} {International Journal of Modern
  Physics D}\ }\textbf {\bibinfo {volume} {26}},\ \bibinfo {eid} {1750051-239}
  (\bibinfo {year} {2017})}\BibitemShut {NoStop}%
\bibitem [{\citenamefont {{Abdujabbarov}}\ \emph
  {et~al.}(2016{\natexlab{a}})\citenamefont {{Abdujabbarov}}, \citenamefont
  {{Juraev}}, \citenamefont {{Ahmedov}},\ and\ \citenamefont
  {{Stuchl{\'{\i}}k}}}]{Abdujabbarov16a}%
  \BibitemOpen
  \bibfield  {author} {\bibinfo {author} {\bibfnamefont {A.}~\bibnamefont
  {{Abdujabbarov}}}, \bibinfo {author} {\bibfnamefont {B.}~\bibnamefont
  {{Juraev}}}, \bibinfo {author} {\bibfnamefont {B.}~\bibnamefont {{Ahmedov}}},
  \ and\ \bibinfo {author} {\bibfnamefont {Z.}~\bibnamefont
  {{Stuchl{\'{\i}}k}}},\ }\href {\doibase 10.1007/s10509-016-2818-9} {\bibfield
   {journal} {\bibinfo  {journal} {Astrophys Space Sci}\ }\textbf {\bibinfo
  {volume} {361}},\ \bibinfo {pages} {226} (\bibinfo {year}
  {2016}{\natexlab{a}})}\BibitemShut {NoStop}%
\bibitem [{\citenamefont {{Abdujabbarov}}\ \emph
  {et~al.}(2016{\natexlab{b}})\citenamefont {{Abdujabbarov}}, \citenamefont
  {{Amir}}, \citenamefont {{Ahmedov}},\ and\ \citenamefont
  {{Ghosh}}}]{Abdujabbarov16b}%
  \BibitemOpen
  \bibfield  {author} {\bibinfo {author} {\bibfnamefont {A.}~\bibnamefont
  {{Abdujabbarov}}}, \bibinfo {author} {\bibfnamefont {M.}~\bibnamefont
  {{Amir}}}, \bibinfo {author} {\bibfnamefont {B.}~\bibnamefont {{Ahmedov}}}, \
  and\ \bibinfo {author} {\bibfnamefont {S.~G.}\ \bibnamefont {{Ghosh}}},\
  }\href {\doibase 10.1103/PhysRevD.93.104004} {\bibfield  {journal} {\bibinfo
  {journal} {Phys. Rev. D}\ }\textbf {\bibinfo {volume} {93}},\ \bibinfo {eid}
  {104004} (\bibinfo {year} {2016}{\natexlab{b}})},\ \Eprint
  {http://arxiv.org/abs/1604.03809} {arXiv:1604.03809 [gr-qc]} \BibitemShut
  {NoStop}%
\bibitem [{\citenamefont {{Mizuno}}\ \emph {et~al.}(2018)\citenamefont
  {{Mizuno}}, \citenamefont {{Younsi}}, \citenamefont {{Fromm}}, \citenamefont
  {{Porth}}, \citenamefont {{De Laurentis}}, \citenamefont {{Olivares}},
  \citenamefont {{Falcke}}, \citenamefont {{Kramer}},\ and\ \citenamefont
  {{Rezzolla}}}]{Mizuno18}%
  \BibitemOpen
  \bibfield  {author} {\bibinfo {author} {\bibfnamefont {Y.}~\bibnamefont
  {{Mizuno}}}, \bibinfo {author} {\bibfnamefont {Z.}~\bibnamefont {{Younsi}}},
  \bibinfo {author} {\bibfnamefont {C.~M.}\ \bibnamefont {{Fromm}}}, \bibinfo
  {author} {\bibfnamefont {O.}~\bibnamefont {{Porth}}}, \bibinfo {author}
  {\bibfnamefont {M.}~\bibnamefont {{De Laurentis}}}, \bibinfo {author}
  {\bibfnamefont {H.}~\bibnamefont {{Olivares}}}, \bibinfo {author}
  {\bibfnamefont {H.}~\bibnamefont {{Falcke}}}, \bibinfo {author}
  {\bibfnamefont {M.}~\bibnamefont {{Kramer}}}, \ and\ \bibinfo {author}
  {\bibfnamefont {L.}~\bibnamefont {{Rezzolla}}},\ }\href {\doibase
  10.1038/s41550-018-0449-5} {\bibfield  {journal} {\bibinfo  {journal} {Nature
  Astronomy}\ } (\bibinfo {year} {2018}),\ 10.1038/s41550-018-0449-5},\ \Eprint
  {http://arxiv.org/abs/1804.05812} {arXiv:1804.05812} \BibitemShut {NoStop}%
\bibitem [{\citenamefont {{Shaikh}}\ \emph {et~al.}(2018)\citenamefont
  {{Shaikh}}, \citenamefont {{Kocherlakota}}, \citenamefont {{Narayan}},\ and\
  \citenamefont {{Joshi}}}]{Shaikh18b}%
  \BibitemOpen
  \bibfield  {author} {\bibinfo {author} {\bibfnamefont {R.}~\bibnamefont
  {{Shaikh}}}, \bibinfo {author} {\bibfnamefont {P.}~\bibnamefont
  {{Kocherlakota}}}, \bibinfo {author} {\bibfnamefont {R.}~\bibnamefont
  {{Narayan}}}, \ and\ \bibinfo {author} {\bibfnamefont {P.~S.}\ \bibnamefont
  {{Joshi}}},\ }\href@noop {} {\bibfield  {journal} {\bibinfo  {journal} {ArXiv
  e-prints}\ } (\bibinfo {year} {2018})},\ \Eprint
  {http://arxiv.org/abs/1802.08060} {arXiv:1802.08060 [astro-ph.HE]}
  \BibitemShut {NoStop}%
\bibitem [{\citenamefont {{Bisnovatyi-Kogan}}\ and\ \citenamefont
  {{Tsupko}}(2017)}]{Kogan17}%
  \BibitemOpen
  \bibfield  {author} {\bibinfo {author} {\bibfnamefont {G.}~\bibnamefont
  {{Bisnovatyi-Kogan}}}\ and\ \bibinfo {author} {\bibfnamefont
  {O.}~\bibnamefont {{Tsupko}}},\ }\href {\doibase 10.3390/universe3030057}
  {\bibfield  {journal} {\bibinfo  {journal} {Universe}\ }\textbf {\bibinfo
  {volume} {3}},\ \bibinfo {pages} {57} (\bibinfo {year} {2017})}\BibitemShut
  {NoStop}%
\bibitem [{\citenamefont {{Perlick}}\ and\ \citenamefont
  {{Tsupko}}(2017)}]{Perlick17}%
  \BibitemOpen
  \bibfield  {author} {\bibinfo {author} {\bibfnamefont {V.}~\bibnamefont
  {{Perlick}}}\ and\ \bibinfo {author} {\bibfnamefont {O.~Y.}\ \bibnamefont
  {{Tsupko}}},\ }\href {\doibase 10.1103/PhysRevD.95.104003} {\bibfield
  {journal} {\bibinfo  {journal} {Phys. Rev. D}\ }\textbf {\bibinfo {volume}
  {95}},\ \bibinfo {eid} {104003} (\bibinfo {year} {2017})},\ \Eprint
  {http://arxiv.org/abs/1702.08768} {arXiv:1702.08768 [gr-qc]} \BibitemShut
  {NoStop}%
\bibitem [{\citenamefont {{Schee}}\ and\ \citenamefont
  {{Stuchl{\'{\i}}k}}(2015)}]{Schee15}%
  \BibitemOpen
  \bibfield  {author} {\bibinfo {author} {\bibfnamefont {J.}~\bibnamefont
  {{Schee}}}\ and\ \bibinfo {author} {\bibfnamefont {Z.}~\bibnamefont
  {{Stuchl{\'{\i}}k}}},\ }\href {\doibase 10.1088/1475-7516/2015/06/048}
  {\bibfield  {journal} {\bibinfo  {journal} {JCAP}\ }\textbf {\bibinfo
  {volume} {6}},\ \bibinfo {eid} {048} (\bibinfo {year} {2015})},\ \Eprint
  {http://arxiv.org/abs/1501.00835} {arXiv:1501.00835 [astro-ph.HE]}
  \BibitemShut {NoStop}%
\bibitem [{\citenamefont {{Schee}}\ and\ \citenamefont
  {{Stuchl{\'{\i}}k}}(2009{\natexlab{a}})}]{Schee09a}%
  \BibitemOpen
  \bibfield  {author} {\bibinfo {author} {\bibfnamefont {J.}~\bibnamefont
  {{Schee}}}\ and\ \bibinfo {author} {\bibfnamefont {Z.}~\bibnamefont
  {{Stuchl{\'{\i}}k}}},\ }\href {\doibase 10.1007/s10714-008-0753-y} {\bibfield
   {journal} {\bibinfo  {journal} {General Relativity and Gravitation}\
  }\textbf {\bibinfo {volume} {41}},\ \bibinfo {pages} {1795} (\bibinfo {year}
  {2009}{\natexlab{a}})},\ \Eprint {http://arxiv.org/abs/0812.3017}
  {arXiv:0812.3017} \BibitemShut {NoStop}%
\bibitem [{\citenamefont {{Stuchl{\'{\i}}k}}\ and\ \citenamefont
  {{Schee}}(2014)}]{Stuchlik14}%
  \BibitemOpen
  \bibfield  {author} {\bibinfo {author} {\bibfnamefont {Z.}~\bibnamefont
  {{Stuchl{\'{\i}}k}}}\ and\ \bibinfo {author} {\bibfnamefont {J.}~\bibnamefont
  {{Schee}}},\ }\href {\doibase 10.1088/0264-9381/31/19/195013} {\bibfield
  {journal} {\bibinfo  {journal} {Classical and Quantum Gravity}\ }\textbf
  {\bibinfo {volume} {31}},\ \bibinfo {eid} {195013} (\bibinfo {year}
  {2014})},\ \Eprint {http://arxiv.org/abs/1402.2891} {arXiv:1402.2891
  [astro-ph.HE]} \BibitemShut {NoStop}%
\bibitem [{\citenamefont {{Schee}}\ and\ \citenamefont
  {{Stuchl{\'{\i}}k}}(2009{\natexlab{b}})}]{Schee09}%
  \BibitemOpen
  \bibfield  {author} {\bibinfo {author} {\bibfnamefont {J.}~\bibnamefont
  {{Schee}}}\ and\ \bibinfo {author} {\bibfnamefont {Z.}~\bibnamefont
  {{Stuchl{\'{\i}}k}}},\ }\href {\doibase 10.1142/S0218271809014881} {\bibfield
   {journal} {\bibinfo  {journal} {International Journal of Modern Physics D}\
  }\textbf {\bibinfo {volume} {18}},\ \bibinfo {pages} {983} (\bibinfo {year}
  {2009}{\natexlab{b}})},\ \Eprint {http://arxiv.org/abs/0810.4445}
  {arXiv:0810.4445} \BibitemShut {NoStop}%
\bibitem [{\citenamefont {{Stuchl{\'{\i}}k}}\ and\ \citenamefont
  {{Schee}}(2010)}]{Stuchlik10}%
  \BibitemOpen
  \bibfield  {author} {\bibinfo {author} {\bibfnamefont {Z.}~\bibnamefont
  {{Stuchl{\'{\i}}k}}}\ and\ \bibinfo {author} {\bibfnamefont {J.}~\bibnamefont
  {{Schee}}},\ }\href {\doibase 10.1088/0264-9381/27/21/215017} {\bibfield
  {journal} {\bibinfo  {journal} {Classical and Quantum Gravity}\ }\textbf
  {\bibinfo {volume} {27}},\ \bibinfo {eid} {215017} (\bibinfo {year}
  {2010})},\ \Eprint {http://arxiv.org/abs/1101.3569} {arXiv:1101.3569 [gr-qc]}
  \BibitemShut {NoStop}%
\bibitem [{\citenamefont {{Mishra}}\ \emph {et~al.}(2019)\citenamefont
  {{Mishra}}, \citenamefont {{Chakraborty}},\ and\ \citenamefont
  {{Sarkar}}}]{Mishra19}%
  \BibitemOpen
  \bibfield  {author} {\bibinfo {author} {\bibfnamefont {A.~K.}\ \bibnamefont
  {{Mishra}}}, \bibinfo {author} {\bibfnamefont {S.}~\bibnamefont
  {{Chakraborty}}}, \ and\ \bibinfo {author} {\bibfnamefont {S.}~\bibnamefont
  {{Sarkar}}},\ }\href@noop {} {\bibfield  {journal} {\bibinfo  {journal}
  {arXiv e-prints}\ } (\bibinfo {year} {2019})},\ \Eprint
  {http://arxiv.org/abs/1903.06376} {arXiv:1903.06376 [gr-qc]} \BibitemShut
  {NoStop}%
\bibitem [{\citenamefont {Eiroa}\ and\ \citenamefont {Sendra}(2018)}]{Eiroa18}%
  \BibitemOpen
  \bibfield  {author} {\bibinfo {author} {\bibfnamefont {E.~F.}\ \bibnamefont
  {Eiroa}}\ and\ \bibinfo {author} {\bibfnamefont {C.~M.}\ \bibnamefont
  {Sendra}},\ }\href {\doibase 10.1140/epjc/s10052-018-5586-6} {\bibfield
  {journal} {\bibinfo  {journal} {The European Physical Journal C}\ }\textbf
  {\bibinfo {volume} {78}},\ \bibinfo {pages} {91} (\bibinfo {year}
  {2018})}\BibitemShut {NoStop}%
\bibitem [{\citenamefont {Giddings}(2019)}]{Giddings19}%
  \BibitemOpen
  \bibfield  {author} {\bibinfo {author} {\bibfnamefont {S.~B.}\ \bibnamefont
  {Giddings}},\ }\href@noop {} {\  (\bibinfo {year} {2019})},\ \Eprint
  {http://arxiv.org/abs/1904.05287} {arXiv:1904.05287 [gr-qc]} \BibitemShut
  {NoStop}%
\bibitem [{\citenamefont {{Bambi}}\ and\ \citenamefont
  {{Malafarina}}(2013)}]{Bambi13d}%
  \BibitemOpen
  \bibfield  {author} {\bibinfo {author} {\bibfnamefont {C.}~\bibnamefont
  {{Bambi}}}\ and\ \bibinfo {author} {\bibfnamefont {D.}~\bibnamefont
  {{Malafarina}}},\ }\href {\doibase 10.1103/PhysRevD.88.064022} {\bibfield
  {journal} {\bibinfo  {journal} {Phys. Rev. D}\ }\textbf {\bibinfo {volume}
  {88}},\ \bibinfo {eid} {064022} (\bibinfo {year} {2013})},\ \Eprint
  {http://arxiv.org/abs/1307.2106} {arXiv:1307.2106 [gr-qc]} \BibitemShut
  {NoStop}%
\bibitem [{\citenamefont {{Ilyas}}\ \emph {et~al.}(2017)\citenamefont
  {{Ilyas}}, \citenamefont {{Yang}}, \citenamefont {{Malafarina}},\ and\
  \citenamefont {{Bambi}}}]{Ilyas17}%
  \BibitemOpen
  \bibfield  {author} {\bibinfo {author} {\bibfnamefont {B.}~\bibnamefont
  {{Ilyas}}}, \bibinfo {author} {\bibfnamefont {J.}~\bibnamefont {{Yang}}},
  \bibinfo {author} {\bibfnamefont {D.}~\bibnamefont {{Malafarina}}}, \ and\
  \bibinfo {author} {\bibfnamefont {C.}~\bibnamefont {{Bambi}}},\ }\href
  {\doibase 10.1140/epjc/s10052-017-5014-3} {\bibfield  {journal} {\bibinfo
  {journal} {European Physical Journal C}\ }\textbf {\bibinfo {volume} {77}},\
  \bibinfo {eid} {461} (\bibinfo {year} {2017})},\ \Eprint
  {http://arxiv.org/abs/1611.03972} {arXiv:1611.03972 [gr-qc]} \BibitemShut
  {NoStop}%
\bibitem [{\citenamefont {{Zipoy}}(1966)}]{Zipoy66}%
  \BibitemOpen
  \bibfield  {author} {\bibinfo {author} {\bibfnamefont {D.~M.}\ \bibnamefont
  {{Zipoy}}},\ }\href {\doibase 10.1063/1.1705005} {\bibfield  {journal}
  {\bibinfo  {journal} {Journal of Mathematical Physics}\ }\textbf {\bibinfo
  {volume} {7}},\ \bibinfo {pages} {1137} (\bibinfo {year} {1966})}\BibitemShut
  {NoStop}%
\bibitem [{\citenamefont {{Voorhees}}(1970)}]{Voorhees70}%
  \BibitemOpen
  \bibfield  {author} {\bibinfo {author} {\bibfnamefont {B.~H.}\ \bibnamefont
  {{Voorhees}}},\ }\href {\doibase 10.1103/PhysRevD.2.2119} {\bibfield
  {journal} {\bibinfo  {journal} {Phys. Rev. D}\ }\textbf {\bibinfo {volume}
  {2}},\ \bibinfo {pages} {2119} (\bibinfo {year} {1970})}\BibitemShut
  {NoStop}%
\bibitem [{\citenamefont {{Erez}}\ and\ \citenamefont {{Rosen}}()}]{Erez59}%
  \BibitemOpen
  \bibfield  {author} {\bibinfo {author} {\bibfnamefont {G.}~\bibnamefont
  {{Erez}}}\ and\ \bibinfo {author} {\bibfnamefont {N.}~\bibnamefont
  {{Rosen}}},\ }\href@noop {} {\bibfield  {journal} {\bibinfo  {journal} {Bull.
  Res. Council Israel}\ }\textbf {\bibinfo {volume} {8}},\ \bibinfo {eid} {47
  (1959)}}\BibitemShut {NoStop}%
\bibitem [{\citenamefont {{Bini}}\ \emph {et~al.}(2013)\citenamefont {{Bini}},
  \citenamefont {{Crosta}}, \citenamefont {{de Felice}}, \citenamefont
  {{Geralico}},\ and\ \citenamefont {{Vecchiato}}}]{Bini13}%
  \BibitemOpen
  \bibfield  {author} {\bibinfo {author} {\bibfnamefont {D.}~\bibnamefont
  {{Bini}}}, \bibinfo {author} {\bibfnamefont {M.}~\bibnamefont {{Crosta}}},
  \bibinfo {author} {\bibfnamefont {F.}~\bibnamefont {{de Felice}}}, \bibinfo
  {author} {\bibfnamefont {A.}~\bibnamefont {{Geralico}}}, \ and\ \bibinfo
  {author} {\bibfnamefont {A.}~\bibnamefont {{Vecchiato}}},\ }\href {\doibase
  10.1088/0264-9381/30/4/045009} {\bibfield  {journal} {\bibinfo  {journal}
  {Classical and Quantum Gravity}\ }\textbf {\bibinfo {volume} {30}},\ \bibinfo
  {eid} {045009} (\bibinfo {year} {2013})},\ \Eprint
  {http://arxiv.org/abs/1408.5264} {arXiv:1408.5264 [gr-qc]} \BibitemShut
  {NoStop}%
\bibitem [{\citenamefont {Turimov}\ \emph {et~al.}(2018)\citenamefont
  {Turimov}, \citenamefont {Ahmedov}, \citenamefont {Kolo{\v s}},\ and\
  \citenamefont {Stuchl{\'\i}k}}]{Turimov18a}%
  \BibitemOpen
  \bibfield  {author} {\bibinfo {author} {\bibfnamefont {B.}~\bibnamefont
  {Turimov}}, \bibinfo {author} {\bibfnamefont {B.}~\bibnamefont {Ahmedov}},
  \bibinfo {author} {\bibfnamefont {M.}~\bibnamefont {Kolo{\v s}}}, \ and\
  \bibinfo {author} {\bibfnamefont {Z.}~\bibnamefont {Stuchl{\'\i}k}},\ }\href
  {\doibase 10.1103/PhysRevD.98.084039} {\bibfield  {journal} {\bibinfo
  {journal} {Phys. Rev. D}\ }\textbf {\bibinfo {volume} {98}},\ \bibinfo
  {pages} {084039} (\bibinfo {year} {2018})},\ \Eprint
  {http://arxiv.org/abs/1810.01460} {arXiv:1810.01460 [gr-qc]} \BibitemShut
  {NoStop}%
\bibitem [{\citenamefont {{\.{Z}}enczykowski}(2018)}]{Zenczykowski18}%
  \BibitemOpen
  \bibfield  {author} {\bibinfo {author} {\bibfnamefont {P.}~\bibnamefont
  {{\.{Z}}enczykowski}},\ }\href {\doibase 10.1007/s10699-018-9562-2}
  {\bibfield  {journal} {\bibinfo  {journal} {Foundations of Science}\ }
  (\bibinfo {year} {2018}),\ 10.1007/s10699-018-9562-2}\BibitemShut {NoStop}%
\bibitem [{\citenamefont {{Malafarina}}(2017)}]{Malafarina17}%
  \BibitemOpen
  \bibfield  {author} {\bibinfo {author} {\bibfnamefont {D.}~\bibnamefont
  {{Malafarina}}},\ }\href {\doibase 10.3390/universe3020048} {\bibfield
  {journal} {\bibinfo  {journal} {Universe}\ }\textbf {\bibinfo {volume} {3}},\
  \bibinfo {pages} {48} (\bibinfo {year} {2017})},\ \Eprint
  {http://arxiv.org/abs/1703.04138} {arXiv:1703.04138 [gr-qc]} \BibitemShut
  {NoStop}%
\bibitem [{\citenamefont {{Bambi}}\ \emph {et~al.}(2013)\citenamefont
  {{Bambi}}, \citenamefont {{Malafarina}},\ and\ \citenamefont
  {{Modesto}}}]{Bambi13f}%
  \BibitemOpen
  \bibfield  {author} {\bibinfo {author} {\bibfnamefont {C.}~\bibnamefont
  {{Bambi}}}, \bibinfo {author} {\bibfnamefont {D.}~\bibnamefont
  {{Malafarina}}}, \ and\ \bibinfo {author} {\bibfnamefont {L.}~\bibnamefont
  {{Modesto}}},\ }\href {\doibase 10.1103/PhysRevD.88.044009} {\bibfield
  {journal} {\bibinfo  {journal} {Phys. Rev. D}\ }\textbf {\bibinfo {volume}
  {88}},\ \bibinfo {eid} {044009} (\bibinfo {year} {2013})},\ \Eprint
  {http://arxiv.org/abs/1305.4790} {arXiv:1305.4790 [gr-qc]} \BibitemShut
  {NoStop}%
\bibitem [{\citenamefont {{Hern{\'a}ndez-Pastora}}\ and\ \citenamefont
  {{Mart{\'{\i}}n}}(1994)}]{Hernandez-Pastora94}%
  \BibitemOpen
  \bibfield  {author} {\bibinfo {author} {\bibfnamefont {J.~L.}\ \bibnamefont
  {{Hern{\'a}ndez-Pastora}}}\ and\ \bibinfo {author} {\bibfnamefont
  {J.}~\bibnamefont {{Mart{\'{\i}}n}}},\ }\href {\doibase 10.1007/BF02107146}
  {\bibfield  {journal} {\bibinfo  {journal} {General Relativity and
  Gravitation}\ }\textbf {\bibinfo {volume} {26}},\ \bibinfo {pages} {877}
  (\bibinfo {year} {1994})}\BibitemShut {NoStop}%
\bibitem [{\citenamefont {{Herrera}}\ \emph {et~al.}(1999)\citenamefont
  {{Herrera}}, \citenamefont {{Paiva}},\ and\ \citenamefont
  {{Santos}}}]{Herrera99}%
  \BibitemOpen
  \bibfield  {author} {\bibinfo {author} {\bibfnamefont {L.}~\bibnamefont
  {{Herrera}}}, \bibinfo {author} {\bibfnamefont {F.~M.}\ \bibnamefont
  {{Paiva}}}, \ and\ \bibinfo {author} {\bibfnamefont {N.~O.}\ \bibnamefont
  {{Santos}}},\ }\href {\doibase 10.1063/1.532943} {\bibfield  {journal}
  {\bibinfo  {journal} {Journal of Mathematical Physics}\ }\textbf {\bibinfo
  {volume} {40}},\ \bibinfo {pages} {4064} (\bibinfo {year} {1999})},\ \Eprint
  {http://arxiv.org/abs/gr-qc/9810079} {gr-qc/9810079} \BibitemShut {NoStop}%
\bibitem [{\citenamefont {{Quevedo}}(2011)}]{Quevedo11}%
  \BibitemOpen
  \bibfield  {author} {\bibinfo {author} {\bibfnamefont {H.}~\bibnamefont
  {{Quevedo}}},\ }\href {\doibase 10.1142/S0218271811019852} {\bibfield
  {journal} {\bibinfo  {journal} {International Journal of Modern Physics D}\
  }\textbf {\bibinfo {volume} {20}},\ \bibinfo {pages} {1779} (\bibinfo {year}
  {2011})},\ \Eprint {http://arxiv.org/abs/1012.4030} {arXiv:1012.4030 [gr-qc]}
  \BibitemShut {NoStop}%
\bibitem [{\citenamefont {{Virbhadra}}(1996)}]{Virbhadra96}%
  \BibitemOpen
  \bibfield  {author} {\bibinfo {author} {\bibfnamefont {K.~S.}\ \bibnamefont
  {{Virbhadra}}},\ }\href@noop {} {\bibfield  {journal} {\bibinfo  {journal}
  {ArXiv General Relativity and Quantum Cosmology e-prints}\ } (\bibinfo {year}
  {1996})},\ \Eprint {http://arxiv.org/abs/gr-qc/9606004} {gr-qc/9606004}
  \BibitemShut {NoStop}%
\bibitem [{\citenamefont {Boshkayev}\ \emph {et~al.}(2016)\citenamefont
  {Boshkayev}, \citenamefont {Gasperin}, \citenamefont {Gutierrez-Pineres},
  \citenamefont {Quevedo},\ and\ \citenamefont {Toktarbay}}]{Boshkayev15}%
  \BibitemOpen
  \bibfield  {author} {\bibinfo {author} {\bibfnamefont {K.}~\bibnamefont
  {Boshkayev}}, \bibinfo {author} {\bibfnamefont {E.}~\bibnamefont {Gasperin}},
  \bibinfo {author} {\bibfnamefont {A.~C.}\ \bibnamefont {Gutierrez-Pineres}},
  \bibinfo {author} {\bibfnamefont {H.}~\bibnamefont {Quevedo}}, \ and\
  \bibinfo {author} {\bibfnamefont {S.}~\bibnamefont {Toktarbay}},\ }\href
  {\doibase 10.1103/PhysRevD.93.024024} {\bibfield  {journal} {\bibinfo
  {journal} {Phys. Rev. D}\ }\textbf {\bibinfo {volume} {93}},\ \bibinfo
  {pages} {024024} (\bibinfo {year} {2016})},\ \Eprint
  {http://arxiv.org/abs/1509.03827} {arXiv:1509.03827 [gr-qc]} \BibitemShut
  {NoStop}%
\bibitem [{\citenamefont {{Chowdhury}}\ \emph {et~al.}(2012)\citenamefont
  {{Chowdhury}}, \citenamefont {{Patil}}, \citenamefont {{Malafarina}},\ and\
  \citenamefont {{Joshi}}}]{Chowdhury12}%
  \BibitemOpen
  \bibfield  {author} {\bibinfo {author} {\bibfnamefont {A.~N.}\ \bibnamefont
  {{Chowdhury}}}, \bibinfo {author} {\bibfnamefont {M.}~\bibnamefont
  {{Patil}}}, \bibinfo {author} {\bibfnamefont {D.}~\bibnamefont
  {{Malafarina}}}, \ and\ \bibinfo {author} {\bibfnamefont {P.~S.}\
  \bibnamefont {{Joshi}}},\ }\href {\doibase 10.1103/PhysRevD.85.104031}
  {\bibfield  {journal} {\bibinfo  {journal} {Phys. Rev. D.}\ }\textbf
  {\bibinfo {volume} {85}},\ \bibinfo {eid} {104031} (\bibinfo {year}
  {2012})},\ \Eprint {http://arxiv.org/abs/1112.2522} {arXiv:1112.2522 [gr-qc]}
  \BibitemShut {NoStop}%
\bibitem [{\citenamefont {{Hernandez}}(1967)}]{Hernandez67}%
  \BibitemOpen
  \bibfield  {author} {\bibinfo {author} {\bibfnamefont {W.~C.}\ \bibnamefont
  {{Hernandez}}},\ }\href {\doibase 10.1103/PhysRev.153.1359} {\bibfield
  {journal} {\bibinfo  {journal} {Physical Review}\ }\textbf {\bibinfo {volume}
  {153}},\ \bibinfo {pages} {1359} (\bibinfo {year} {1967})}\BibitemShut
  {NoStop}%
\bibitem [{\citenamefont {{Stewart}}\ \emph {et~al.}(1982)\citenamefont
  {{Stewart}}, \citenamefont {{Papadopoulos}}, \citenamefont {{Witten}},
  \citenamefont {{Berezdivin}},\ and\ \citenamefont {{Herrera}}}]{Stewart82}%
  \BibitemOpen
  \bibfield  {author} {\bibinfo {author} {\bibfnamefont {B.~W.}\ \bibnamefont
  {{Stewart}}}, \bibinfo {author} {\bibfnamefont {D.}~\bibnamefont
  {{Papadopoulos}}}, \bibinfo {author} {\bibfnamefont {L.}~\bibnamefont
  {{Witten}}}, \bibinfo {author} {\bibfnamefont {R.}~\bibnamefont
  {{Berezdivin}}}, \ and\ \bibinfo {author} {\bibfnamefont {L.}~\bibnamefont
  {{Herrera}}},\ }\href {\doibase 10.1007/BF00756201} {\bibfield  {journal}
  {\bibinfo  {journal} {General Relativity and Gravitation}\ }\textbf {\bibinfo
  {volume} {14}},\ \bibinfo {pages} {97} (\bibinfo {year} {1982})}\BibitemShut
  {NoStop}%
\bibitem [{\citenamefont {{Herrera}}\ \emph {et~al.}(2005)\citenamefont
  {{Herrera}}, \citenamefont {{Magli}},\ and\ \citenamefont
  {{Malafarina}}}]{Herrera05}%
  \BibitemOpen
  \bibfield  {author} {\bibinfo {author} {\bibfnamefont {L.}~\bibnamefont
  {{Herrera}}}, \bibinfo {author} {\bibfnamefont {G.}~\bibnamefont {{Magli}}},
  \ and\ \bibinfo {author} {\bibfnamefont {D.}~\bibnamefont {{Malafarina}}},\
  }\href {\doibase 10.1007/s10714-005-0120-1} {\bibfield  {journal} {\bibinfo
  {journal} {General Relativity and Gravitation}\ }\textbf {\bibinfo {volume}
  {37}},\ \bibinfo {pages} {1371} (\bibinfo {year} {2005})},\ \Eprint
  {http://arxiv.org/abs/gr-qc/0407037} {gr-qc/0407037} \BibitemShut {NoStop}%
\bibitem [{\citenamefont {Benavides-Gallego}\ \emph {et~al.}(2019)\citenamefont
  {Benavides-Gallego}, \citenamefont {Abdujabbarov}, \citenamefont
  {Malafarina}, \citenamefont {Ahmedov},\ and\ \citenamefont
  {Bambi}}]{Benavides-Gallego18}%
  \BibitemOpen
  \bibfield  {author} {\bibinfo {author} {\bibfnamefont {C.~A.}\ \bibnamefont
  {Benavides-Gallego}}, \bibinfo {author} {\bibfnamefont {A.}~\bibnamefont
  {Abdujabbarov}}, \bibinfo {author} {\bibfnamefont {D.}~\bibnamefont
  {Malafarina}}, \bibinfo {author} {\bibfnamefont {B.}~\bibnamefont {Ahmedov}},
  \ and\ \bibinfo {author} {\bibfnamefont {C.}~\bibnamefont {Bambi}},\ }\href
  {\doibase 10.1103/PhysRevD.99.044012} {\bibfield  {journal} {\bibinfo
  {journal} {Phys. Rev. D}\ }\textbf {\bibinfo {volume} {99}},\ \bibinfo
  {pages} {044012} (\bibinfo {year} {2019})},\ \Eprint
  {http://arxiv.org/abs/1812.04846} {arXiv:1812.04846 [gr-qc]} \BibitemShut
  {NoStop}%
\bibitem [{\citenamefont {{Ayzenberg}}\ and\ \citenamefont
  {{Yunes}}(2018)}]{Ayzenberg18}%
  \BibitemOpen
  \bibfield  {author} {\bibinfo {author} {\bibfnamefont {D.}~\bibnamefont
  {{Ayzenberg}}}\ and\ \bibinfo {author} {\bibfnamefont {N.}~\bibnamefont
  {{Yunes}}},\ }\href {\doibase 10.1088/1361-6382/aae87b} {\bibfield  {journal}
  {\bibinfo  {journal} {Classical and Quantum Gravity}\ }\textbf {\bibinfo
  {volume} {35}},\ \bibinfo {eid} {235002} (\bibinfo {year} {2018})},\ \Eprint
  {http://arxiv.org/abs/1807.08422} {arXiv:1807.08422 [gr-qc]} \BibitemShut
  {NoStop}%
\bibitem [{\citenamefont {{Gott}}\ \emph {et~al.}(2019)\citenamefont {{Gott}},
  \citenamefont {{Ayzenberg}}, \citenamefont {{Yunes}},\ and\ \citenamefont
  {{Lohfink}}}]{Gott19}%
  \BibitemOpen
  \bibfield  {author} {\bibinfo {author} {\bibfnamefont {H.}~\bibnamefont
  {{Gott}}}, \bibinfo {author} {\bibfnamefont {D.}~\bibnamefont {{Ayzenberg}}},
  \bibinfo {author} {\bibfnamefont {N.}~\bibnamefont {{Yunes}}}, \ and\
  \bibinfo {author} {\bibfnamefont {A.}~\bibnamefont {{Lohfink}}},\ }\href
  {\doibase 10.1088/1361-6382/ab01b0} {\bibfield  {journal} {\bibinfo
  {journal} {Classical and Quantum Gravity}\ }\textbf {\bibinfo {volume}
  {36}},\ \bibinfo {eid} {055007} (\bibinfo {year} {2019})},\ \Eprint
  {http://arxiv.org/abs/1808.05703} {arXiv:1808.05703 [gr-qc]} \BibitemShut
  {NoStop}%
\bibitem [{\citenamefont {{Psaltis}}\ and\ \citenamefont
  {{Johannsen}}(2012)}]{Psaltis12}%
  \BibitemOpen
  \bibfield  {author} {\bibinfo {author} {\bibfnamefont {D.}~\bibnamefont
  {{Psaltis}}}\ and\ \bibinfo {author} {\bibfnamefont {T.}~\bibnamefont
  {{Johannsen}}},\ }\href {\doibase 10.1088/0004-637X/745/1/1} {\bibfield
  {journal} {\bibinfo  {journal} {Astrophys. J.}\ }\textbf {\bibinfo {volume}
  {745}},\ \bibinfo {eid} {1} (\bibinfo {year} {2012})},\ \Eprint
  {http://arxiv.org/abs/1011.4078} {arXiv:1011.4078 [astro-ph.HE]} \BibitemShut
  {NoStop}%
\bibitem [{\citenamefont {{V{\'a}zquez}}\ and\ \citenamefont
  {{Esteban}}(2004)}]{Vazquez04}%
  \BibitemOpen
  \bibfield  {author} {\bibinfo {author} {\bibfnamefont {S.~E.}\ \bibnamefont
  {{V{\'a}zquez}}}\ and\ \bibinfo {author} {\bibfnamefont {E.~P.}\ \bibnamefont
  {{Esteban}}},\ }\href {\doibase 10.1393/ncb/i2004-10121-y} {\bibfield
  {journal} {\bibinfo  {journal} {Nuovo Cim. B}\ }\textbf {\bibinfo {volume}
  {119}},\ \bibinfo {pages} {489} (\bibinfo {year} {2004})}\BibitemShut
  {NoStop}%
\bibitem [{\citenamefont {{Tsukamoto}}\ \emph {et~al.}(2014)\citenamefont
  {{Tsukamoto}}, \citenamefont {{Li}},\ and\ \citenamefont
  {{Bambi}}}]{Tsukamoto14}%
  \BibitemOpen
  \bibfield  {author} {\bibinfo {author} {\bibfnamefont {N.}~\bibnamefont
  {{Tsukamoto}}}, \bibinfo {author} {\bibfnamefont {Z.}~\bibnamefont {{Li}}}, \
  and\ \bibinfo {author} {\bibfnamefont {C.}~\bibnamefont {{Bambi}}},\ }\href
  {\doibase 10.1088/1475-7516/2014/06/043} {\bibfield  {journal} {\bibinfo
  {journal} {Journal of Cosmology and Astroparticles}\ }\textbf {\bibinfo
  {volume} {6}},\ \bibinfo {eid} {043} (\bibinfo {year} {2014})},\ \Eprint
  {http://arxiv.org/abs/1403.0371} {arXiv:1403.0371 [gr-qc]} \BibitemShut
  {NoStop}%
\bibitem [{\citenamefont {{Arifov}}(1981)}]{Arifov81}%
  \BibitemOpen
  \bibfield  {author} {\bibinfo {author} {\bibfnamefont {L.~Y.}\ \bibnamefont
  {{Arifov}}},\ }\href {\doibase 10.1007/BF00898270} {\bibfield  {journal}
  {\bibinfo  {journal} {Soviet Physics Journal}\ }\textbf {\bibinfo {volume}
  {24}},\ \bibinfo {pages} {346} (\bibinfo {year} {1981})}\BibitemShut
  {NoStop}%
\bibitem [{\citenamefont {{Goddi}}\ \emph {et~al.}(2017)\citenamefont
  {{Goddi}}, \citenamefont {{Falcke}}, \citenamefont {{Kramer}}, \citenamefont
  {{Rezzolla}}, \citenamefont {{Brinkerink}}, \citenamefont {{Bronzwaer}},
  \citenamefont {{Davelaar}}, \citenamefont {{Deane}}, \citenamefont {{de
  Laurentis}}, \citenamefont {{Desvignes}}, \citenamefont {{Eatough}},
  \citenamefont {{Eisenhauer}}, \citenamefont {{Fraga-Encinas}}, \citenamefont
  {{Fromm}}, \citenamefont {{Gillessen}}, \citenamefont {{Grenzebach}},
  \citenamefont {{Issaoun}}, \citenamefont {{Jan{\ss}en}}, \citenamefont
  {{Konoplya}}, \citenamefont {{Krichbaum}}, \citenamefont {{Laing}},
  \citenamefont {{Liu}}, \citenamefont {{Lu}}, \citenamefont {{Mizuno}},
  \citenamefont {{Moscibrodzka}}, \citenamefont {{M{\"u}ller}}, \citenamefont
  {{Olivares}}, \citenamefont {{Pfuhl}}, \citenamefont {{Porth}}, \citenamefont
  {{Roelofs}}, \citenamefont {{Ros}}, \citenamefont {{Schuster}}, \citenamefont
  {{Tilanus}}, \citenamefont {{Torne}}, \citenamefont {{van Bemmel}},
  \citenamefont {{van Langevelde}}, \citenamefont {{Wex}}, \citenamefont
  {{Younsi}},\ and\ \citenamefont {{Zhidenko}}}]{Goddi17}%
  \BibitemOpen
  \bibfield  {author} {\bibinfo {author} {\bibfnamefont {C.}~\bibnamefont
  {{Goddi}}}, \bibinfo {author} {\bibfnamefont {H.}~\bibnamefont {{Falcke}}},
  \bibinfo {author} {\bibfnamefont {M.}~\bibnamefont {{Kramer}}}, \bibinfo
  {author} {\bibfnamefont {L.}~\bibnamefont {{Rezzolla}}}, \bibinfo {author}
  {\bibfnamefont {C.}~\bibnamefont {{Brinkerink}}}, \bibinfo {author}
  {\bibfnamefont {T.}~\bibnamefont {{Bronzwaer}}}, \bibinfo {author}
  {\bibfnamefont {J.~R.~J.}\ \bibnamefont {{Davelaar}}}, \bibinfo {author}
  {\bibfnamefont {R.}~\bibnamefont {{Deane}}}, \bibinfo {author} {\bibfnamefont
  {M.}~\bibnamefont {{de Laurentis}}}, \bibinfo {author} {\bibfnamefont
  {G.}~\bibnamefont {{Desvignes}}}, \bibinfo {author} {\bibfnamefont {R.~P.}\
  \bibnamefont {{Eatough}}}, \bibinfo {author} {\bibfnamefont {F.}~\bibnamefont
  {{Eisenhauer}}}, \bibinfo {author} {\bibfnamefont {R.}~\bibnamefont
  {{Fraga-Encinas}}}, \bibinfo {author} {\bibfnamefont {C.~M.}\ \bibnamefont
  {{Fromm}}}, \bibinfo {author} {\bibfnamefont {S.}~\bibnamefont
  {{Gillessen}}}, \bibinfo {author} {\bibfnamefont {A.}~\bibnamefont
  {{Grenzebach}}}, \bibinfo {author} {\bibfnamefont {S.}~\bibnamefont
  {{Issaoun}}}, \bibinfo {author} {\bibfnamefont {M.}~\bibnamefont
  {{Jan{\ss}en}}}, \bibinfo {author} {\bibfnamefont {R.}~\bibnamefont
  {{Konoplya}}}, \bibinfo {author} {\bibfnamefont {T.~P.}\ \bibnamefont
  {{Krichbaum}}}, \bibinfo {author} {\bibfnamefont {R.}~\bibnamefont
  {{Laing}}}, \bibinfo {author} {\bibfnamefont {K.}~\bibnamefont {{Liu}}},
  \bibinfo {author} {\bibfnamefont {R.-S.}\ \bibnamefont {{Lu}}}, \bibinfo
  {author} {\bibfnamefont {Y.}~\bibnamefont {{Mizuno}}}, \bibinfo {author}
  {\bibfnamefont {M.}~\bibnamefont {{Moscibrodzka}}}, \bibinfo {author}
  {\bibfnamefont {C.}~\bibnamefont {{M{\"u}ller}}}, \bibinfo {author}
  {\bibfnamefont {H.}~\bibnamefont {{Olivares}}}, \bibinfo {author}
  {\bibfnamefont {O.}~\bibnamefont {{Pfuhl}}}, \bibinfo {author} {\bibfnamefont
  {O.}~\bibnamefont {{Porth}}}, \bibinfo {author} {\bibfnamefont
  {F.}~\bibnamefont {{Roelofs}}}, \bibinfo {author} {\bibfnamefont
  {E.}~\bibnamefont {{Ros}}}, \bibinfo {author} {\bibfnamefont
  {K.}~\bibnamefont {{Schuster}}}, \bibinfo {author} {\bibfnamefont
  {R.}~\bibnamefont {{Tilanus}}}, \bibinfo {author} {\bibfnamefont
  {P.}~\bibnamefont {{Torne}}}, \bibinfo {author} {\bibfnamefont
  {I.}~\bibnamefont {{van Bemmel}}}, \bibinfo {author} {\bibfnamefont {H.~J.}\
  \bibnamefont {{van Langevelde}}}, \bibinfo {author} {\bibfnamefont
  {N.}~\bibnamefont {{Wex}}}, \bibinfo {author} {\bibfnamefont
  {Z.}~\bibnamefont {{Younsi}}}, \ and\ \bibinfo {author} {\bibfnamefont
  {A.}~\bibnamefont {{Zhidenko}}},\ }\href {\doibase 10.1142/S0218271817300014}
  {\bibfield  {journal} {\bibinfo  {journal} {International Journal of Modern
  Physics D}\ }\textbf {\bibinfo {volume} {26}},\ \bibinfo {eid} {1730001-239}
  (\bibinfo {year} {2017})},\ \Eprint {http://arxiv.org/abs/1606.08879}
  {arXiv:1606.08879 [astro-ph.HE]} \BibitemShut {NoStop}%
\bibitem [{\citenamefont {{Falcke}}(2017)}]{Falcke17}%
  \BibitemOpen
  \bibfield  {author} {\bibinfo {author} {\bibfnamefont {H.}~\bibnamefont
  {{Falcke}}},\ }in\ \href {\doibase 10.1088/1742-6596/942/1/012001} {\emph
  {\bibinfo {booktitle} {Journal of Physics Conference Series}}},\ \bibinfo
  {series} {Journal of Physics Conference Series}, Vol.\ \bibinfo {volume}
  {942}\ (\bibinfo {year} {2017})\ p.\ \bibinfo {pages} {012001},\ \Eprint
  {http://arxiv.org/abs/1801.03298} {arXiv:1801.03298 [astro-ph.HE]}
  \BibitemShut {NoStop}%
\end{thebibliography}%

\end{document}